\documentclass[aps, prc, twocolumn, tightenlines, letterpaper, amsmath, amssymb, preprintnumbers, superscriptaddress, floatfix, longbibliography, nofootinbib]{revtex4-2}
\usepackage[T1]{fontenc}
\usepackage{bm}
\usepackage{xspace}
\usepackage[dvipsnames]{xcolor}
\usepackage{graphicx}
\usepackage{tabularx}
\usepackage{enumitem}
\usepackage{dsfont}
\usepackage[normalem]{ulem}

\usepackage{amsmath, nccmath}
\usepackage{lipsum}
\usepackage{cancel}
\usepackage{bm}

\usepackage[hypertexnames=false]{hyperref}
\hypersetup{
    colorlinks=true,       
    linkcolor=blue,          
    citecolor=blue,        
    filecolor=blue,      
    urlcolor=blue           
}

\newcolumntype{Y}{>{\centering\arraybackslash}X}

\usepackage{mathtools}
\usepackage{orcidlink}

\DeclarePairedDelimiter\ket{\lvert}{\rangle}

\usepackage{stackengine}
\stackMath
\newcommand\tenq[2][1]{%
 \def\useanchorwidth{T}%
  \ifnum#1>1%
    \stackunder[0pt]{\tenq[\numexpr#1-1\relax]{#2}}{\scriptscriptstyle\sim}%
  \else%
    \stackunder[1pt]{#2}{\scriptscriptstyle\sim}%
  \fi%
}

\begin{document}

\begin{figure}
\vskip -1.cm
\leftline{\includegraphics[width=0.15\textwidth]{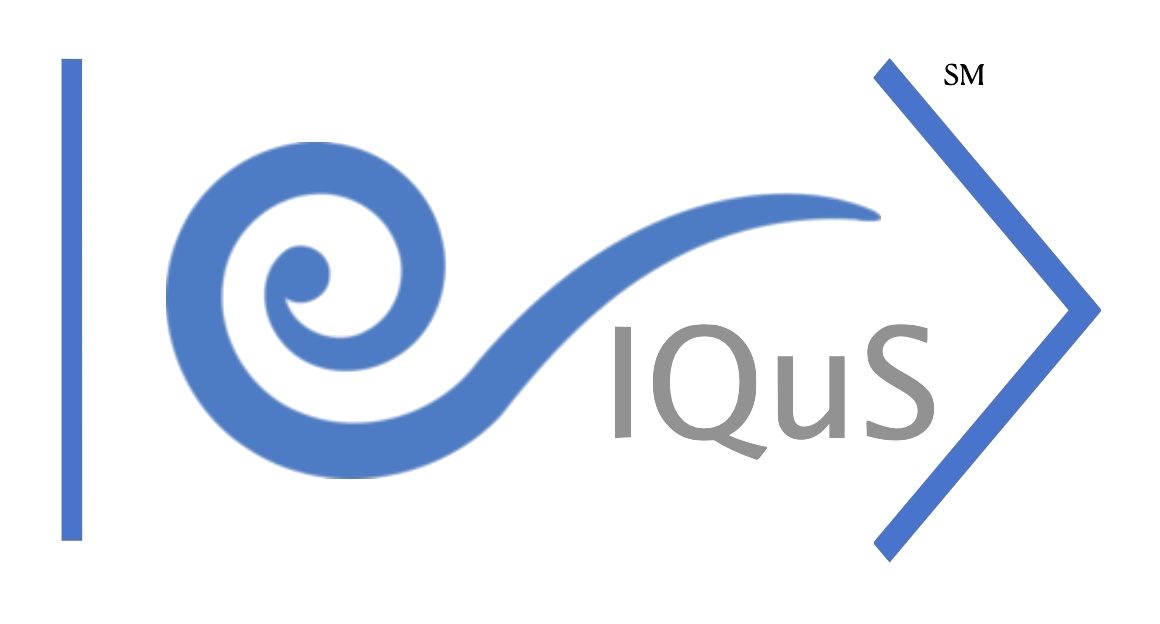}}
\vskip -2.cm
\end{figure}

\title{Quantum Simulations of SO(5) Many-Fermion Systems using Qudits}

\author{Marc Illa\,\orcidlink{0000-0003-3570-2849}}
\email{marcilla@uw.edu}
\affiliation{InQubator for Quantum Simulation (IQuS), Department of Physics, University of Washington, Seattle, WA 98195}

\author{Caroline E. P. Robin\,\orcidlink{0000-0001-5487-270X}}
\email{crobin@physik.uni-bielefeld.de}
\affiliation{Fakult\"at f\"ur Physik, Universit\"at Bielefeld, D-33615, Bielefeld, Germany}
\affiliation{GSI Helmholtzzentrum f\"ur Schwerionenforschung, Planckstra{\ss}e 1, 64291 Darmstadt, Germany}

\author{Martin J.~Savage\,\orcidlink{0000-0001-6502-7106}}
\email{mjs5@uw.edu}
\thanks{On leave from the Institute for Nuclear Theory.}
\affiliation{InQubator for Quantum Simulation (IQuS), Department of Physics, University of Washington, Seattle, WA 98195}

\preprint{IQuS@UW-21-049}
\date{\today}

\begin{abstract}
\begin{description}
\item[Background]
The structure and dynamics of many-body systems are the result of a delicate interplay between underlying interactions.
Fermionic pairing, for example, plays a central role in various physical systems, ranging from condensed matter to nuclear systems, where it
can lead to collective phenomena such as superconductivity and superfluidity. 
In atomic nuclei, the interplay between pairing and particle-hole interactions leads to a high degree of complexity and intricate entanglement structures.
Despite this apparent complexity, symmetries emerge and manifest themselves in observable regular patterns. These symmetries and their breakings have long been used to determine relevant degrees of freedom and simplify classical descriptions of many-body systems.
\item[Purpose] 
This work explores the potential utility of quantum computers with arrays of qudits  
in simulating interacting fermionic systems,
when the qudits can naturally map the relevant degrees of freedom determined by an underlying symmetry group.
\item[Method]
The Agassi model of fermions interacting via particle-hole and pairing interactions is based on an underlying $so(5)$ algebra. Such systems can intuitively be partitioned into pairs of modes with five basis states, which thus naturally map to arrays of $d=5$ qudits (qu5its).
Classical noiseless simulations of the time evolution of systems with up to twelve qu5its are performed, 
by implementing quantum circuits that are developed herein, using {\tt python} 
codes invoking Google's {\tt cirq} software.
The resource requirements of the qu5it circuits are analyzed and compared with
two different mappings to qubit systems, 
a physics-aware Jordan-Wigner mapping requiring four qubits per mode-pair 
and a state-to-state mapping requiring three qubits per mode-pair.
\item[Results]
While the dimensionality of Hilbert spaces in mappings to qu5it systems are less than those for 
the corresponding qubit systems,
the number of entangling operations, depending on the available hardware, can either be greater or smaller than for the physics-aware Jordan-Wigner mapping.
The state-to-state mapping, 
while having a smaller Hilbert space than Jordan-Wigner mappings, 
appears to be the least efficient in gate counts. Further,
a previously unknown
sign problem has been identified from Trotterization errors in time evolving high-energy excitations.
\item[Conclusions] 
There appear to be advantages in employing quantum computers with arrays of qudits to perform simulations of 
many-body dynamics that exploit the role of underlying symmetries,
specifically in lowering the required quantum resources and in reducing anticipated errors that take the simulation out of the physical space. If the necessary entangling gates are not directly supported by the hardware, physics-aware mappings to qubits may, however, be advantageous for other aspects.

\end{description}
\end{abstract}

\maketitle

\section{Introduction}
\label{sec:intro}
\noindent

Pairing of fermionic particles plays a special role in the structure and dynamics 
of quantum many-body systems of physical importance in our everyday lives.
This phenomenon provides the underlying mechanism responsible for superconductivity and superfluidity~\cite{PhysRev.104.1189,PhysRev.106.162,PhysRev.108.1175}, and 
it is crucial to the description of exotic materials~\cite{Nica_2021,PhysRevB.104.L060503,PhysRevResearch.3.043105,bahari2023new}, 
structure, reaction and decay properties of nuclei~\cite{RevModPhys.75.607},
and matter at extreme densities~\cite{Alford:2007xm}.
Such systems are typically challenging to simulate with classical computers
due to their fundamentally quantum nature, 
and the role that entanglement and quantum correlations play in their structure, 
see, for example, Ref.~\cite{Kraus_2009}.
From a nuclear physics perspective, 
there is growing interest in better understanding the entanglement structure of hadrons and nuclei~\cite{Ho:2015rga,Kharzeev:2017qzs,Baker:2017wtt,Beane:2018oxh,Beane:2018oxh,Beane:2019loz,Tu:2019ouv,Beane:2020wjl,Iskander:2020rkb,Kruppa:2020rfa,Beane:2021zvo,Klco:2020rga,Kharzeev:2021yyf,Kruppa:2021yqs,Klco:2021cxq,Robin:2020aeh,Low:2021ufv,PhysRevD.106.L031501,Klco:2021lap,Johnson_2023,Bai:2022hfv,Ehlers:2022oal,Tichai:2022bxr,Pazy:2022mmg,Bulgac:2022ygo,PhysRevA.105.062449,Bulgac:2022cjg,Jafarizadeh:2022kcq,Robin:2023fow,Gu:2023aoc,Sun:2023nzj},
with the potential to advance our ability to compute the properties and dynamics of 
nuclear many-body systems through entanglement re-organization~\cite{Robin:2020aeh,Robin:2023fow}.
This is driven, in part, by the emergence of early quantum computers and simulators, 
and their accessibility to domain scientists who, 
in co-design partnership with quantum information scientists,
are progressing toward 
understanding the capabilities, algorithms, and software required to 
establish a quantum advantage for scientific applications.
The capabilities that quantum computers offer beyond classical 
computing is the control of entanglement and coherence, 
the same elements that limit classical simulations of 
quantum many-body systems, including those involving fermionic pairing.

Various quantum algorithms aiming at solving fermionic pairing Hamiltonians have been proposed in symmetry-breaking plus restoration frameworks 
and particle-number preserving approaches at zero~\cite{PhysRevA.93.062301,Lacroix:2020nhy,RuizGuzman:2021qyj,RuizGuzman:2022enk,Khamoshi_2021} and finite temperature~\cite{Jiang:2022lub}.
In addition to pairing phenomena, nuclei also display important particle-hole correlations.
The Agassi model~\cite{AGASSI196849}, which we have chosen to study
in this work, 
is an exactly solvable model which, in part, reveals
the interplay between the pairing force and particle-hole interactions of the monopole-monopole type.
This model constitutes an extension of the Lipkin-Meshkov-Glick (LMG) model~\cite{LIPKIN1965188}, 
which was initially formulated in the context of nuclear physics studies.
The LMG model can be considered to be a system of spin-$\tfrac{1}{2}$ particles distributed on a simplex in the presence of a background magnetic field and with equal strength interactions that create and annihilate even numbers of particle-hole pairs.
The occupation of each mode is fixed, and particles can be promoted or demoted within a mode 
by the action of the interactions.
The LMG model has also been the subject of a large number of studies in condensed matter, as it was subsequently found to be relevant for describing the Josephson effect~\cite{PhysRevA.90.063617} and two-mode Bose-Einstein condensates~\cite{PhysRevA.70.062304}. 
Further, a more general class of LMG models are known to provide 
a means to prepare spin-squeezed states~\cite{PhysRevA.47.5138} 
through time-evolution away from a tensor-product state, 
of utility in quantum sensing~\cite{Ma_2011,Salvatori2014QuantumMI,Garbe_2022}.
These many features have motivated intense work aiming at investigating the entanglement properties of 
the LMG model, in particular the entanglement and quantum correlations of the ground state near the phase transitions, see, {\it e.g.}, Refs.~\cite{PhysRevA.69.054101,PhysRevA.69.022107,PhysRevA.71.064101,PhysRevA.77.052105,PhysRevA.100.062104,PhysRevA.103.032426,PhysRevA.104.032428,PhysRevA.105.062449}, as well as the dynamics of entanglement measures~\cite{PhysRevA.70.062304}.
Given that the building blocks of the model are SU(2) spins, 
it has been the focus of recent studies using currently available simulators and quantum computers 
to determine ground-state energies using the variational quantum eigensolver (VQE)~\cite{PhysRevC.104.024305,Chikaoka:2022kff}, or ADAPT-VQE~\cite{PhysRevC.105.064317} with different wave-function ans\"atze.
Recently, we have shown~\cite{Robin:2023fow} that the LMG model can be efficiently implemented throughout much of the Hilbert space via rearrangement of entanglement induced by global SU(2) rotations. The rotation angle can be learned in the VQE process of determining the ground state, with the Hamiltonian Learning-VQE  (HL-VQE) that we introduced in Ref.~\cite{Robin:2023fow}, for a given truncation of the model space. Exponential convergence  was demonstrated throughout large fractions of the Hilbert space~\cite{Robin:2023fow}.
Further, recent work used the quantum equation of motion (qEOM) algorithm~\cite{PhysRevResearch.2.043140}, a modification of VQE, to determine excited states in the LMG model~\cite{Hlatshwayo:2022yqt}.

The importance of pairing in nuclear physics, and other areas, 
motivated an extension to the LMG model by the inclusion of a pairing term, 
known as the Agassi model.
In this model, pairs of particles from mode pairs, can scatter by changing both their states within the mode-pair, or 
can scatter into pairs of unoccupied states in other mode pairs.
As a result, this model has a more interesting phase diagram than the LMG model. 
Specifically, with two dimensionless parameters, this model has a phase diagram that exhibits a spherical phase, 
a deformed phase and a superfluid Bardeen-Cooper-Schrieffer (BCS) phase~\cite{E_D_Davis_1986,Garc_a_Ramos_2018}.
The nature of the pairing interaction is such that the number of particles 
is preserved throughout time evolution,
and a five dimensional Hilbert space is required for an even number of particles in two modes, 
forming the fundamental representation of an SO(5) symmetry in the absence of interactions.
Generally speaking, models exhibiting an SO(5) symmetry are ubiquitous in quantum many-body physics, 
{\it e.g.}, Refs.~\cite{Klein:1981ra,PhysRevB.104.L060503}.

Having recognized the potential role of future quantum simulations in advancing our understanding of,
and establishing predictive capabilities for, such pairing models, the first pioneering works on simulating 
the dynamics of the Agassi model for small systems have recently been performed~\cite{P_rez_Fern_ndez_2022,Saiz:2022rof}.
In those works, a Jordan-Wigner (JW) mapping was used to simulate systems with 
two and four modes using one qubit per site to define 
its occupation, without appealing to their inherent SO(5) symmetries.
In the present work, we find advantages in making use of the SO(5) symmetry 
in mapping pairs of modes (4 sites) to five dimensional qudits (which we call qu5its),
to enable more efficient quantum simulations of such systems.
While qubits form the quantum registers of many digital quantum computers that are presently available, significant progress is being made toward qudit quantum registers, particularly using superconducting radio frequency (SRF) cavities~\cite{Alam:2022crs,2022APS..MARY34008R}, superconducting qudit devices~\cite{doi:10.1126/science.1173440,2018PhRvL.120m0503T,2016NatCo...712930J,Liu_2017,Blok:2020may,Wu:2020,Seifert:2023ous,seifert2022timeefficient}, nitrogen vacancy (NV) centers in diamond~\cite{Choi:2017,Keeffe:2019,Zhou:2023xnx} and most recently trapped ion quantum computers~\cite{PhysRevResearch.2.033128,Ringbauer:2021lhi,Hrmo:2022bvo,Sun:2023dsf}. 
There are well-known motivations for advancing beyond qubits to qudits, including in computation~\cite{PhysRevA.62.052309,Bullock_2005,Ralph:2007,Gedik:2015,10.1145/3307650.3322253,Baker:2020arXiv}, cryptography~\cite{Bru__2002,Cerf_2002}, and error correction~\cite{2021PhRvL.127a0504G,Lim:2023hkb,Omanakuttan:2023ftv}.
We expect that early arrays of qu5its will become available in the near future, at which point simulations of the Agassi model can be performed using qu5its.
In another arena, the utility of qudit systems in simulating (non-)Abelian lattice gauge theories, including lattice QCD, is being explored,  see, {\it e.g.}, Refs.~\cite{Ciavarella:2021nmj,Gustafson:2021qbt,Gustafson:2022xlj,Gonzalez-Cuadra:2022hxt,zache2023fermionqudit},
where the qudits provide a commensurate Hilbert space for each truncated gauge link space.
Further, the LMG model, which described the dynamics of systems of coupled spin-${1\over 2}$ particles or spins, 
simulated with arrays of qubits, has been generalized to systems of coupled spin-$d$ particles or spins simulated with arrays of qudits~\cite{Calixto_2021}.

We note that the Agassi model, or pairing-plus-monopole model, is one physical interpretation of the 
$so(5)$ ($sp(4)$) algebra. 
As is well known, there are at least two other interpretations of this algebra, which lead to other models of interest for nuclear physics, see, {\it e.g.}, Refs.~\cite{Klein:1981ra,RevModPhys.63.375,Sviratcheva:2001mg}. 
These include the charge-independent pairing model~\cite{Sviratcheva:2002jz,PhysRevLett.96.072503} describing like-particle and proton-neutron pairing coupled to total isospin $T=1$, and the vibration-rotation model in two dimensions. In each of these models the Hamiltonian is described by a subset of the generators of SO(5), and different sub-algebras correspond to the limits of the model.
Although in this work  we restrict ourselves to the Agassi model, the techniques developed here can also be applied to these other interpretations of the SO(5) symmetry group.

This paper is organized as follows. In Sec.~\ref{sec:agassi}, the main aspects of the Agassi model are reviewed.
In Sec.~\ref{sec:qu5its},
quantum circuits for preparing initial states on a qu5it register, 
and for a subsequent Trotterized time evolution~\cite{Trotter:1959a}, are constructed. 
Results obtained from classical simulations of the time evolution of systems up to 24 modes, 
using {\tt Mathematica}, {\tt Julia}, and {\tt python} 
code built on {\tt cirq}'s qudit capabilities, are presented.
In Sec.~\ref{sec:QC2}, the complexity of the quantum circuits 
for qu5its are compared to those of circuits using mappings onto qubits.

\section{The Agassi Model: Lipkin-Meshkov-Glick Hamiltonian, Pairing Interactions, and SO(5) Symmetry.}
\label{sec:agassi}
\noindent
In this section, we review the Agassi model, starting from the limit without pairing, which constitutes the LMG model.
The low-lying energy spectra of the Agassi model are examined for 
selected sets of Hamiltonian couplings, given in Table~\ref{tab:EvalsEVGsets},
to establish the inputs for quantum simulations with 
both qubits and qu5its.

\subsection{The Lipkin-Meshkov-Glick Model}
\label{sec:LGM_ex}
\begin{figure}[!tb]
\centering{\includegraphics[width=\columnwidth] {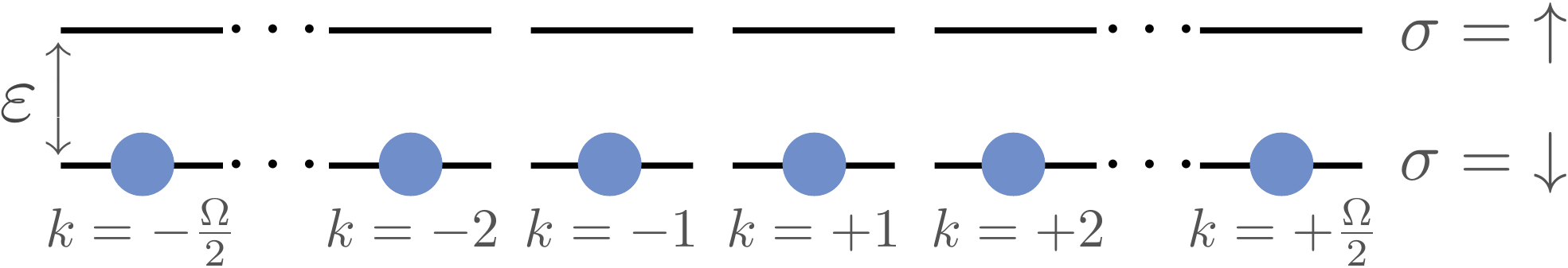}} 
\caption{The lowest-energy non-interacting (0p-0h) configuration of a system of $\Omega$ particles in the LMG model.
}
\label{fig:Lipkin}
\end{figure}

\noindent
In its original formulation, the LMG model~\cite{LIPKIN1965188} 
describes a system of $N=\Omega$ fermions, each
distributed on two levels with energy $\pm \varepsilon/2$, as shown in Fig.~\ref{fig:Lipkin}.
We denote the single-particle states (``sites'') by $(k,\sigma)$, where $\sigma =\; \uparrow, \downarrow$ refers to the upper and lower level, respectively, and $k=\pm 1, \pm 2,\dots, \pm \Omega/2$ denotes the modes. 
The Hamiltonian\footnote{The LMG Hamiltonian in principle also contains a ``swap'' term that can lower a particle from $(k,\uparrow)$ to $(k,\downarrow)$ while exciting another from $(k',\downarrow)$ to $(k',\uparrow)$.
We do not include this term as it does not appear in the Agassi model~\cite{AGASSI196849}.} governing the system is
%
%
%
\begin{align}
\hat H \ & = \ 
\frac{\varepsilon}{2} 
\sum_{k} \left( 
\hat c_{k,\uparrow}^\dagger \hat c_{k,\uparrow}
-
\hat c_{k,\downarrow}^\dagger \hat c_{k,\downarrow}
\right)
\nonumber \\
    & -
\frac{V}{2} \sum_{k, k'} 
\left(
\hat c^\dagger_{k,\uparrow} \hat c^\dagger_{k',\uparrow} 
\hat c_{k', \downarrow} \hat c_{k, \downarrow}
+
\hat c^\dagger_{k,\downarrow} \hat c^\dagger_{k',\downarrow} 
\hat c_{k',\uparrow} \hat c_{k,\uparrow}
\right) \ ,
\label{eq:H_unrot}
\end{align}
where the operators $\hat c^\dagger_{k,\sigma}$ and $\hat  c_{k,\sigma}$ 
create and destroy a particle in level $\sigma$ of state $k$, respectively.
The first term accounts for the single-particle energies, while the interaction term scatters 
 particles from the lower to the upper level, or {\it vice versa}, within the same modes.
Each of the particles interact with equal strength with the 
others~\cite{PhysRevA.69.022107,PhysRevA.70.062304}.
This is illustrated in Fig.~\ref{fig:Lipkin}, 
which shows the lowest-energy, or ``zero-particle-zero-hole (0p-0h)'', 
non-interacting configuration of the system.

The LMG model can be mapped into a system of $N$ interacting spins in a background magnetic field by introducing the collective spin operators
\begin{align}
\hat{S}_z & \ = \  
\frac{1}{2} 
\sum_{k=-\Omega/2}^{\Omega/2} \left( 
\hat c_{k,\uparrow}^\dagger \hat c_{k,\uparrow}
-
\hat c_{k,\downarrow}^\dagger \hat c_{k,\downarrow}
\right)
\ = \ 
\sum_{k=-\Omega/2}^{\Omega/2}  \hat S_{k,z} \ ,
\nonumber \\
\hat{S}_+ & \ = \ \sum_{k=-\Omega/2}^{\Omega/2}   \hat c_{k,\uparrow}^\dagger \hat c_{k,\downarrow}
\ =\ 
\sum_{k=-\Omega/2}^{\Omega/2}  \hat S_{k,+} \ ,
\nonumber \\
\hat{S}_- & \ = \ (\hat{S}_+)^\dag \ ,
\label{eq:Ss}
\end{align}
transforming in the adjoint representation of SU(2).
The Hamiltonian in Eq.~(\ref{eq:H_unrot}) can then be written as
\begin{equation}
\hat H \ = \ \varepsilon \hat{S}_z - \frac{V}{2} \left( \hat{S}_+^2 + \hat{S}_-^2 \right) \ .
\label{eq:H_spin}
\end{equation}
%

\subsection{The Agassi Model and SO(5) Symmetry}
\label{sec:Agassi_ex}
\noindent
The Agassi model~\cite{AGASSI196849} adds a pairing interaction to the 
previous LMG model, with a Hamiltonian of the form
\begin{equation}
\hat{H} \ = \
 \varepsilon\ \hat S_z
\ - \ \frac{V}{2} \left( \hat{S}_+^2 + \hat{S}_-^2 \right) 
\ -\ 
g \hat B^\dagger \hat B
\ .
\label{eq:agassiDEF}
\end{equation}
It is convenient to write the collective operators in Eq.~\eqref{eq:agassiDEF} 
in terms of operators acting on pairs of modes $(k,-k)$,
\begin{equation}
\hat S_\alpha \ = \ \sum_{k=1}^{\Omega/2} \hat T_{k,\alpha}
\ ,\ \ 
\hat B \ =\ \sum_{k=1}^{\Omega/2} (\hat b_{k,\uparrow} + \hat b_{k,\downarrow}) \ ,
\end{equation}
with
\begin{align}
\hat T_{k,+} & \ = \
\hat c_{k,\uparrow}^\dagger \hat c_{k,\downarrow} + 
\hat c_{-k,\uparrow}^\dagger \hat c_{-k,\downarrow}
\ ,\ \
\hat T_{k,-} \ = \ \hat T_{k,+}^\dagger
\ ,
\nonumber\\
\hat T_{k,z} & \ = \
\frac{1}{2}( 
\hat c_{k,\uparrow}^\dagger \hat c_{k,\uparrow}
+ \hat c_{-k,\uparrow}^\dagger \hat c_{-k,\uparrow}
- \hat c_{k,\downarrow}^\dagger \hat c_{k,\downarrow}
- \hat c_{-k,\downarrow}^\dagger \hat c_{-k,\downarrow} ) \ ,
\nonumber\\
\hat b_{k, \uparrow} & \ = \
\hat c_{-k,\uparrow} \hat c_{k,\uparrow}
 \ ,\ \
\hat b_{k, \downarrow} \ = \ \hat c_{-k,\downarrow} \hat c_{k,\downarrow} \ ,
\nonumber\\
\hat b_{k,z} & \ = \
\frac{1}{\sqrt{2}} (
\hat c_{-k,\uparrow} \hat c_{k,\downarrow}
+ \hat c_{-k,\downarrow} \hat c_{k,\uparrow}
 ) \ ,
\nonumber \\
\hat N_{k} & \ = \
\hat c_{k,\uparrow}^\dagger \hat c_{k,\uparrow}
+ \hat c_{-k,\uparrow}^\dagger \hat c_{-k,\uparrow}
+ \hat c_{k,\downarrow}^\dagger \hat c_{k,\downarrow}
+ \hat c_{-k,\downarrow}^\dagger \hat c_{-k,\downarrow}
\ ,
\label{eq:SO5gensops}
\end{align}
which, along with $\hat b_{k,\uparrow}^\dagger$, $\hat b_{k,\downarrow}^\dagger$ 
and $\hat b_{k,z}^\dagger$, constitute a set of 10 generators of SO(5) 
(see Appendix~\ref{app:repSO5}).  
The additional pairing term, with coupling constant $g$, furnishes interactions such as 
$\hat c_{k,\downarrow} \hat c_{-k,\downarrow}
\left(\hat c_{q,\downarrow} \hat c_{-q,\downarrow}
\right)^\dagger$
and
$\hat c_{k,\uparrow} \hat c_{-k,\uparrow}
\left(\hat c_{k',\downarrow} \hat c_{-k',\downarrow}
\right)^\dagger$, which, unlike the LMG model, can change the occupation number of a given mode.
As in the LMG model, the Agassi model has all modes interacting with equal strength, 
and hence there is no intrinsic concept of distance.

\begin{figure}[!tb]
\centering{\includegraphics[width=\columnwidth] {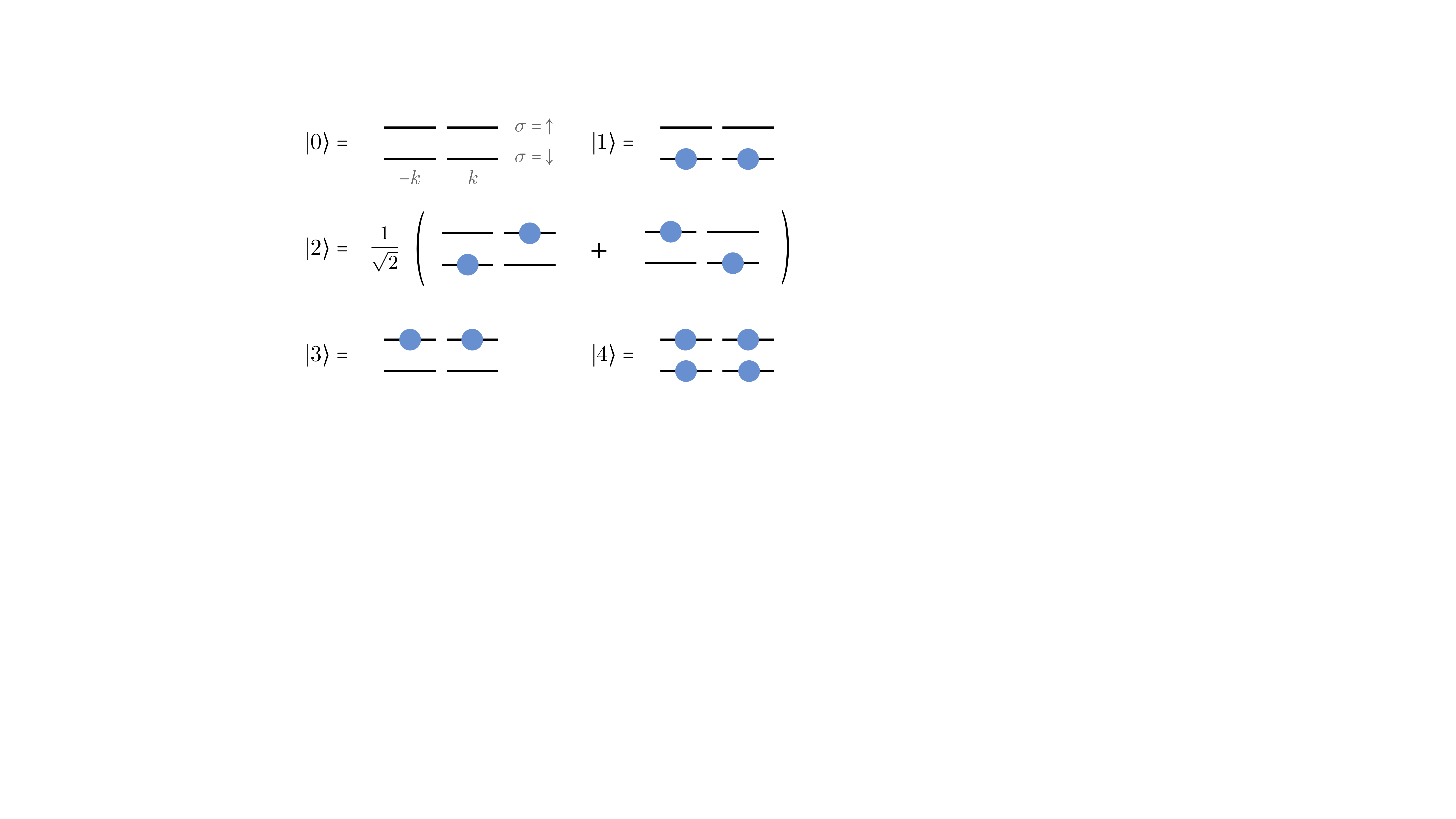}} 
\caption{Basis states for a wavefunction of one pair of modes $(k,-k)$, as defined in Eq.~\eqref{eq:5basis}.} 
\label{fig:config}
\end{figure}

A five dimensional set of basis states supporting a wavefunction on one pair 
of modes $(k,-k)$, can be defined in terms of single particle occupation numbers as
\begin{align}
|n_{k \downarrow} & ,  n_{k \uparrow} , n_{-k \downarrow} , n_{-k \uparrow} \rangle \nonumber \\
& \in \ 
\{\ 
|0000\rangle 
\ ,\  
|1010\rangle \  ,\  
\frac{1}{\sqrt{2}}\left( 
|0110\rangle  + |1001\rangle 
  \right) ,\ \nonumber \\
  & \ \ \
|0101\rangle\ ,\  
|1111\rangle \ 
\}
\nonumber \\ 
& = \
\{\ 
|0\rangle , |1\rangle , |2\rangle , |3\rangle , |4\rangle \ \}
\ ,
\label{eq:5basis}
\end{align}
which form a fundamental representation of SO(5) generated by the operators in Eq.~\eqref{eq:SO5gensops}.
The basis states in Eq.~\eqref{eq:5basis} are schematically represented in Fig.~\ref{fig:config}. Each of them has a well defined
particle number $(0,2,2,2,4)$,
pair number $(0,1,0,1,2)$, 
third component of spin $(0,-1,0,1,0)$, 
and parity $(1,1,-1,1,1)$.
\\ \\ \indent
The phase diagram of the Agassi model is interesting, and has been previously investigated via mean-field calculations of the ground state in the $N=\Omega$ sector, see, for example, Refs.~\cite{E_D_Davis_1986,Garc_a_Ramos_2018}.
The hierarchies in the parameters $\varepsilon$, $V$ and $g$ define the three phases of the system: 
a {\it symmetric phase}, for which a spherical mean-field description is adapted, 
and two symmetry-broken phases, a {\it deformed phase} (parity-broken) and 
a {\it superfluid BCS phase}. 
It has been found helpful to form
dimensionless parameters when discussing the structure of the theory,
\begin{align}
\bar{v} & \ = \ \frac{(\Omega - 1) V}{\varepsilon} \ , \nonumber \\
\bar{g} & \ = \ \frac{(\Omega - 1) g}{\varepsilon} \ , \quad \bar{g}_0 \ = \ \bar{g} + \frac{V}{\varepsilon} \ .
\end{align}
In the large-$\Omega$ limit, 
the mean-field analysis has
shown~\cite{E_D_Davis_1986} that 
for $\bar{v} < 1$ and $\bar{g}_0 < 1$,
the system is in the symmetric  phase (spherical non-superfluid),
and the ground state gapped below the excited states.
For $\bar{v} > 1$ and $\bar{g}_0 < \bar{v}$, 
the system is in a deformed phase, 
with a doubly degenerate ground state of mixed parity, which are
gapped to the excited states.
Finally, 
for $\bar{v} < \bar{g}_0$, $\bar{g}_0 > 1$, 
the system is in a BCS phase, 
with degenerate states (having different particle numbers) that are
gapped to excited states.

For the small- and modest-$\Omega$ systems that are solved exactly in this work, 
the structure of the systems are found to approach those of the large-$\Omega$ limit, 
modified by finite-size effects.
For $\bar{v}, \bar{g}_0 \lesssim 1$, as is the case in the large-$\Omega$ limit, 
the system has an isolated even-parity ground state that is gapped  below  excited states,
with the wavefunction dominated by a single configuration with all $N$ particles in the lower spin state, $|1\rangle$.
For $\bar{v} \gtrsim \bar{g}_0 \gtrsim 1$, there are two low-lying states of opposite parity that are gapped to higher excited states, with wavefunctions dominated by configurations where the particles occupy states
$|1\rangle , |2\rangle , |3\rangle$.
Finally, for $\bar{v} \lesssim \bar{g}_0$, $\bar{g}_0 \gtrsim 1$, 
there are two low-lying states of even parity that are gapped to excited states,
with wavefunctions dominated by configurations with different particle numbers, where particles occupy states $\ket{0}$,
$\ket{1}$, $\ket{3}$ and $\ket{4}$.
This phase structure has also been investigated in terms of the quantum discord in Ref.~\cite{PhysRevA.103.032426}.
\\ \\ \indent
In the quantum simulations performed in the next sections, we select, for demonstrative purposes, the following numerical values of the couplings:
\begin{eqnarray}
(\varepsilon , V , g) & : & {\rm set}\text{-}0  = (1.0,0.0,0.0)\ ,
\nonumber\\
& : & {\rm set}\text{-}1  = (1.0,0.5,0.5)\ ,
\nonumber\\
& : & {\rm set}\text{-}2  = (1.0,1.5,0.5)\ ,
\nonumber\\
& : & {\rm set}\text{-}3  = (1.0,0.5,1.5)\ ,
\nonumber\\
& : & {\rm set}\text{-}4  = (1.0,1.5,1.5)
\ .
\label{eq:setdefs}
\end{eqnarray}
Set-0 corresponds the non-interacting symmetric case.
Set-1, set-3 and set-4, lead to a superfluid phase for $N=\Omega > 2$, while set-2 corresponds to a deformed phase for $N=\Omega > 2$.
As mentioned in, {\it e.g.}, Ref.~\cite{P_rez_Fern_ndez_2022}, the case $\Omega=2$ has only one symmetry-broken phase (see also Sec.~\ref{sec:one_qu5it}). In that case, the system is in the symmetry-broken phase with parameters set-2, set-3 and set-4, and at the critical point with set-1. 
In Table~\ref{tab:EvalsEVGsets} of Appendix~\ref{app:spectrum} we show the
lowest-lying energies of systems 
(specifically, the average energy per mode, ${\cal E}_i = E_i/\Omega$)
obtained with the aforementioned sets of couplings, for $\Omega=2,4,6,8$, 
in different particle-number sectors.

\section{Quantum Simulations using Qu5its (d=5 Qudits)}
\label{sec:qu5its}
\noindent
In the first pioneering quantum simulation of the Agassi model~\cite{P_rez_Fern_ndez_2022}, 
a JW mapping to qubits was employed to define the occupation of 
each site in the Hilbert space of a four-site (one-mode-pair) system.
This simulation was able to describe a two particle system evolving throughout the $2^4$ dimensional 
Hilbert space, and was later expanded to four particles using eight qubits~\cite{Saiz:2022rof}. 
With such a mapping, systems with $2 \Omega$ sites, {\it i.e.}, $\Omega/2$ mode-pairs, would be embedded into a $2^{2\Omega}$ dimensional space.

The basis for each pair of modes in Eq.~\eqref{eq:5basis}  
naturally embeds into a five-dimensional 
($d=5$) qudit --- a ``qu5it''. 
With the expectation that qudit quantum computers become accessible in the near future, particularly, 
those that support qu5its, we develop the quantum circuits required for qu5it simulations 
and compare with equivalent circuits for qubit systems.
In developing quantum circuits for state preparation and time evolution,
it is most convenient to work with Givens rotations between states within a qu5it, 
and tensor-products of Givens rotations for entangling operations between qu5its.
As the Hamiltonian in Eq.~\eqref{eq:agassiDEF} is quadratic in operations between SO(5) irreps, 
establishing the quantum circuits for systems of one- and two-mode-pair (qu5its)
systems is sufficient for generalizing to systems with arbitrary numbers of mode-pairs.
 \\ \\ \indent
 Consequently, in this section we present qu5it circuits for the preparation of arbitrary initial states and 
 implementation of their time evolution, for one- and two-mode-pair systems. 
 The generalization to higher number of mode pairs is discussed and implemented onto simulators of arrays of qu5its, 
 with comparison to exact time evolution. 
For the numerical simulations performed using a qu5it simulator, 
as well as  exact matrix exponentiations,
two tensor-product initial states (that can be straightforwardly prepared) 
are employed. These are
\begin{align}
|\psi(0)\rangle_A &\ = \ |1\rangle^{\otimes {\Omega/2}}
\ ,
\nonumber\\
|\psi(0)\rangle_B &\ = \ |4\rangle^{\otimes {\Omega/4}} \otimes |0\rangle^{\otimes {\Omega/4}}
\ .
\label{eq:psi0s}
\end{align}
Both of these states correspond to half-filled configurations, meaning that the number of particles $N$ equals the number of modes $\Omega$, 
which equals twice the number of qu5its $N_{q5}$.
State $|\psi(0)\rangle_A$  is the ground state of the non-interacting systems, where all of the particles are in the 
spin-$\downarrow$ single-particle state.
State $|\psi(0)\rangle_B$, on the other hand,
is the state in which all of the particles are concentrated into half of the 
mode pairs only, 
{\it i.e.}, they occupy single-particle states $(k,\sigma)$ with $|k|=\{1,\ldots,\Omega/4\}$ and $\sigma =\{ \uparrow,\downarrow\}$.
$|\psi(0)\rangle_B$ energetically lies far above the non-interacting ground state.

\subsection{One Mode-Pair (One Qu5it)} \label{sec:one_qu5it}
\noindent
The Hamiltonian for a single mode pair ($\Omega=2$)
in the basis in Eq.~\eqref{eq:5basis} is given by
\begin{equation}
\hat{H}_1 
\ = \
\varepsilon  
\hat{T}_z
- (V+g)   \hat{\cal X}_{13} - g  \hat{N}_{\rm pairs}
\ ,
\label{eq:onemodepairH}
\end{equation}
where $\hat{T}_z$ is
\begin{align}
T_{z} = &
\begin{pmatrix}
0&0&0&0&0 \\
0&-1&0&0&0 \\
0&0&0&0&0 \\
0&0&0&1&0 \\
0&0&0&0&0
\end{pmatrix} \ ,
\end{align}
$\hat{N}_{\rm pairs}$ is the operator 
that counts the number of pairs 
$\left[(1,\sigma),(-1,\sigma)\right]$,
\begin{align}
N_{\rm pairs} = &
\begin{pmatrix}
0&0&0&0&0 \\
0&1&0&0&0 \\
0&0&0&0&0 \\
0&0&0&1&0 \\
0&0&0&0&2
\end{pmatrix} \ ,
\end{align}
and the Givens operator $\hat{\cal X}_{13}$ is 
\begin{equation}
{\cal X}_{13} \ = \
\begin{pmatrix}
0&0&0&0&0 \\
0&0&0&1&0 \\
0&0&0&0&0 \\
0&1&0&0&0 \\
0&0&0&0&0
\end{pmatrix}\ .
\end{equation}
The indices of ${\cal X}_{ij}$ are defined by the labels of the states in Eq.~\eqref{eq:5basis},
$i,j,\in \{ 0,1,2,3,4 \}$.

\subsubsection{State Preparation of One Qu5it}
\noindent
Preparing an initial state on a single qu5it is straightforward.  
Arbitrary transformations among the states of a qu5it can be accomplished with SU(5) unitary operations, with the associated 24 generators. For the SO(5) transformations relevant for the LMG and Agassi models, this 
number is reduced to 10.  
However, to prepare a real wavefunction from a given initial state of a qu5it, only 4 parameters are required 
to establish the 5 real numbers, one for each state in the Hilbert space, subject to the probability constraint.
Therefore, the transformation matrices produced by the Givens operators 
$\hat {\cal X}_{01}$, $\hat {\cal X}_{12}$, $\hat {\cal X}_{23}$, $\hat {\cal X}_{34}$, 
given in Appendix~\ref{app:repGIVENS},
or any other spanning set,
are sufficient to prepare any real initial state of a qu5it from, say, $|1\rangle$,
\begin{equation}
    |\psi\rangle \ = \
    e^{-i \theta_3 \hat {\cal X}_{34} }\ 
    e^{-i \theta_2 \hat {\cal X}_{23} }\ 
    e^{-i \theta_1 \hat {\cal X}_{12} }\ 
    e^{-i \theta_0 \hat {\cal X}_{01} }\ 
    |1\rangle
    \ .
    \label{eq:1qstateprep}
\end{equation}
This same wavefunction parameterization of the single qu5it can be used in finding the ground-state wavefunction, 
using, for instance, VQE~\cite{Peruzzo:2014,McClean:2016} or variants thereof~\cite{Higgott:2019,McArdle:2019,Yuan:2019,Grimsley:2019NatCo,Tang:2021PRXQ,Stokes:2020Quant,Koczor:2019arXiv,PhysRevResearch.2.043140,Gomes:2021ckn},
by minimizing the ground-state energy with respect to the $\theta_{0,1,2,3}$.
The expression in Eq.~\eqref{eq:1qstateprep} is a general form for initializing a state, and it can be simplified by utilizing symmetries.  The number of particles is preserved in the time evolution of the system, and as such, if the device can be prepared in a state of the target particle number, then the number of required rotation angles can be reduced.
For example, the states $\ket{0}$ and $\ket{4}$ are the ground states of the $N=0$ and $N=4$ sectors of one qu5it.

\subsubsection{Time Evolution of One Qu5it}
\begin{figure}[!tb]
\centering{\includegraphics[width=\columnwidth]{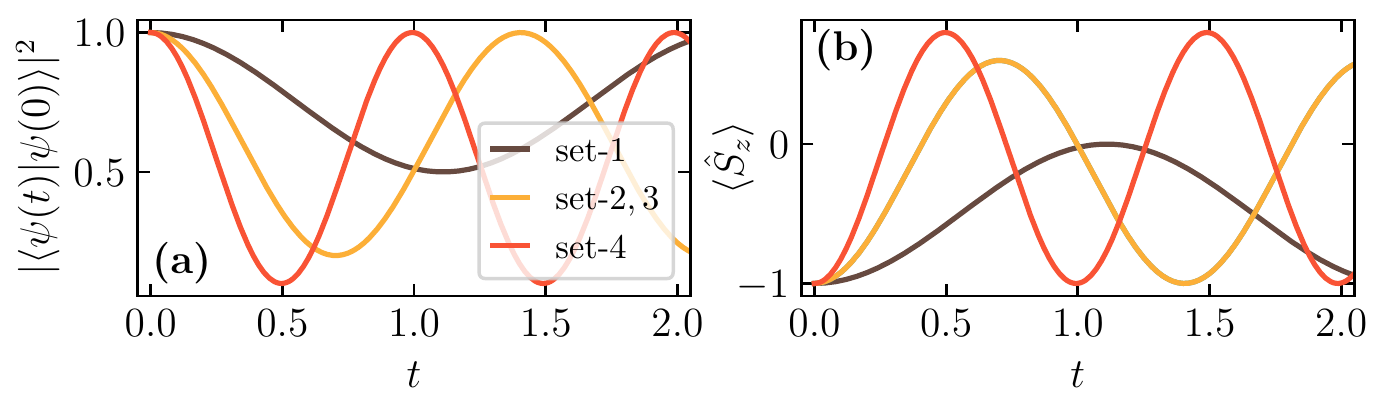}} 
\caption{The exact time dependence of \textbf{(a)} $|\langle \psi (t)|\psi(0)\rangle|^2$ and 
\textbf{(b)} $\langle \hat S_z \rangle$ for a system with $\Omega=2$ starting in the initial state $|\psi(0)\rangle_A=|1\rangle$,
for four sets of couplings, set-1 to set-4, given in Eq.~(\ref{eq:setdefs}).
}
\label{fig:N2all}
\end{figure}

\noindent
As is well known, the time evolution of a state prepared on a qu5it can be evolved forward in time by application of the evolution operator $\hat U(t) = e^{-i t \hat H}$.
For the system we are considering, the Hamiltonian is given in Eq.~\eqref{eq:onemodepairH},
and involves contributions from two diagonal phase operators and one Givens operator, $\hat {\cal X}_{13}$.
This evolution operator can be determined exactly, 
\begin{equation}
    \scalebox{0.8}{$
    U(t) \ = \
    \begin{pmatrix}
    1 & 0 & 0 & 0 & 0 \\
    0 & e^{i g t} \left(   a(t) + i  b(t) \right) & 0 & i e^{i g t} c(t) & 0\\
    0 & 0 & 1 & 0 & 0 \\
    0 & i e^{i g t} c(t) & 0 & e^{i g t} \left(   a(t) - i b(t) \right) & 0 \\
    0&0&0&0& e^{i 2 g t}
    \end{pmatrix}$}
    \ ,
    \label{eq:EVOL1}
\end{equation}
with
\begin{align}
& a(t)  \ =\   \cos \alpha t \ ,\ 
    b(t)  = \beta \sin \alpha t 
    \ ,\ 
    c(t)  = \gamma \sin \alpha t\ ,
    \nonumber\\
&\alpha \ =\  \sqrt{\varepsilon^2 + (g+V)^2} \ ,\ 
    \beta \ =\ \frac{\varepsilon}{\alpha}\ ,\ 
    \gamma \ =\ \frac{g+V}{\alpha}
    \ .
\end{align}
The exact time evolution of $\langle \hat S_z \rangle$ 
and the amplitude for survival of the ground state 
of one qu5it, prepared initially in state $|1\rangle$ 
(corresponding to $\ket{\psi(0)}_A$ in Eq.~\eqref{eq:psi0s}),
and obtained with the parameter sets defined in Eq.~\eqref{eq:setdefs},
are shown in Fig.~\ref{fig:N2all}. 
The Hamiltonian is such that state $|1\rangle$ only couples to state $|3\rangle$
under time evolution, rendering the number of pairs to be time independent in the $\Omega=2$ system.
It is clear from Eqs.~\eqref{eq:onemodepairH} and \eqref{eq:EVOL1} 
that the structure and dynamics of systems on a single mode-pair are governed by 
one dimensionless parameter $(g+V)/\varepsilon$. 
Such systems thus only present two phases: a symmetric phase for $g+V < \varepsilon$ and a symmetry-broken phase for $g+V > \varepsilon$. The sets of parameters set-2 and set-3 are thus equivalent, and the set-1 corresponds to the critical point of phase transition.

Trotterization~\cite{Trotter:1959a} of the evolution operator does not exactly recover the form 
given in Eq.~\eqref{eq:EVOL1}
because, although
$[ \hat {\cal X}_{13} , \hat N_{\rm pairs} ] = [\hat{T}_z , \hat N_{\rm pairs}] = 0$,
the operators $\hat {\cal X}_{13}$ and  $\hat{T}_z $ do not commute. More precisely, 
\begin{align}
& \left[ 
\varepsilon \hat{T}_z - g  \hat{N}_{\rm pairs} , -(g+V)  \hat{\cal X}_{13} \right] 
\ =\  
i 2 (g+V)\varepsilon\   \hat{\cal Y}_{13} 
\ ,
\nonumber\\
& {\cal Y}_{13} 
\ =\ 
i
\begin{pmatrix}
0&0&0&0&0 \\
0&0&0&-1&0 \\
0&0&0&0&0 \\
0&1&0&0&0 \\
0&0&0&0&0
\end{pmatrix}
\ .
\label{eq:commut_1}
\end{align}
As such, there will be Trotter errors in digitized time evolution performed using quantum circuits,
for which a leading-order (LO) Trotterization can take the form,
\begin{eqnarray}
    &&\hat{U} (t) \simeq \left[ \hat U(\Delta t) \right]^{n_{Trot}} \; , \nonumber \\
    &&\hat U(\Delta t)  \simeq  e^{ i \Delta t (g+V) \hat {\cal X}_{13} } \ e^{ -i \Delta t \varepsilon \hat T_z} \ e^{ i \Delta t g \hat N_{\rm pairs}} \; ,
\end{eqnarray}
where  $\Delta t$ is the Trotter time step and $n_{Trot} = t /\Delta t$ is the number of  steps.

\subsection{Two Mode-Pairs (Two Qu5its)}
\noindent
Systems with two or more mode-pairs allow for the possibility of moving pairs of particles between modes, and for the possibility of breaking pairs (through the monopole interaction $V$) while conserving the total number of particles. This introduces a richness that is not present for a single mode pair.
Because there  is an even number of fermions per mode-pair, 
mappings to systems with more that one qu5it are effectively bosonic, 
without potential negative signs associated with operators acting on separated qu5its.
For two mode pairs ($\Omega=4$), the Hamiltonian in Eq.~\eqref{eq:agassiDEF} reduces to 
\begin{align}
 & \hat H_2
\ = \ \varepsilon \left( \hat T_{1,z} +  \hat T_{2,z}  \right)
 \nonumber \\
& - V \left( 
\hat T^2_{1,+} + \hat T^2_{2,+} +\hat T^2_{1,-} + \hat T^2_{2,-}\right.  \nonumber\\ 
 & \left. \quad \quad \quad + \ \{ \hat T_{1,+} , \hat T_{2,+} \} + \{ \hat T_{1,-} , \hat T_{2,-} \} \right)
\nonumber\\
& -
g\left[
(  \hat b_{1,\uparrow} + \hat b_{1,\downarrow})^\dagger 
(  \hat b_{1,\uparrow} + \hat b_{1,\downarrow}) +
(  \hat b_{2,\uparrow} + \hat b_{2,\downarrow})^\dagger 
(  \hat b_{2,\uparrow} + \hat b_{2,\downarrow}) \right.\nonumber \\
&+ \left.
(  \hat b_{1,\uparrow} + \hat b_{1,\downarrow})^\dagger 
(  \hat b_{2,\uparrow} + \hat b_{2,\downarrow}) + 
(  \hat b_{2,\uparrow} + \hat b_{2,\downarrow})^\dagger 
(  \hat b_{1,\uparrow} + \hat b_{1,\downarrow})
\right]
\ ,
\label{eq:twomodepairH}
\end{align}
which, after reduction to Givens operators, becomes\footnote{
The notation of the summations is such that, for example,
\begin{equation}
\sum_{r, s \in \{(12),(23)\} }  f_{r,s}
\ = \
f_{12,12} + f_{12,23} + f_{23,12} + f_{23,23} 
\ .
\end{equation}
}
\begin{align}
\hat H_2 
= & 
\left[\varepsilon \hat T_z - (V+g)  \hat {\cal X}_{13} - g \hat N_{\rm pairs}\right] \otimes \hat I_5
\nonumber\\
& + 
\hat I_5 \otimes \left[\varepsilon \hat T_z - (V+g)  \hat {\cal X}_{13} - g \hat N_{\rm pairs}
\right]
 \nonumber\\
 & -
 V \sum_{r, s \in \{(12),(23)\} } 
 \left( \hat {\cal X}_r\otimes \hat {\cal X}_s - \hat {\cal Y}_r\otimes \hat {\cal Y}_s\right)
\nonumber\\
 & -
 \frac{g}{2} 
 \sum_{\substack{r, s \in \{(01),(03), \\ \ \ -(14), -(34) \} }} 
  \left( \hat {\cal X}_r\otimes \hat {\cal X}_s + \hat {\cal Y}_r\otimes \hat  {\cal Y}_s\right)
\ ,
\label{eq:twomodepairHGIVENS}
\end{align}
where the negative signs in the second summation indices mean that the operator with the corresponding index has an additional minus sign.
The energy densities of
the two-qu5it systems, 
for the parameter sets defined in Eq.~\eqref{eq:setdefs},
are given in Table~\ref{tab:EvalsEVGsets}.
While the ground state energy density is somewhat larger for $N=4$ compared with $N=2$, 
the energy densities of the excited states are substantially different.

The extreme case $V \gg g$ approaches the LMG model, in which the particles are distributed on mode pairs in states
$\{ |1\rangle , |2\rangle , |3\rangle \}$. On the contrary, the case $V \ll g$ approaches the result of the two-level pairing model with
particles being distributed among the 
states $\{ |0\rangle , |1\rangle , |3\rangle , |4\rangle \}$ (states with paired spins).

\subsubsection{State Preparation on Two Qu5its}
\begin{figure}[!t]
	\centering
    \includegraphics[width=0.5\columnwidth]{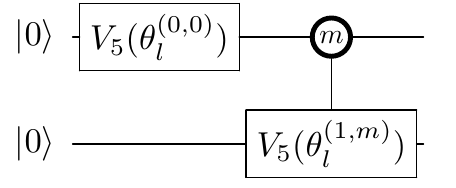}
		\caption{A quantum circuit that prepares
 an arbitrary real wavefunction on two qu5its from an initial state 
 $|0\rangle\otimes |0\rangle$.
 Four angles, $\theta^{(0,0)}_l$ are required to define the wavefunction on the first qu5it, 
 and 20 angles, $\theta^{(1,m)}_l$, are required in the controlled rotation operator 
 to define the four states of the second qu5it for each of the five states of the first qu5it (where $m=0,1,2,3,4$ and $l=0,1,2,3$).
 The decomposition of the controlled-rotation operator is given in Fig.~\ref{fig:2Q5SPcontrol}.
		}
		\label{fig:2Q5SP}
\end{figure}

\begin{figure}[!t]
	\centering
    \includegraphics[width=\columnwidth]{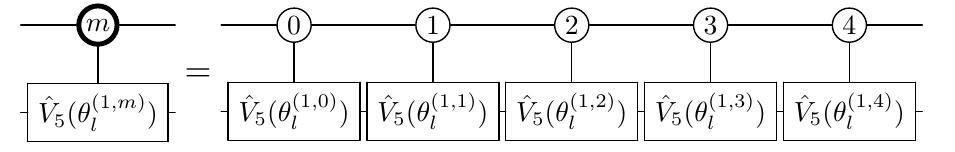}
	\caption{The explicit construction of the controlled rotation operator 
    shown in Fig.~\ref{fig:2Q5SP} in terms of projectors on the first qu5it.}
	\label{fig:2Q5SPcontrol}
\end{figure}

\noindent
Preparing an arbitrary real wavefunction on the two-qu5it system is a
straightforward extension of preparing such a state on two qubits, 
as can be found, {\it e.g.}, in Ref.~\cite{Klco:2019xro}.
A quantum circuit for preparing a state on two
qu5its is shown in Fig.~\ref{fig:2Q5SP}, 
and is given in terms of single-qu5it rotations 
$\hat V_5(\theta^{(0,0)}_l)\otimes \hat I_5$ and controlled 
rotations, $\hat\Lambda^{(m)}\otimes\hat V_5(\theta^{(1,m)}_l)$, where 
$\hat\Lambda^{(m)}$ are projectors onto the $m^{\rm th}$ state in the qu5it Hilbert space.
The form of the controlled rotation operator in terms of projectors on the first qu5it is given in Fig.~\ref{fig:2Q5SPcontrol}.
With 25 states in the Hilbert space, 24 angles are required to prepare a 
real wavefunction from some default state.
With the quantum circuit shown in Fig.~\ref{fig:2Q5SP}, 
we see that qu5it-0 (top qu5it in the figure) can be prepared with 4 angles 
(defining the corresponding 5 amplitudes), 
while 
preparing  qu5it-1 (bottom qu5it in Fig.~\ref{fig:2Q5SP}) 
requires 4 angles per qu5it-0 state, hence a total of 20 angles,
which provides the requisite 24 angles.
Correspondingly, variational algorithms employed to establish the ground state (or excited states) of this system will, in the absence of 
further assumptions, require 24 angles.

The qu5it rotation operators $\hat{V}_5$ have the same form as 
Eq.~\eqref{eq:1qstateprep}. For example,
\begin{align}
    \hat{V}_5({\theta^{(1,3)}_l}) \ = \ & \
    e^{-i \theta^{(1,3)}_3 \hat {\cal X}_{34} }\ 
    e^{-i \theta^{(1,3)}_2 \hat {\cal X}_{23} } \nonumber \\
    & \ \times e^{-i \theta^{(1,3)}_1 \hat {\cal X}_{12} }\ 
    e^{-i \theta^{(1,3)}_0 \hat {\cal X}_{01} }
    \ ,
    \label{eq:V5eg}
\end{align}
that enables the preparation of arbitrary real amplitudes on qu5it-1 from 
an initial state $|0\rangle$ (and 
with angles associated with a projection onto state $|3\rangle$ of qu5it-0).

\subsubsection{Time Evolution of Two Qu5its}
\noindent
The time evolution of the two-qu5it system is determined by the Hamiltonian in Eq.~\eqref{eq:twomodepairHGIVENS}.
As in the case of one qu5it, it contains non-commuting terms 
so that
Trotterization will only approximate the time evolution.
The resulting errors can, however, be systematically reduced by decreasing the time step $\Delta t$, or by using higher-order formulas.
Considering the LO Trotterization of the evolution operator $\hat{U}(\Delta t)$, 
with one- and two-body contributions from the different interaction terms, gives
\begin{equation}
\hat U(\Delta t) \ = \ \hat U_2(\Delta t) \ \hat U_1(\Delta t)
\  .
\label{eq:trotOp}
\end{equation}
where 
\begin{align}
\hat U_1(\Delta t) & \ = \
\hat U_1^{(\hat T_z)}(\Delta t)\   \hat U_{1}^{(\hat {\cal X}_{13})}(\Delta t) \  \hat U_1^{(\hat N_{\rm pair})}(\Delta t)
\  ,
\nonumber\\
\hat U_1^{(\hat T_z)}(\Delta t) & \ = \
\prod_{j=1,2}  e^{-i  \Delta t  \varepsilon  \hat T_{z,j}}
\ ,
\nonumber\\
\hat U_{1}^{(\hat {\cal X}_{13})}(\Delta t) & \ = \
\prod_{j=1,2}  e^{+i \Delta t (V+g)   {\hat {\cal X}_{13,j}} } 
\ ,
\nonumber\\
\hat U_1^{(\hat N_{\rm pair})}(\Delta t) & \ = \
\prod_{j=1,2}  e^{+i \Delta t g  {\hat N_{\rm pair, j}} }
\ ,
\label{eq:1bodQ2}
\end{align}
and
\begin{align}
\hat U_2(\Delta t) & \ = \ \hat U_2^{(g)}(\Delta t) \ \hat U_{2}^{(V)}(\Delta t)
\  ,
\nonumber\\
\hat U_{2}^{(V)}(\Delta t) & \ = 
  \prod_{r, s \in \{(12),(23)\} } \ 
   e^{+i V \Delta t {\hat {\cal X}_{r,1}}\otimes {\hat {\cal X}_{s,2}}} \ 
   e^{-i V \Delta t {\hat {\cal Y}_{r,1}}\otimes {\hat {\cal Y}_{s,2}}} 
   \ ,
\nonumber\\
\hat U_{2}^{(g)}(\Delta t) & \ = 
 \prod_{\substack{r, s \in \{(01),(03), \\ \ \  -(14),-(34) \} }}  
e^{+i \frac{1}{2} g \Delta t {\hat {\cal X}_{r,1}}\otimes {\hat {\cal X}_{s,2}} } \
e^{+i \frac{1}{2} g \Delta t {\hat {\cal Y}_{r,1}}\otimes {\hat {\cal Y}_{s,2}} }
\ ,
\label{eq:2bodQ2}
\end{align}
As these expressions are LO in the Trotter expansion, the presented order of operators in Eqs.~\eqref{eq:trotOp}, \eqref{eq:1bodQ2}, and \eqref{eq:2bodQ2} 
has been chosen at random. In principle,
all possible sequences of operator products could be explored 
in order to minimize Trotter errors in observables of interest for a given Trotter time step.
\\ \\ \indent
While the one-body 
terms can be implemented with phase operators and Givens rotations as discussed previously, 
the two-body entangling interactions are more complex. 
We discuss two possible implementations, based on 
gates that are available on current devices. 
One possibility is that qudit systems are able to perform 
two-qudit Givens rotation gates, 
as, for example, M{\o}lmer-S{\o}rensen gates in trapped-ion systems~\cite{Ringbauer:2021lhi}.
The other possibility is that 
generalized CNOT gates are available, such as
in transmon-qudit systems~\cite{Goss:2022bqd,Fischer:2022fgv}. This latter approach was studied in Ref.~\cite{Ciavarella:2021nmj} for a qutrit-based system, and here we generalize it to higher-dimension qudits.

\begin{figure}[!t]
    \raggedright 
    \textbf{(a)}
	\includegraphics[width=\columnwidth]{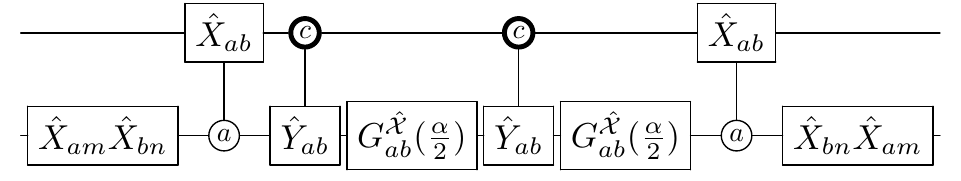}
    \textbf{(b)}
    \includegraphics[width=\columnwidth]{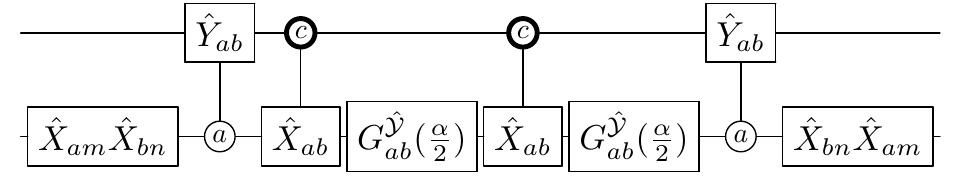}
	\caption{Circuits for the two-qu5it entangling gates, with \textbf{(a)} implementing $G^{\hat{\mathcal{X}}\hat{\mathcal{X}}}_{abmn}(\alpha)=e^{-i\alpha{\hat{\mathcal{X}}_{ab}}\otimes{\hat{\mathcal{X}}_{mn}}}$, and \textbf{(b)} implementing $G^{\hat{\mathcal{Y}}\hat{\mathcal{Y}}}_{abmn}(\alpha)=e^{-i\alpha{\hat{\mathcal{Y}}_{ab}}\otimes{\hat{\mathcal{Y}}_{mn}}}$. 
    The index $c$ on the controlled gates runs over the complement states of $a$ and $b$ (in a similar fashion to Fig.~\ref{fig:2Q5SPcontrol}, but here $c\neq\{a,b\}$), and the Givens rotation $G_{ab}^{\hat{\mathcal{X}}}(\frac{\alpha}{2})$ is defined as $e^{-i\frac{\alpha}{2}\hat{\mathcal{X}}_{ab}}$ (similarly for $G_{ab}^{\hat{\mathcal{Y}}}(\frac{\alpha}{2})$).
		}
		\label{fig:2Q5XXYY}
\end{figure}

\begin{figure}[!h]
    \centering
	\includegraphics[width=0.6\columnwidth]{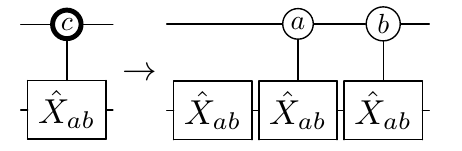}
	\caption{Simplification for the multi-controlled $C \hat X$ gates in Fig.~\ref{fig:2Q5XXYY}, where the 3 controlled rotations ($c$ runs over $c\neq\{a,b\}$) are replaced by 2 controlled rotations plus a single qu5it gate. The same simplification can be applied to the $C \hat Y$ gates.}
		\label{fig:CX_simplification}
\end{figure}

\begin{figure}[!t]
    \raggedright 
    \textbf{(a)}
	\includegraphics[width=\columnwidth]{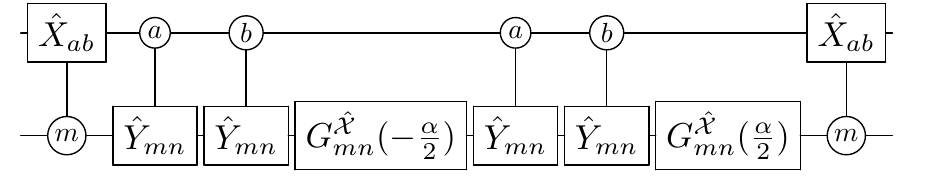}
    \textbf{(b)}
    \includegraphics[width=\columnwidth]{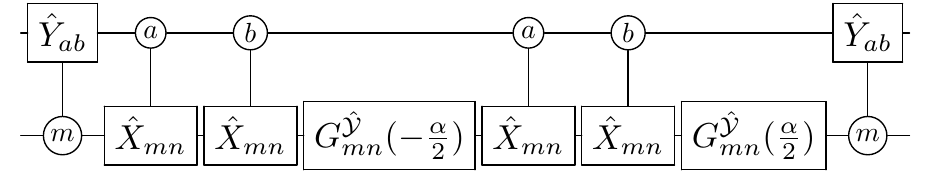}
	\caption{Improved circuits for the two-qu5it entangling gates derived from those in Fig.~\ref{fig:2Q5XXYY}, with \textbf{(a)} implementing $G^{\hat{\mathcal{X}}\hat{\mathcal{X}}}_{abmn}(\alpha)=e^{-i\alpha{\hat{\mathcal{X}}_{ab}}\otimes{\hat{\mathcal{X}}_{mn}}}$, and \textbf{(b)} implementing $G^{\hat{\mathcal{Y}}\hat{\mathcal{Y}}}_{abmn}(\alpha)=e^{-i\alpha{\hat{\mathcal{Y}}_{ab}}\otimes{\hat{\mathcal{Y}}_{mn}}}$. 
		}
		\label{fig:2Q5XXYYcomp}
\end{figure}

The quantum circuits require rotations on one of the qu5its controlled by the state of the other qu5it. 
These can be implemented via controlled $\hat X_{ab}$ and $\hat Y_{ab}$ gates, with
\begin{subequations}
\begin{align}
    \hat X_{ab} \ &= \ |a\rangle\langle b| + |b\rangle\langle a| \ + \ \sum_{c \neq \{a,b\}} |c\rangle\langle c| \ , \\
    \hat Y_{ab}\  &= \ i(|a\rangle\langle b| - |b\rangle\langle a|) \ + \ \sum_{c \neq \{a,b\}} |c\rangle\langle c| \ .
\end{align}
\end{subequations}
To be concrete, consider the action of an $\hat X_{14}$ gate acting on qu5it-1 controlled by qu5it-0,
which is implemented when qu5it-0 is in the state 
$\ket{2}$. This operation takes the form
\begin{equation}
\left[C\hat X\right]_{14}^{2}
\ = \ 
\hat\Lambda^{(2)}\otimes \hat X_{14}\ +\ \sum_{l\ne 2} \hat \Lambda^{(l)}\otimes \hat I_5
\  .
\end{equation}
The generalization to multiple control states is straightforward.
For example,
$\hat X_{23}$ controlled by the states 
$\ket{0},\ket{1},\ket{4}$ reads
\begin{align}
\left[C\hat X\right]_{23}^{0,1,4}
& =  
\sum_{l=\{0,1,4\}} \hat\Lambda^{(l)}\otimes \hat X_{23}
\ +\ 
\sum_{l\ne \{0,1,4\}} \hat \Lambda^{(l)}\otimes \hat I_5
\nonumber\\
& =  
\left[C\hat X\right]_{23}^{0} \ \left[C\hat X\right]_{23}^{1} \ \left[C\hat X\right]_{23}^{4}
\ .
\end{align}
The two types of required two-qu5it gates are $e^{-i\alpha{\hat{\mathcal{X}}_{ab}}\otimes{\hat{\mathcal{X}}_{mn}}}$ 
and $e^{-i\alpha{\hat{\mathcal{Y}}_{ab}}\otimes{\hat{\mathcal{Y}}_{mn}}}$, 
which can be 
implemented using 
the quantum circuits shown in Fig.~\ref{fig:2Q5XXYY}.
Additional simplifications, as shown in Fig.~\ref{fig:CX_simplification}, can be made to reduce the number of controlled gates (this further reduces the number of controlled gates as the dimension of the qudit increases).
After including these simplifications, and explicitly including the basis change, the  circuits in Fig.~\ref{fig:2Q5XXYY} can be reduced to those in Fig.~\ref{fig:2Q5XXYYcomp}.
While the circuits that are shown in Fig.~\ref{fig:2Q5XXYY} involve 
2 $G$ gates, 8 $C\hat X$ or $C\hat Y$ gates and 4 single qu5it gates,
those in Fig.~\ref{fig:2Q5XXYYcomp} contain 
2 $G$ gates and 6 $C\hat X$ or $C\hat Y$ gates.
In the resource estimates that follow, we use the circuits in Fig.~\ref{fig:2Q5XXYYcomp}.
These one- and two-body operators are sufficient to evolve a prepared state forward in time.

\subsection{Arbitrary Numbers of Mode-Pairs}
\label{subsec:ManyModes}
\noindent
State preparation requires quantum circuits with a number  of control operators that scales with the number of qu5its.
Without the concept of a distance between qu5its in this model, {\it i.e.}, the all-to-all nature of the interactions, 
designs of localized circuits~\cite{Klco:2019yrb} 
with exponentially-converging circuit truncations are not obvious.
The high-degree of symmetry in the Hamiltonian and the fact that the model is exactly solvable suggests, however, 
that simplifications in the state-preparation circuit and relations between the angles of the controlled operators seem likely.
Re-organization of entanglement structures via application of, for example,
the HL-VQE algorithm~\cite{Robin:2023fow} could also potentially allow for efficient simulations of truncated systems, with exponential convergence towards the exact solution, as was obtained in the case of the LMG model.
On the other hand, due to the quadratic form of the Agassi model Hamiltonian, only one- and two-body interactions 
are required in evolving the quantum state of an arbitrary number of modes forward in time.  
Thus, if the entangled states of interest can be reached from a tensor-product initial state,
the one- and two-body circuits used to evolve two mode-pair systems forward, 
when implemented between all pairs of modes, are sufficient.

The required number of circuit elements per Trotter step can be determined for a system with $\Omega/2$ mode pairs.
The number of one-body operators, 
assuming that the phase gates and Givens rotation can be implemented as fundamental gates, 
is $n_{1\rm mode-pair}=3$.
Counting gates in the quantum circuits shown in 
Fig.~\ref{fig:2Q5XXYYcomp}
(to implement the two-body operators) 
reveals that the total number of gates required is 
\begin{align}
   & n_{\tfrac{\Omega}{2}\rm mode-pairs}  = \frac{\Omega}{2}\left[ 
    1\  G^{\hat{\mathcal{X}}} , \  
    2\  {\rm Phase} 
     \right] +  \frac{1}{2}\frac{\Omega}{2}\left(  \frac{\Omega}{2}-1 \right) \nonumber\\
    & \qquad \qquad \times \left[ 
    120\  { C{\hat X}} , \  
    120\  { C{\hat Y}} , \  
    40\  G^{\hat{\mathcal{X}}} , \ 
    40\  G^{\hat{\mathcal{Y}}}
     \right] ,
    \label{eq:5resources}
\end{align}
scaling, as expected, quadratically with the number of qu5its, and with sizable coefficients.
This equation should be read as $ 20 \Omega\left(  \Omega-2 \right)$ controlled-$\hat X$ gates, 
$10 \Omega\left(  \Omega-2 \right)$ single-qu5it Givens rotations, etc. 
The requirements are reduced significantly if two-qu5it Givens rotations are available, becoming
\begin{align}
    n_{\tfrac{\Omega}{2}\rm mode-pairs}  =&\; \frac{\Omega}{2}\left[ 
    1\  G^{\hat{\mathcal{X}}} , \  
    2\  {\rm Phase} 
     \right] +  \frac{1}{2}\frac{\Omega}{2}\left(  \frac{\Omega}{2}-1 \right) \nonumber\\
    &\qquad \times \left[ 
    20\  {G^{\hat{\mathcal{X}}\hat{\mathcal{X}}}} , \  
    20\  {G^{\hat{\mathcal{Y}}\hat{\mathcal{Y}}}} 
     \right] .
    \label{eq:5resources-2}
\end{align}
For modest-size systems, the time dependence of an initial state can be 
determined exactly by matrix exponentiation.  
\begin{figure}[!t]
\centering{\includegraphics[width=\columnwidth]{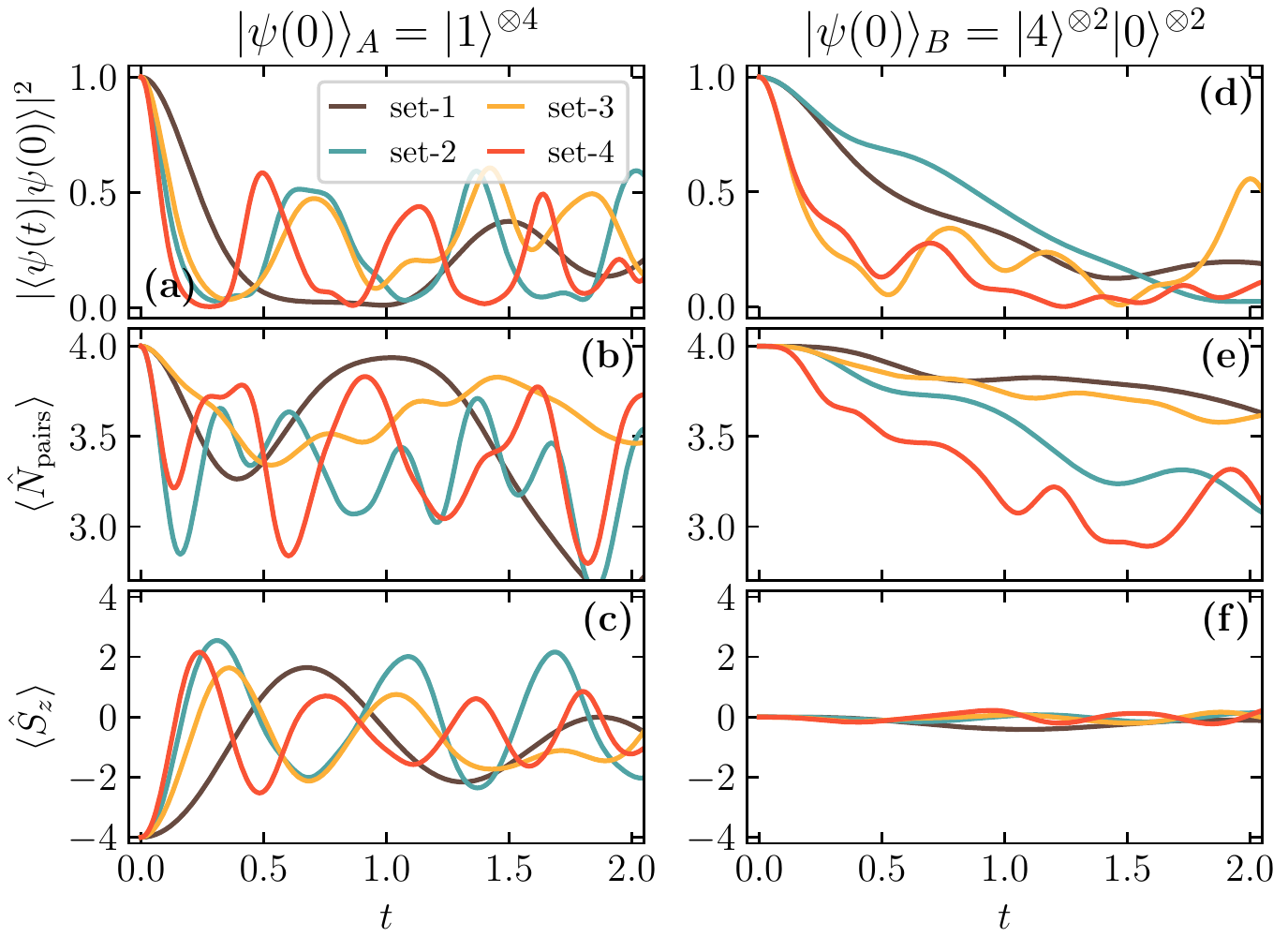}} 
\caption{The exact time dependence of \textbf{(a,d)} $|\langle \psi (t)|\psi (0)\rangle|^2$, \textbf{(b,e)} number of pairs, 
and \textbf{(c,f)} 
$\langle \hat S_z \rangle$
for a system with 
$N=\Omega=8$ particles
for four sets of couplings, set-1 to set-4, given in Eq.~(\ref{eq:setdefs}).
The left column is associated with the initial state 
$|\psi(0)\rangle_A$,
while the right column with 
$|\psi(0)\rangle_B$, as given in Eq.~(\ref{eq:psi0s}).
}
\label{fig:N8all}
\end{figure}
As an example, in Fig.~\ref{fig:N8all}, the exact
time dependence of vacuum-to-vacuum survival amplitude, number of pairs, and
$\langle \hat S_z \rangle$
from four sets of couplings, set-1 to set-4,\footnote{The set-0 is not displayed since it leads to a trivial time evolution.} are displayed in the $\Omega=8$ system, 
for quenches from the two different tensor-product initial states given in Eq.~(\ref{eq:psi0s}):
$|\psi(0)\rangle_A = |1\rangle^{\otimes 4}$ and 
$|\psi(0)\rangle_B = |4\rangle^{\otimes 2}\otimes |0\rangle^{\otimes 2}$.
However, exact time evolution soon becomes unwieldy with increasing system size, 
and quantum simulations are required to extend to larger systems.
A quantum simulation of this system can be accomplished with $\Omega/2=4$ qu5its 
(corresponding to a $625$ dimensional Hilbert space)
with quantum circuits implementing Trotter evolution, or variants thereof.  
If performed with previous JW qubit mappings~\cite{P_rez_Fern_ndez_2022,Saiz:2022rof}, 
the analogous quantum simulation would require $4^{\Omega/2} = 16$ qubits
(corresponding to a $65{,}536$ dimensional Hilbert space).

\begin{figure}[!tb]
\centering{\includegraphics[width=\columnwidth]{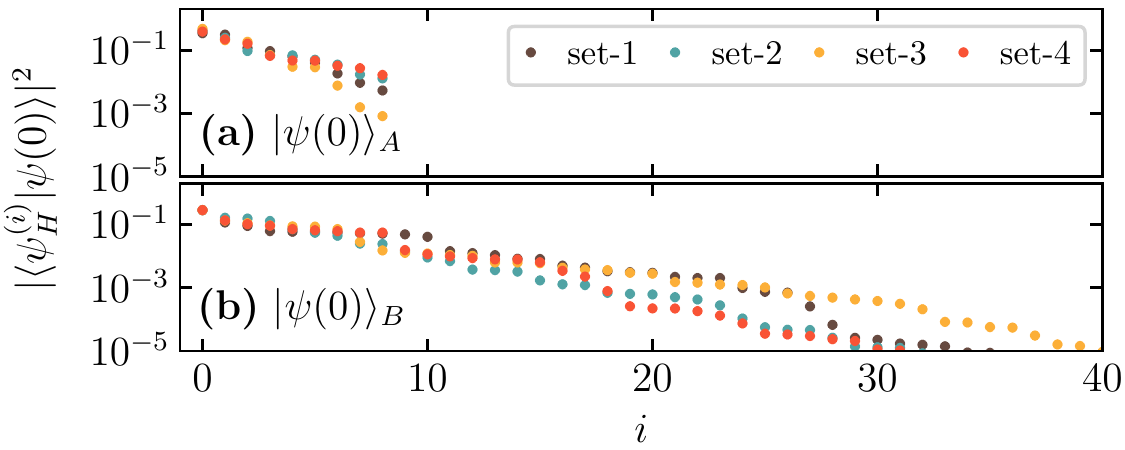}} 
\caption{Sorted non-zero overlaps between the initial state $|\psi(0)\rangle_{A/B}$ and the eigenstates of the Hamiltonian $|\psi_H^{(i)}\rangle$ for a system with $N=\Omega=8$ particles and for the four sets of couplings. Top panel \textbf{(a)} shows $|\psi(0)\rangle_A$, and bottom panel \textbf{(b)} shows $|\psi(0)\rangle_B$.
}
\label{fig:N8overlaps}
\end{figure}

It is informative to consider the exact time evolution over longer time intervals,
beyond those shown in Fig.~\ref{fig:N8all}, as 
displayed in Appendix~\ref{app:longtime}.
Observables evolving from $|\psi(0)\rangle_A$ are found to continue to rapidly oscillate at later times,
while those from $|\psi(0)\rangle_B$ tend to fall more slowly toward their average long-time values, about
which they slowly oscillate with amplitudes that decrease with increasing interaction strength.
This behavior is consistent with $|\psi(0)\rangle_A$ being a spin-stretched ground state of the non-interacting Hamiltonian, with dominant overlap only onto a small number of states, as shown in Fig.~\ref{fig:N8overlaps}a.
These are effectively {\it scar} states for this quenched system~\cite{PhysRevLett.53.1515}.
In contrast, the evolution from $|\psi(0)\rangle_B$, the initial state with energy that is in the middle of the excitation spectrum, is consistent with comparable overlap onto a large number of states, as shown in Fig.~\ref{fig:N8overlaps}b, giving multiple coherent contributions in time evolution. 
The exact wavefunctions for both initial configurations confirm these described structures.
Interestingly, these different overlap structures suggest that they will incur different systematic errors in Trotterized time-evolution when simulated with quantum computers and simulators.  
With multiple 
states contributing to time evolution, the errors from the system initially in $|\psi(0)\rangle_B$ are expected to be significantly larger than from starting in $|\psi(0)\rangle_A$, 
because of a ``sign problem'' in the form of incomplete cancellations between multiple amplitudes,
as opposed to a few, in the unitary evolution.
As we will show in subsequent sections, this is indeed the situation.

\subsection{Quantum Simulations using {\tt cirq}-based Classical Simulation}
\label{subsec:trott}
\noindent
With the developments presented in the previous sections, 
quantum simulations of these fundamentally five-dimensional systems can be implemented on quantum simulators 
of arrays of qu5its.\footnote{In principle, 
they can also be straightforwardly implemented on devices with qudits with $d>5$.}
Without access to a qu5it quantum computer at present, we have used Google's {\tt cirq} software~\cite{cirq_developers_2022_7465577} to perform classical noiseless simulations for a selection of parameter sets and system sizes.  
The qudit capabilities of {\tt cirq} form the foundation of a {\tt python}-based~\cite{python3} simulation code
that furnishes single-thread\footnote{The multi-threaded high-performance version of {\tt cirq}, {\tt qsim}~\cite{qsim}, is still a qubit-based simulator, and does not support qudits (as of this writing).} single-CPU-node instances for qudit systems.

\begin{figure}[!t]
\centering{\includegraphics[width=\columnwidth]{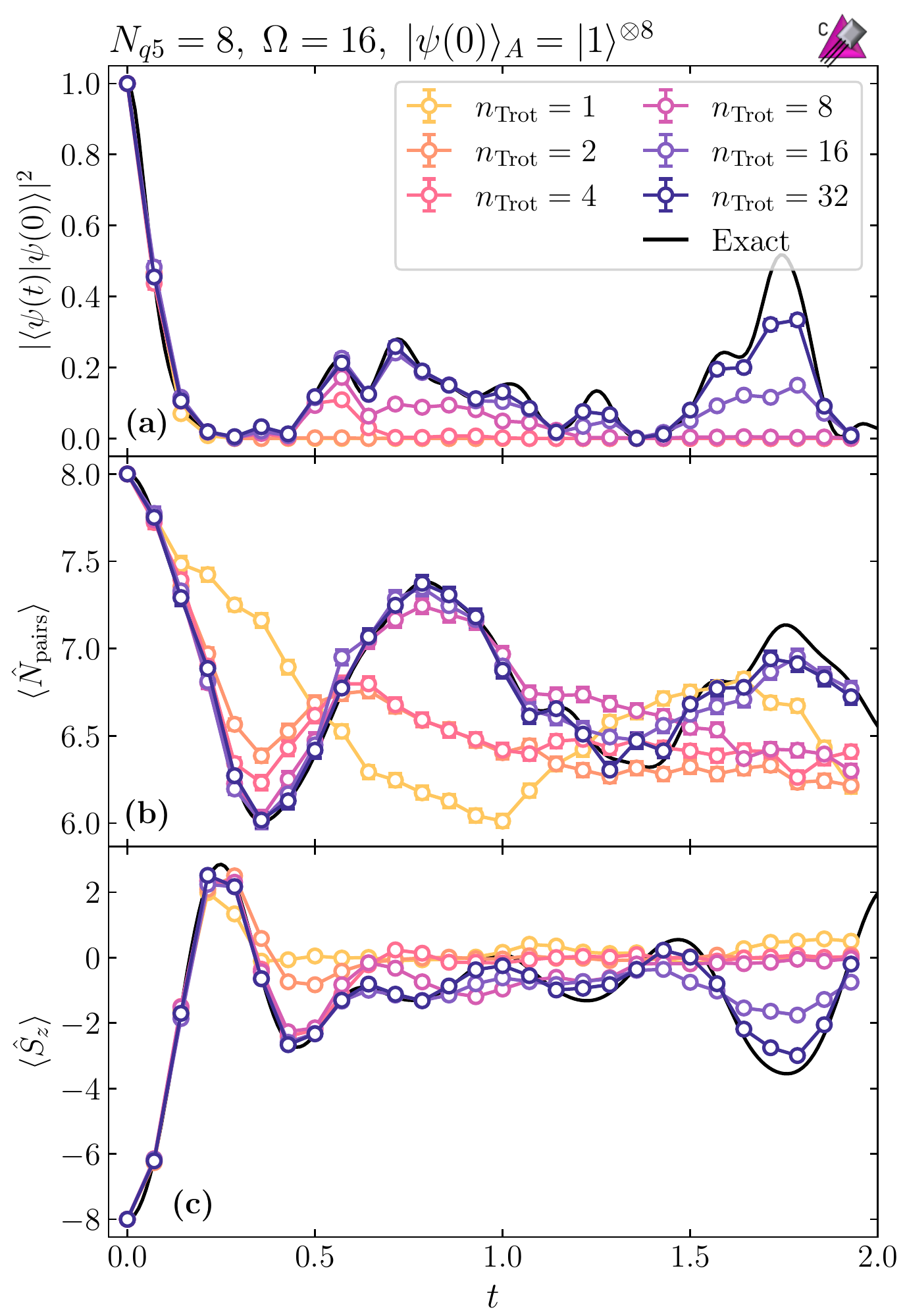}} 
\caption{The time dependence of \textbf{(a)} $|\langle \psi (t)|\psi (0)\rangle|^2$, \textbf{(b)} number of pairs, 
and \textbf{(c)}  
$\langle \hat S_z \rangle$,
for set-3 of couplings (given in Eq.~(\ref{eq:setdefs}))
in a system with $N=16$ particles initially in the state 
$|\psi(0)\rangle_A = |1\rangle^{\otimes 8}$.
The different colors indicate different numbers of Trotter steps ($n_{\rm Trot}$).
The points are the results of noiseless classical simulation of the qu5it system with $10^3$ shots.
The icon denoting classical computation is defined in Ref.~\cite{Klco:2019xro}.
}
\label{fig:SimulationN8wvf2}
\end{figure}
\begin{figure}[!t]
\centering{\includegraphics[width=\columnwidth]{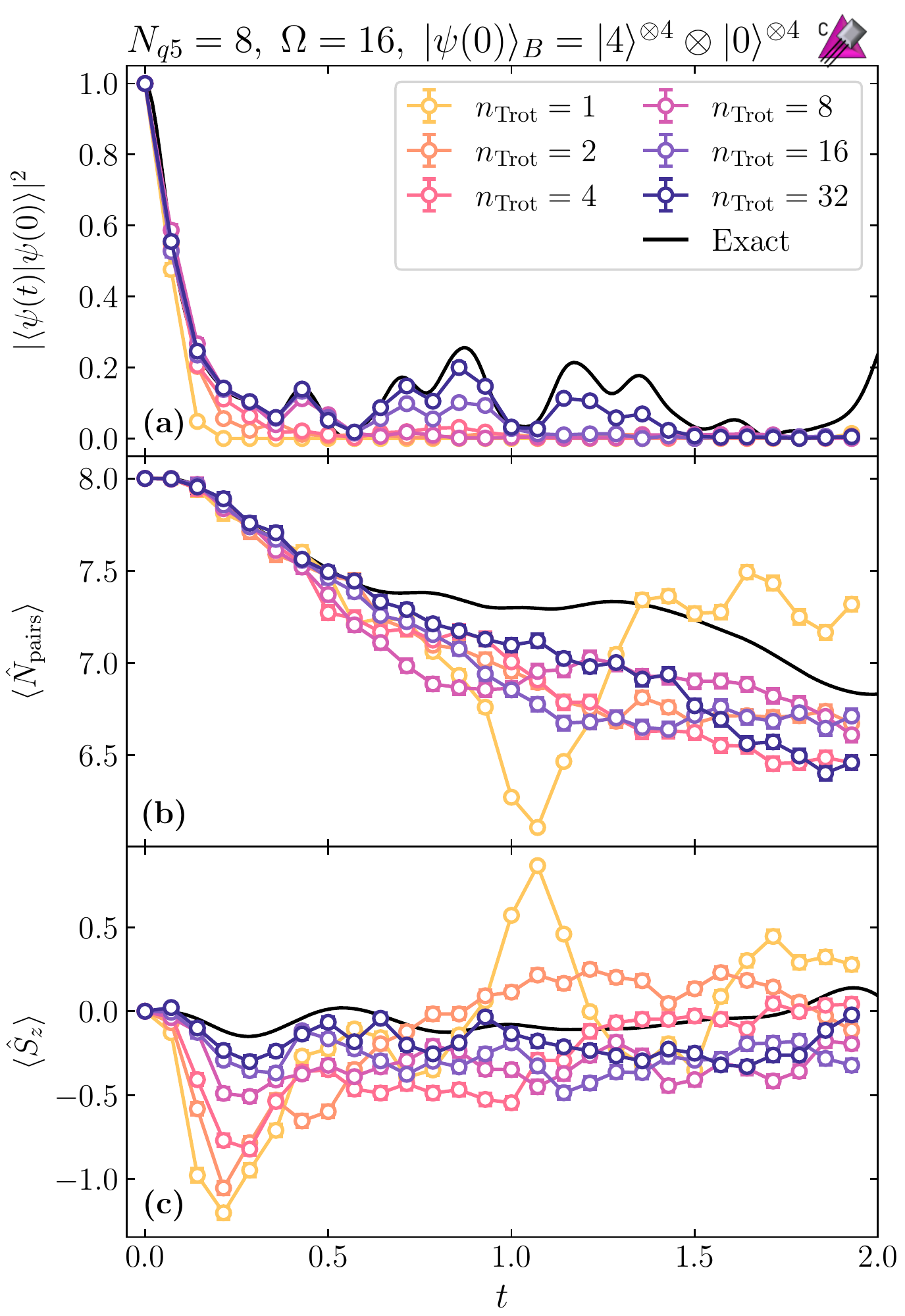}}
\caption{The time dependence of \textbf{(a)} $|\langle \psi (t)|\psi (0)\rangle|^2$, \textbf{(b)} number of pairs, and \textbf{(c)} $\langle \hat S_z \rangle$,
for set-3 of couplings (given in Eq.~(\ref{eq:setdefs}))
in 
a system with $N=16$ particles initially in the state 
$|\psi(0)\rangle_B = |4\rangle^{\otimes 4}\otimes|0\rangle^{\otimes 4}$.
The different colors indicate different numbers of Trotter steps ($n_{\rm Trot}$).
The points are the results of noiseless classical simulation of the qu5it system with $10^3$ shots.
}
\label{fig:SimulationN8wvf1}
\end{figure}
\begin{figure}[!t]
\centering{\includegraphics[width=\columnwidth]{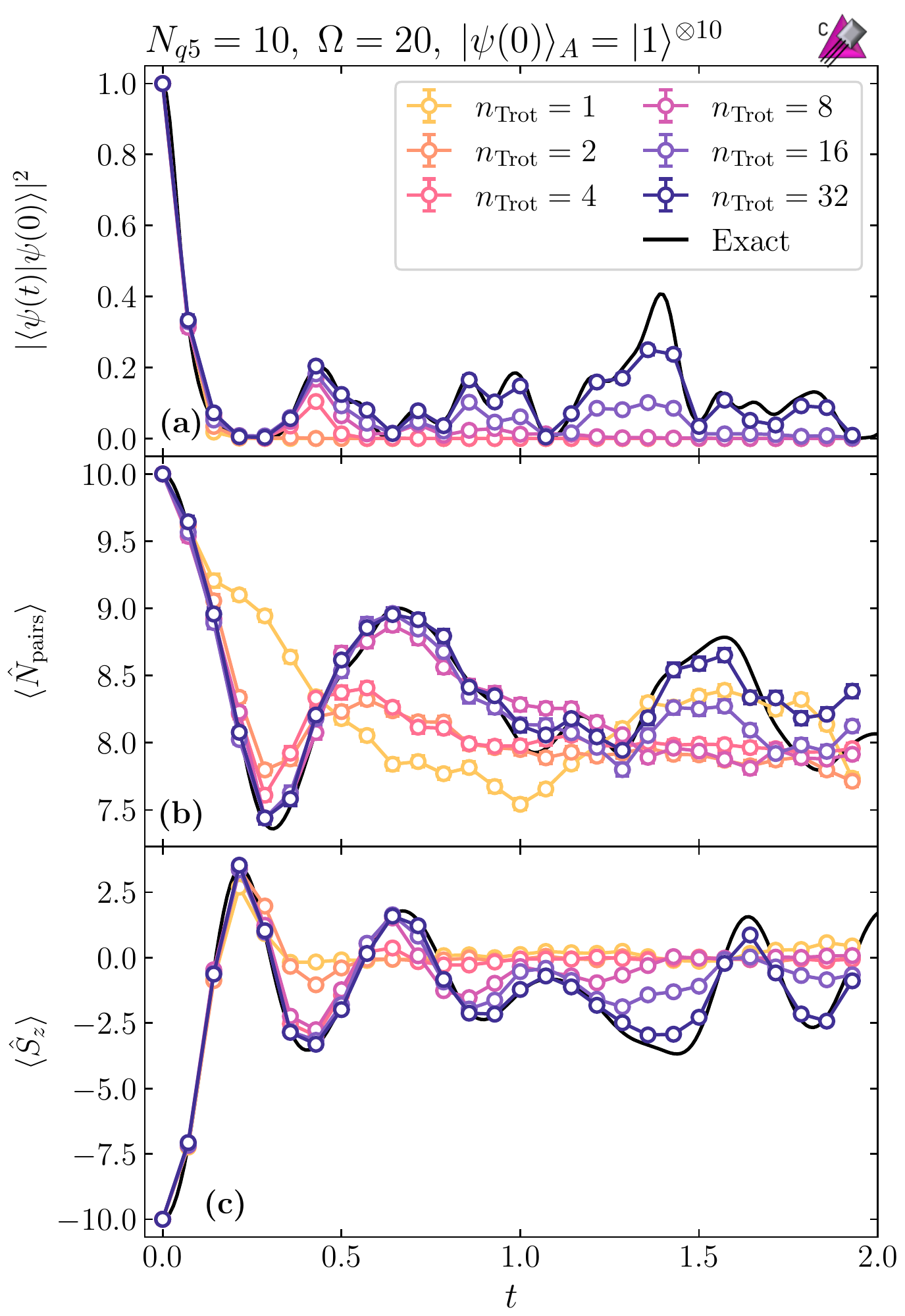}} 
\caption{The time dependence of \textbf{(a)} $|\langle \psi (t)|\psi (0)\rangle|^2$, \textbf{(b)} number of pairs, and \textbf{(c)} $\langle \hat S_z \rangle$,
for set-3 of couplings (given in Eq.~(\ref{eq:setdefs}))
in 
a system with $N=20$ particles initially in the state 
$|\psi(0)\rangle_A = |1\rangle^{\otimes 10}$.
The different colors indicate different numbers of Trotter steps ($n_{\rm Trot}$).
The points are the results of noiseless classical simulation of the qu5it system with $10^3$ shots.
}
\label{fig:SimulationN10wvf2}
\end{figure}
\begin{figure}[!t]
\centering{\includegraphics[width=\columnwidth]{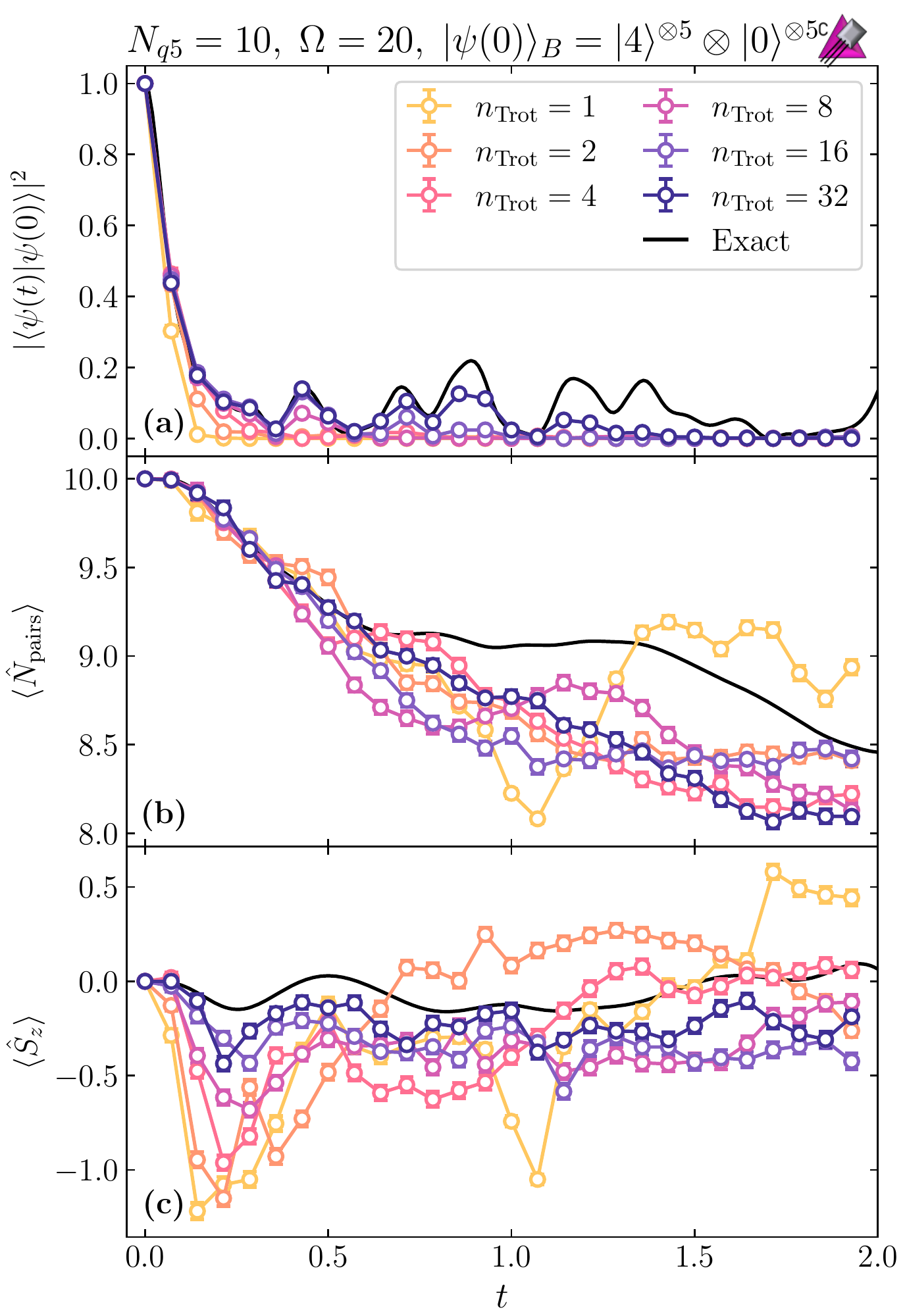}}
\caption{The time dependence of \textbf{(a)} $|\langle \psi (t)|\psi (0)\rangle|^2$, \textbf{(b)} number of pairs, and \textbf{(c)} $\langle \hat S_z \rangle$,
for set-3 of couplings (given in Eq.~(\ref{eq:setdefs}))
in 
a system with $N=20$ particles initially in the state 
$|\psi(0)\rangle_B = |4\rangle^{\otimes 5}\otimes|0\rangle^{\otimes 5}$.
The different colors indicate different numbers of Trotter steps ($n_{\rm Trot}$).
The points are the results of noiseless classical simulation of the qu5it system with $10^3$ shots.
}
\label{fig:SimulationN10wvf1}
\end{figure}
\begin{figure}[!t]
\centering{\includegraphics[width=\columnwidth]{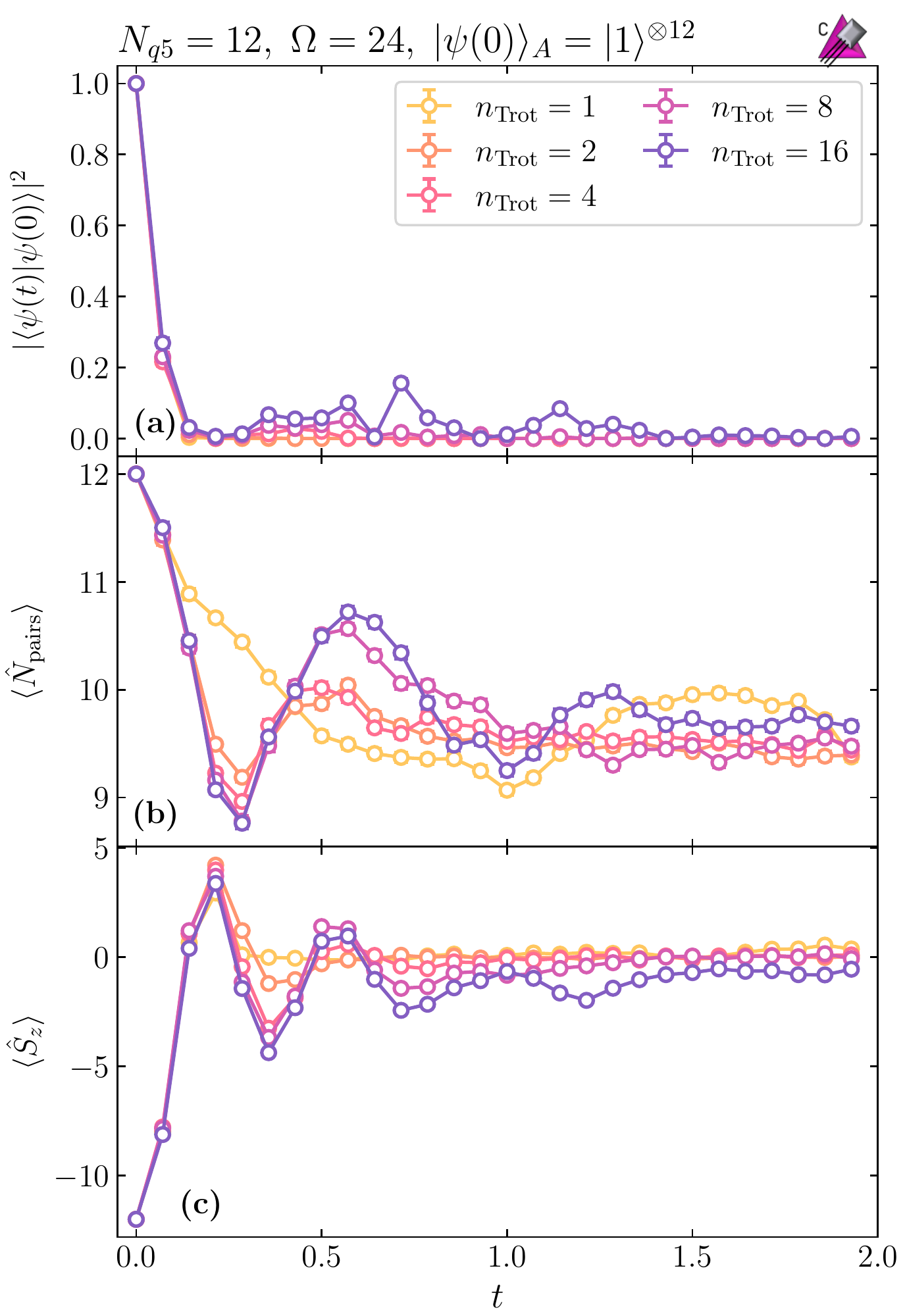}} 
\caption{The time dependence of \textbf{(a)} $|\langle \psi (t)|\psi (0)\rangle|^2$, \textbf{(b)} number of pairs, and \textbf{(c)} $\langle \hat S_z \rangle$,
for set-3 of couplings (given in Eq.~(\ref{eq:setdefs}))
in 
a system with $N=24$ particles initially in the state 
$|\psi(0)\rangle_A = |1\rangle^{\otimes 12}$.
The different colors indicate different numbers of Trotter steps ($n_{\rm Trot}$).
The points are the results of noiseless classical simulation of the qu5it system with $10^3$ shots.
}
\label{fig:SimulationN12wvf2}
\end{figure}
\begin{figure}[!t]
\centering{\includegraphics[width=\columnwidth]{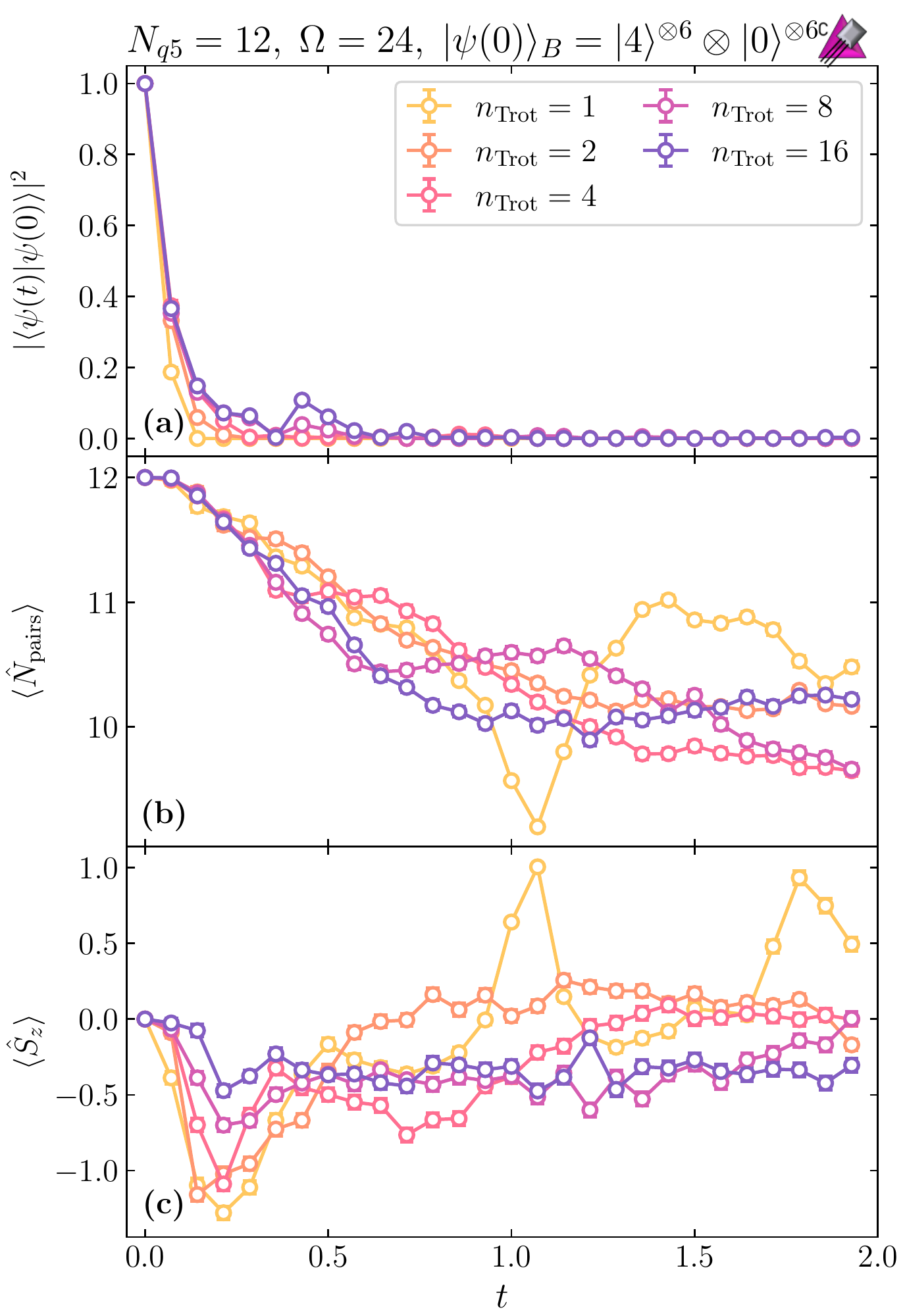}}
\caption{The time dependence of \textbf{(a)} $|\langle \psi (t)|\psi (0)\rangle|^2$, \textbf{(b)} number of pairs, and \textbf{(c)} $\langle \hat S_z \rangle$, 
for set-3 of couplings (given in Eq.~(\ref{eq:setdefs}))
in 
a system with $N=24$ particles initially in the state 
$|\psi(0)\rangle_B = |4\rangle^{\otimes 6}\otimes|0\rangle^{\otimes 6}$.
The different colors indicate different numbers of Trotter steps ($n_{\rm Trot}$).
The points are the results of noiseless classical simulation of the qu5it system with $10^3$ shots.
}
\label{fig:SimulationN12wvf1}
\end{figure}

For the purposes of demonstration, 
we present results 
for the observables discussed above in selection of system sizes
for a range of LO Trotterizations of the evolution operators.
In the simulations, 
we have employed 1000 shots per run using the set-3 Hamiltonian couplings 
given in Eq.~(\ref{eq:setdefs})
and the two different initial states from Eq.~\eqref{eq:psi0s}, 
for systems of size $\{N_{q5}=8, N=\Omega=16\}$, $\{N_{q5}=10, N=\Omega=20\}$, 
corresponding to 16 and 20 particles distributed among 8 and 10 qu5its, respectively.
These are compared with the time evolution of the full state-vector computed via exact exponentiation of the Hamiltonian, using the {\tt Expokit}~\cite{expokit} package in {\tt Julia}~\cite{Julia-2017}.\footnote{For the $\Omega=20$ system, the Hamiltonian was projected to the $N=20$ sector before exact exponentiation, reducing the dimension of the Hamiltonian matrix from 9{,}765{,}625 to 1{,}936{,}881.}
The results obtained for the $\Omega=16$ system are shown in 
Figs.~\ref{fig:SimulationN8wvf2} and~\ref{fig:SimulationN8wvf1}, 
and those for $\Omega=20$ are shown in 
Figs.~\ref{fig:SimulationN10wvf2} and~\ref{fig:SimulationN10wvf1}.
We have also performed the simulation for the $\{N_{q5}=12, \Omega=24\}$ system, shown in 
Figs.~\ref{fig:SimulationN12wvf2} and~\ref{fig:SimulationN12wvf1}. 
For this case, a comparison with the exact time evolution 
could not be accomplished using a single node, as it requires
exponentiating $10^8 \times 10^8$ matrices. 
Simulations of larger systems ($\Omega>24$) were not possible 
using the {\tt cirq}-based {\tt python} code on a single compute node of Hyak (mox)
due to memory limitations.

As expected, the Trotterized evolution curves for observables converge to the expected values as the number of Trotter steps becomes large over the explored time intervals. However, the convergence is faster for the initial state 
in which all the spin-$\downarrow$ states are occupied, $|\psi(0)\rangle_A$. 
This can be understood to arise from Trotter errors being larger for the systems starting in
$|\psi(0)\rangle_B$ than $|\psi(0)\rangle_A$ due to the larger number of comparable 
contributions to the unitary evolution (see Fig.~\ref{fig:N8overlaps}), because of being higher in the energy spectrum, 
and the more substantial {\it sign problem} in forming their sum, 
as mentioned above.  This is discussed in more detail in Appendix~\ref{app:stnprob}.
A potential way to improve the convergence of the Trotterization 
may be found by examining the breaking of symmetries. One of those is the exchange symmetry, since, for the original Hamiltonian $\hat{H}$, $[e^{-it\hat{H}},\hat{P}_{ij}]=0$, with $\hat{P}_{ij}$ being the exchange operator of mode-pairs $i$ and $j$, 
but with the LO Trotterized operator $\hat{U}(t)$, as in Eq.~\eqref{eq:trotOp}, $[\hat{U}(t),\hat{P}_{ij}] \neq 0$. Similar issues have been found in other systems. 
For example, the Hamiltonian governing the evolution of dense coherent neutrino systems exhibits a similar exchange symmetry, and while Trotterization might break the symmetry, it can be recovered~\cite{Illa:2022zgu}. 
A similar situation arises in simulations of $1+1$-dimensional SU(3) lattice gauge theory~\cite{Farrell:2022wyt}
with $n_f$ flavors of quarks, where Trotterization breaks global SU(3) symmetry.

Regarding the performance of the qudit simulator, Table~\ref{tab:TsSscaling} shows the time-to-solution (TtS) values for the different system sizes and number of Trotter steps considered in this work. 
The TtS is found to scale linearly with the number of Trotter steps, as expected, since the circuit depth increases linearly with $n_{\rm Trot}$. 
For a fixed number of Trotter steps, when the system size is changed, we observe two different behaviors.
For $\Omega\lesssim 16$, the TtS scales better than $5^{\Omega/2}$, while for $\Omega \gtrsim 16$, it scales approximately as $5^{\Omega/2}$.
\begin{table}[!htb]
\centering
\begin{tabularx}{0.95\columnwidth}{cYYYYY} 
\hline\hline
$n_{\rm Trot}$ & $\Omega=8$ & $\Omega=12$ & $\Omega=16$ & $\Omega=20$ & $\Omega=24$ \\
\hline
1 & 2 & 7 & 58 & 1921 & 66255 \\
2 & 4 & 14 & 135 & 4803 & 166384 \\
4 & 9 & 28 & 285 & 10448 & 364871 \\
8 & 18 & 56 & 657 & 24005 & 759583 \\
16 & 36 & 113 & 1347 & 43953 & 1539031 \\
32 & 71 & 228 & 2683 & 89047 & - \\
\hline\hline
\end{tabularx}
\caption{
Time-to-solution (TtS) values (in seconds) for the qudit simulator
for different system sizes ($\Omega$) and Trotter steps ($n_{\rm Trot}$). The code was executed on the University of Washington cluster Hyak (mox) on a single node and a single thread.
}
\label{tab:TsSscaling}
\end{table}
As the Hilbert space dimensionality of a 22-qu5it system is approximately that of a 50-qubit system, 
one anticipates that a quantum advantage may become accessible in systems with more than approximately 22 qu5its 
for certain observables.
Gate-based classical simulations of systems approaching this size are beyond the capability of a single-node simulation code,
and will require large-scale parallel computation, as has already been accomplished in classical simulations of gate-based qubit systems~\cite{Smith:2016quil,Steiger:2016projectq,Vlad:2018quan++,Bergholm:2018cyq,Chen:2018qubit,Jones:2019QuEST,Wu:2019compr,qsim,Efthymiou:2020qibo,Guerreschi_2020,Luo:2020yaojl,Suzuki:2021qulacs,Li:2021svsim,leo:2023nvidia}.
The next step toward large-system simulations of qu5it systems is multi-node parallelization that utilizes
multiple GPUs per node.

\section{Comparison with Qubits}
\label{sec:QC2}
\noindent
As mentioned in the previous sections, the Agassi model was first mapped to qubits in Refs.~\cite{P_rez_Fern_ndez_2022,Saiz:2022rof}. 
In those works, a particular JW mapping assignment to qubits was considered.
In order to compare the resource requirements for qubits 
with the requirements for qu5its, given in Sec.~\ref{subsec:ManyModes},
two different mappings are considered here: (i) a physics-aware JW mapping (paJW)
(that differs from that used in Refs.~\cite{P_rez_Fern_ndez_2022,Saiz:2022rof})
and (ii) a mapping using five levels of three qubits as a qu5it.
In the paJW mapping, we have utilized the bosonic nature of mode-pairs from the occupation involving only an even number of fermions to minimize JW strings of $\hat Z$ operators between spins.

\subsection{Physics-aware Jordan-Wigner Mapping to Qubits}
\noindent
A system of $\Omega$ modes can be simulated with $2\Omega$ qubits using a 
JW mapping, in which an
unoccupied state corresponds to a spin-down qubit, while an occupied state corresponds to a spin-up qubit.
Such a mapping was used in the works of Refs.~\cite{P_rez_Fern_ndez_2022,Saiz:2022rof} to explore the one-mode-pair and 
two-mode-pair systems.
We have found it convenient to implement a different 
ordering of states than used in those works, 
where we localize states in the same mode pair to be adjacent, 
in order to take 
better advantage of the approximate SO(5) symmetry.
As the number of particles per mode-pair is always even, 
there are no fermionic phases from operators moving across the lattice. 
The paJW mapping we have employed naturally encodes this feature, 
thereby minimizing the length of Pauli strings between operators.

For the one mode-pair system, requiring four qubits, 
the five states of the qu5it are mapped to the four qubits in the paJW mapping as,\footnote{
In this mapping, an unoccupied state corresponds to spin-down which is identified as  $|1\rangle$, 
while an occupied state is denoted by spin-up, denoted by $|0\rangle$.
}
\begin{align}
|0\rangle \ & \rightarrow \ |1111\rangle
\ , \ 
|1\rangle \ \rightarrow \ |0101\rangle \ , \nonumber\\ 
|2\rangle \ & \rightarrow \ \frac{1}{\sqrt{2}}\left[ |0110\rangle + |1001\rangle \right]
\ ,\nonumber\\
|3\rangle \ & \rightarrow \ |1010\rangle
\ , \ 
|4\rangle \ \rightarrow \  |0000\rangle
\ .
\label{eq:JWstates}
\end{align}
Figure~\ref{fig:JW1pair} in Appendix~\ref{app:JW1mp} shows the 
quantum circuits for one step of LO Trotterized time evolution for one mode-pair, 
which are well-known from quantum chemistry and elsewhere~\cite{Stetina:2020abi,Farrell:2022wyt}.
The resource requirements of one Trotter step, in terms of number of Hadamard-gates, $R_Z$-gates and CNOT-gates, are
\begin{equation}
    n_{\rm 1  mode-pair} \ = \ \left[ 2 \; H, 14 \; R_Z, 14 \; \rm{CNOT} \right] 
    \ .
\end{equation}
The time-evolution circuits for two mode-pairs are  
given in Appendix~\ref{app:JW2mp}.
The resources required to implement Trotterized time evolution from the
terms in the Hamiltonian in Eq.~\eqref{eq:twomodepairHGIVENS} that act only on one mode pair 
are the same as given in the previous paragraph, 
and those required to implement the fundamentally two-mode-pair evolution 
are,
\begin{equation}
    n_{\rm 2  mode-pairs} \ = \ \left[ 16 \; H, 64 \; R_Z, 128 \; \rm{CNOT} \right] 
    \ .
\end{equation}
The resources for an arbitrary number, $\Omega/2$,
of mode pairs follows straightforwardly, 
\begin{align}
    n_{\tfrac{\Omega}{2} \rm mode-pairs} \ = \  \frac{\Omega}{2}\left[ 2 \; H, 14 \; R_Z, 14 \; \rm{CNOT} \right]& \nonumber \\
    +\frac{1}{2}\frac{\Omega}{2}\left(\frac{\Omega}{2} - 1\right) 
    \left[ 16 \; H, 64 \; R_Z, 128 \; \rm{CNOT} \right]&
    \ ,
\end{align}
which will be distributed across a quantum register with $2\Omega$ qubits (four qubits per mode pair).

\subsection{State-to-State Mapping to Qubits}
\noindent
The Hilbert space of a qudit can be mapped to the space of multiple qubits, as is well known, in a state-to-state (StS) map.
In our case, specifically for $d=5$, the five states of a qu5it can be mapped onto five of the eight states of three qubits.
We have chosen the binary mapping, 
\begin{align}
&|0\rangle = |000\rangle \ , \quad |1\rangle = |001\rangle \ , \quad |2\rangle = |010\rangle \ , 
\nonumber \\
&|3\rangle = |011\rangle \ , \quad |4\rangle = |100\rangle 
\ ,
\label{eq:3qmap1}
\end{align}
which best suits  our physical system  (other mappings are discussed in Appendix~\ref{app:Q2Q5}).

The circuits required to perform time evolution 
of a single mode pair 
are given in Appendix~\ref{app:Q2Q51mp}, 
with the 
following number of Hadamard-gates, $R_Z$-gates and CNOT-gates,
\begin{equation}
    n_{\rm 1 mode-pair} \ = \ \left[ 2 \; H, 9 \; R_Z, 10 \; \rm{CNOT} \right] 
    \ .
\end{equation}
These gate requirements correspond to implementing 
one Givens rotation and two phase rotations for a single qu5it.

In the case of two mode pairs, the time-evolution circuits are  
given in Appendix~\ref{app:Q2Q52mp}.
The resource requirements in Table~\ref{tab:costsQ2Q5}
show that this qubit mapping of two mode-pairs requires a substantial number of CNOT-gates per Trotter step, in addition to comparable numbers of single qubit gates and rotations.
From the discussions above, particularly the need for only two mode-pair interactions in a many-body system, the number of gates is found to be
\begin{equation}
   n_{\rm 2 mode-pairs} \ = \ \left[ 32 \; H, 512 \; R_Z, 688 \; \rm{CNOT} \right]
    \ ,
\end{equation}
which is approximately a factor of five greater in the number of entangling gates than the paJW mapping.
The total resource requirements to simulate $\Omega/2$ pairs of modes with this mapping are
\begin{align}
    n_{\tfrac{\Omega}{2}\rm mode-pair} \ = \ \frac{\Omega}{2}\left[ 2 \; H, 9 \; R_Z, 10 \; \rm{CNOT} \right]& \nonumber \\
    +\frac{1}{2}\frac{\Omega}{2}\left(\frac{\Omega}{2} - 1\right) 
    \left[ 32 \; H, 512 \; R_Z, 688 \; \rm{CNOT} \right]&
    \ .
\end{align}
%

\subsection{Comparison Between Qu5it and Qubit Mappings}
\noindent
It is interesting to further compare the three mappings considered in this work, especially the size of the Hilbert space and number of entangling gates. 
Although we have not considered the effect of noisy hardware, these comparisons indicate how prone the different mappings are to errors.

\begin{figure}[!t]
\centering{\includegraphics[width=\columnwidth]{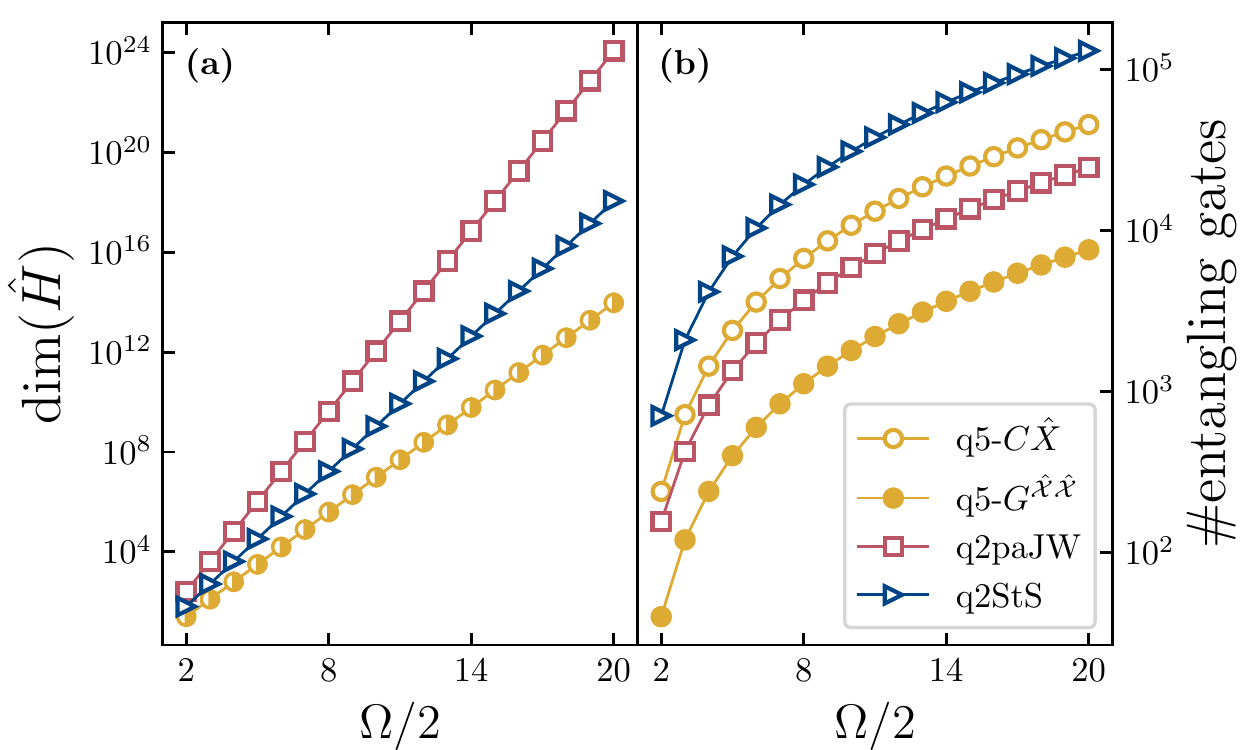}} 
\caption{A comparison of different mappings: qu5its (q5), qubits with paJW (q2paJW) and qubits with StS (q2StS). \textbf{(a)} Dimension of the Hilbert space as a function of the number of mode-pairs. 
\textbf{(b)} Total number of entangling gates (for a single Trotter step) as a function of the number of 
mode-pairs, with $C\hat{X}$-type or $G^{\hat{\mathcal{X}}\hat{\mathcal{X}}}$-type for qu5its, and CNOTs for qubits.
}
\label{fig:mappingComparison}
\end{figure}

Figure~\ref{fig:mappingComparison} shows the dimension of Hilbert space of different mappings as a function of the number of mode-pairs $\Omega/2$, 
as well as the total number of entangling gates, with CNOTs for qubits and $C\hat{X}+C\hat{Y}$ or $G^{\hat{\mathcal{X}}\hat{\mathcal{X}}}+G^{\hat{\mathcal{Y}}\hat{\mathcal{Y}}}$ for qu5its.
The reduced dimensionality of the Hilbert space for qu5its compared to qubits, as clearly shown in Fig.~\ref{fig:mappingComparison}a,
means that qu5its are much less susceptible to errors taking the system to unwanted regions of the Hilbert space, than are qubits.
Importantly, all of the states in the qu5its are physically-allowed states 
(state-changing errors, the equivalent of bit-flip errors, could move the state to a different particle number sector, which is still physically allowed, but prohibited due to the symmetries of the Hamiltonian), 
while the number of unphysical states in both qubit mappings grows exponentially with system size.
Figure~\ref{fig:mappingComparison}b displays the number of entangling gates required for the simulation of systems with increasing system size. 
If two-qu5it Givens rotations of the type $G^{\hat{\mathcal{X}}\hat{\mathcal{X}}}$ 
are directly implemented on the quantum device (plain yellow circles), the mapping to qu5its is the most advantageous in all aspects. Interestingly, if the two-qu5it Givens rotations are implemented via generalized $C\hat{X}$-type gates, the pa-JW mapping requires fewer entangling gates than the mapping to qu5its. 
In that case, there would be a trade-off to be made between the errors due to noise that take the system to unwanted parts of the Hilbert space, and those due to  entangling gates of limited fidelity.

To quantify the impact of noise in the different simulation configurations, 
a realistic noisy simulator of qudit arrays~\cite{10.1145/3307650.3322253,Gustafson:2022xlj}, in addition to the existing ones for qubit systems,
is required.

\section{Summary and Conclusions}
\label{sec:Concl}
\noindent
Pairing interactions between fermions in quantum many-body systems 
play a central role in important physical phenomena, ranging from structure of materials to 
the stability and decay of nuclei.
Building upon our previous work, and pioneering studies by other authors,
we have considered the quantum simulation of the Agassi model, 
which is a model of quantum systems with pairing and particle-hole interactions that extends 
the Lipkin-Meshkov-Glick model.
The nature of the interactions is such that systems involving even numbers of particles are
naturally embedded in an array of $d=5$ qudits (qu5its) that make manifest the underlying SO(5) symmetry.
The quantum circuits required to prepare entangled states and for time evolution of these systems are developed.
They have been classically simulated using a code that we have developed 
on top of Google's {\tt cirq} to examine the time dependence of spin, pairing and persistance probabilities, starting from two distinct initial tensor-product states, one that is the non-interacting ground state, and the other a highly excited configuration.
The time evolutions of these states were found to exhibit quite different behaviors, 
originating from distinct decompositions onto eigenstates of the full Hamiltonian.

A comparison between resource requirements for simulations with qu5its and qubits is revealing.
For the Agassi model, and for the mappings we have identified, 
there are advantages and disadvantages in choosing one over the other.
The nature of the pairing is such that each mode-pair can be mapped to a qu5it, 
and with an even number of fermions occupying each qu5it at all times, 
the system can be considered to be a lattice simulation of a collection of bosons.
While the qu5it array has a significantly smaller Hilbert space than for corresponding systems of qubits, 
depending on the available entangling gates, the number of such gates required for time evolution 
can be larger than for our paJW mapping onto four qubits-per-mode-pair,
but less than for a StS mapping onto 3 qubits-per-mode-pair.

A simulation of the Agassi model with 20 mode-pairs (for a single Trotter step) requires
either 
10 qu5its and 45.6K entangling gates (or 7.6K two-qu5it Givens rotations), 
or 
80 qubits and 24.6K entangling gates (paJW),  
or 
60 qubits and 130.9K entangling gates (StS).
These resource requirements indicate that there is no ``clear winner'' 
between qu5its and qubits for simulating the Agassi model.
If both types of quantum computers were available, 
the choice of which to utilize depends on the overall performance of the hardware, 
which could be assessed with a series of tests and benchmarks~\cite{Jankovic:2023bnw}.
There have been impressive recent advances in developing qudit systems across a number of platforms, 
{\it e.g.}, trapped-ion, NV centers, superconducting and SRF-cavity systems.
It will be interesting to assess the performance of emerging qudit systems in simulations of the types of systems 
considered in this work.  
The advantages gained through a greatly reduced Hilbert space in qu5it systems will be partially off-set by the increased number of entangling gates required for time evolution. 
Benchmarking such devices with the Agassi model will establish another co-design vector for improving 
available qudit hardware (which is anticipated to mostly consist of improving the qudit entangling gates),
to evolve and eventually outperform available qubit systems.

There is a clear need for improving classical simulation capabilities of qudit systems, 
including in the number of qudits that can be addressed, in the speed at which Trotter evolution can be executed, 
and in the quality of the noise models.
With groups developing classical simulation codes for qubit systems, the analogous capabilities for qudit arrays is lagging, largely because the quantum devices are not readily accessible, but this is expected to change in the future.
We intend to extend  the performance of our code to heterogeneous-node parallelization.

There is a more general lesson that can be learned from our analysis of the time evolution starting from a low-lying state compared to a high-lying one.  
The Trotter errors associated with the high-lying state are larger because of the larger number of states making comparable contributions to the evolution.  
In particular, at any given time, there are contributions from multiple amplitudes,
with cancellations that are disturbed by the Trotter errors.  The more amplitudes with cancellations, the more significant are such errors. 
More generally, such sign problems will also manifest themselves 
as signal-to-noise problems due to device noise.
These sign problems will be important in studies of, for example, high-energy fragmentation~\cite{Banerjee:2012pg,Bauer:2019qxa,Verdel:2019chj,Florio:2023dke} in QCD, for
which there are many energetically accessible states, or coupled channels, 
that will contribute with comparable amplitudes to the time evolution.

Finally,
while the present work has focused on the use of symmetry considerations to guide mappings to qudits, we have not attempted to investigate and optimize the entanglement structures of the systems under study.
While a quantitative understanding of entanglement between systems of qubits 
is now well advanced and studies of entanglement between systems of qutrits continues to develop, our work motivates further studies related to the entanglement of qudit systems.
Early pioneering work on entanglement in fermionic paired systems~\cite{Kraus_2009} has shown that pairing is not equivalent to entanglement of the whole state, nor its two-particle reduced density matrix, but manifests as a different type of quantum correlation.  
The model we have studied here and mapped to systems of qu5its, 
provides an explicit quantum system to further study such quantum correlations.
An optimization of the entanglement structure between mode-pairs could potentially be achieved by developing entanglement-driven algorithms, such as the HL-VQE algorithm~\cite{Robin:2023fow} which we plan to extend to qudit systems in the near future. Such entanglement rearrangement, combined with the symmetry-informed mappings developed in this work, could further improve the efficiency of the quantum simulations.

The results shown in figures can be found in a 
{\tt github} repository~\cite{githubdata}.

\section*{Acknowledgments}

We would like to thank 
Anthony Ciavarella,
Roland Farrell and
Momme Hengstenberg
for helpful discussions,
and for all of our other 
colleagues and collaborators that provide the platform from which this work has emerged.
Martin Savage would like to thank 
the High-Energy Physics group
at Universit\"at Bielefeld for kind hospitality during some of this work.
The authors also acknowledge support by the Deutsche Forschungsgemeinschaft (DFG, German Research Foundation) through the CRC-TR 211 'Strong-interaction matter under extreme conditions'– project number 315477589 – TRR 211.
This work was supported, in part,  
by Universit\"at Bielefeld and ERC-885281-KILONOVA Advanced Grant (Robin), 
by U.S. Department of Energy, Office of Science, Office of Nuclear Physics, Inqubator for Quantum Simulation (IQuS) under Award Number DOE (NP) Award DE-SC0020970 (Savage), 
and the Quantum Science Center (QSC),\footnote{\url{https://qscience.org}} 
a National Quantum Information Science Research Center of the U.S.\ Department of Energy (Illa).
This work was enabled, in part, by
the use of advanced computational, storage and networking infrastructure provided by the Hyak supercomputer system at the University of Washington,\footnote{\url{https://hyak.uw.edu}}
and was also supported, in part, through the Department of Physics\footnote{\url{https://phys.washington.edu}}
and the College of Arts and Sciences\footnote{\url{https://www.artsci.washington.edu}}
at the University of Washington. 
We are very appreciative of comments made by a referee for 
{\it Physical Review C}, regarding a simplification of controlled-qu5it gates.
The circuit compression resulting from this observation is shown in 
Figs.~\ref{fig:CX_simplification} and \ref{fig:2Q5XXYYcomp}, 
and led to a modest reduction in quantum resources required for time-evolution.

\clearpage
\onecolumngrid
\appendix

\section{Matrix Representation of the SO(5) Generators} 
\label{app:repSO5}
\noindent
The basis states defined in Eq.~\eqref{eq:5basis}, 
\begin{equation}
\{\ 
|0000\rangle 
\ ,\  
|1010\rangle \  ,\  
\frac{1}{\sqrt{2}}\left( 
|0110\rangle  + |1001\rangle 
  \right)\  ,\  
|0101\rangle\ ,\  
|1111\rangle \ 
\} \ = \ \{\ 
|0\rangle , |1\rangle , |2\rangle , |3\rangle , |4\rangle \ \}
\ ,
\label{eq:5basisAPP}
\end{equation}
are given in terms of the occupation of the two states per mode encapsulated in one qu5it,
$|n_{k \downarrow} ,  n_{k \uparrow} , n_{-k \downarrow} , n_{-k \uparrow} \rangle$.
When written in terms of the action of fermionic creation operators, they have the form
\begin{align}
    \ket{0} \ =\ & \ \ket{\phi} \ ,\quad 
    \ket{1} \ =\  c^\dagger_{k,\downarrow}  c^\dagger_{-k,\downarrow}\ket{\phi} \ ,\quad 
    \ket{2} \ =\  \frac{1}{\sqrt{2}} \left( c^\dagger_{k,\uparrow} c^\dagger_{-k,\downarrow} + c^\dagger_{k,\downarrow} c^\dagger_{-k,\uparrow}  \right) \ket{\phi} \ , \nonumber \\
    \ket{3} \ =\ & \ c^\dagger_{k,\uparrow} c^\dagger_{-k,\uparrow} \ket{\phi} \ ,\quad 
    \ket{4} \ =\  c^\dagger_{k,\downarrow} c^\dagger_{k,\uparrow} c^\dagger_{-k,\downarrow} c^\dagger_{-k,\uparrow} \ket{\phi} \ ,
\end{align}
where $\ket{\phi}$ denotes the vacuum state of the qu5it (all states are unoccupied).
In this basis, the matrix representations of the operators defined in Eq.~\eqref{eq:SO5gensops} are 
(omitting the $k$-label)
\begin{align}
T_+ = &
\sqrt{2} \, 
\begin{pmatrix}
0&0&0&0&0 \\
0&0&0&0&0 \\
0&1&0&0&0 \\
0&0&1&0&0 \\
0&0&0&0&0
\end{pmatrix}
, \ \ T_- = 
\sqrt{2} \, 
\begin{pmatrix}
0&0&0&0&0 \\
0&0&1&0&0 \\
0&0&0&1&0 \\
0&0&0&0&0 \\
0&0&0&0&0
\end{pmatrix}
, \ \ T_z = 
\begin{pmatrix}
0&0&0&0&0 \\
0&-1&0&0&0 \\
0&0&0&0&0 \\
0&0&0&1&0 \\
0&0&0&0&0
\end{pmatrix}
, \nonumber \\
b_\uparrow = & 
\begin{pmatrix}
0&0&0&1&0 \\
0&0&0&0&-1 \\
0&0&0&0&0 \\
0&0&0&0&0 \\
0&0&0&0&0
\end{pmatrix}
, \ \ b_\downarrow =  
\begin{pmatrix}
0&1&0&0&0 \\
0&0&0&0&0 \\
0&0&0&0&0 \\
0&0&0&0&-1 \\
0&0&0&0&0
\end{pmatrix}
, \ \ b_z =
\begin{pmatrix}
0&0&1&0&0 \\
0&0&0&0&0 \\
0&0&0&0&1 \\
0&0&0&0&0 \\
0&0&0&0&0
\end{pmatrix}
,  \nonumber\\
b^\dagger_\uparrow = & 
\begin{pmatrix}
0&0&0&0&0 \\
0&0&0&0&0 \\
0&0&0&0&0 \\
1&0&0&0&0 \\
0&-1&0&0&0
\end{pmatrix}
, \ \ 
b^\dagger_\downarrow =  
\begin{pmatrix}
0&0&0&0&0 \\
1&0&0&0&0 \\
0&0&0&0&0 \\
0&0&0&0&0 \\
0&0&0&-1&0
\end{pmatrix}
, \ \ 
b^\dagger_z = 
\begin{pmatrix}
0&0&0&0&0 \\
0&0&0&0&0 \\
1&0&0&0&0 \\
0&0&0&0&0 \\
0&0&1&0&0
\end{pmatrix}
,  \nonumber\\
N =  &
\begin{pmatrix}
0&0&0&0&0 \\
0&2&0&0&0 \\
0&0&2&0&0 \\
0&0&0&2&0 \\
0&0&0&0&4
\end{pmatrix} \; ,
\label{eq:SO5genmats}
\end{align}
which constitute a set of ten generators of SO(5).
The commutation relations between these operators are given in Table~\ref{tab:commut1}.
\begin{table}[!ht]
\centering
\begin{tabular}{c|c|c|c|c|c|c|c|c|c|c} 
\hline
\hline
                      & $T_+$ & $T_-$    & $T_z$   & $b_\uparrow$     & $b_\downarrow$   & $b_z$                       & $b^\dagger_\uparrow$   & $b^\dagger_\downarrow$   & $b^\dagger_z$                   & $N$ \\
\hline
$T_+$                 & 0     & $2 T_z$  & $-T_+  $  & $-\sqrt{2}b_z$   & 0                & $-\sqrt{2}b_{\downarrow}$   & 0                      & $\sqrt{2} b^\dagger_z$   & $\sqrt{2} b^\dagger_\uparrow$   & 0 \\  
$T_-$                 &       & 0        & $T_-$   & 0                & $-\sqrt{2}b_z$   & $-\sqrt{2}b_{\uparrow}$     & $\sqrt{2} b^\dagger_z$ & 0                        & $\sqrt{2} b^\dagger_\downarrow$ & 0 \\
$T_z$                 &       &          & 0       & $-b_{\uparrow}$  & $b_{\downarrow}$ & 0                           & $b^\dagger_\uparrow$   & $- b^\dagger_\downarrow$ & 0                               & 0 \\ 
$b_\uparrow$          &       &          &         & 0                & 0                & 0                           & $- \widetilde{N}_\uparrow$ & 0                        & $- T_- / \sqrt{2}$              & $2b_\uparrow$ \\    
$b_\downarrow$        &       &          &         &                  & 0                & 0                           & 0                      & $-\widetilde{N}_\downarrow$  & $- T_+ / \sqrt{2}$              & $2b_\downarrow$ \\
$b_z$                 &       &          &         &                  &                  & 0                           & $- T_+ / \sqrt{2}$     & $- T_- / \sqrt{2}$       & $- \widetilde{N}$                   & $2b_z$  \\
$b^\dagger_\uparrow$  &       &          &         &                  &                  &                             & 0                      & 0                        & 0                               & $-2b^\dagger_\uparrow$ \\         
$b^\dagger_\downarrow$&       &          &         &                  &                  &                             &                        & 0                        & 0                               & $-2b^\dagger_\downarrow$ \\
$b^\dagger_z$         &       &          &         &                  &                  &                             &                        &                          & 0                               & $-2b^\dagger_z$ \\
$N$                   &       &          &         &                  &                  &                             &                        &                          &                                 & 0 \\ 
\hline
\hline
\end{tabular}
\caption{
The commutation relations  between the ten SO(5) generators with matrix representations given in Eq.~\eqref{eq:SO5genmats}.
The commutator at the intersection of row(a) and column(b) is ordered as [row(a),column(b)].
}
\label{tab:commut1}
\end{table}
Some of the entries in Table~\ref{tab:commut1} make use of the operators given in Eq.~\eqref{eq:newops},
\begin{align}
  \hat{N}_\uparrow & \ = \ \frac{\hat{N}}{2} + \hat{T}_z \ ,\quad 
  \hat{N}_\downarrow \ =\  \frac{\hat{N}}{2} -\hat{T}_z \ , \nonumber \\ 
 \widetilde{N}_\uparrow &\ = \   \hat{N}_\uparrow  - \frac{\hat{\Omega}_5}{2} \ ,\quad 
  \widetilde{N}_\downarrow \ =\  \hat{N}_\downarrow -\frac{\hat{\Omega}_5}{2} \ ,\quad 
  \widetilde{N} \ =\  \frac{1}{2} \left( \widetilde{N}_\uparrow + \widetilde{N}_\downarrow \right)= \frac{1}{2} \left( \hat{N} - \hat{\Omega}_5 \right) \ , \nonumber \\ 
  \hat{\Omega}_5 & \ = \ \frac{1}{5}\text{Tr}(\hat{N})\hat{I}_5 
  \ .
  \label{eq:newops}
\end{align}

Generators in their standard form, normalized to 
$\text{Tr}( T_a T_b ) = 2 \delta_{ab}$,
are defined through linear combinations of the generators 
in Eq.~\eqref{eq:SO5genmats},
\begin{align}
T_1 =\ & \frac{1}{2}(T_+ + T_-) = \frac{1}{\sqrt{2}}
\begin{pmatrix}
0&0&0&0&0 \\
0&0&1&0&0 \\
0&1&0&1&0 \\
0&0&1&0&0 \\
0&0&0&0&0
\end{pmatrix}
, \ \ T_2 = \frac{1}{2i}(T_+ - T_-) = \frac{i}{\sqrt{2}}
\begin{pmatrix}
0&0&0&0&0 \\
0&0&1&0&0 \\
0&-1&0&1&0 \\
0&0&-1&0&0 \\
0&0&0&0&0
\end{pmatrix}
, \nonumber \\ 
T_3 = & \ T_z = 
\begin{pmatrix}
0&0&0&0&0 \\
0&-1&0&0&0 \\
0&0&0&0&0 \\
0&0&0&1&0 \\
0&0&0&0&0
\end{pmatrix}
, \nonumber \\
T_4 =\ & \frac{1}{\sqrt{2}}( b^\dagger_\downarrow + b_\downarrow) = \frac{1}{\sqrt{2}}
\begin{pmatrix}
0&1&0&0&0 \\
1&0&0&0&0 \\
0&0&0&0&0 \\
0&0&0&0&-1 \\
0&0&0&-1&0
\end{pmatrix}
, \ \ T_5 =  \frac{1}{i\sqrt{2}}( b^\dagger_\downarrow - b_\downarrow) = \frac{i}{\sqrt{2}}
\begin{pmatrix}
0&1&0&0&0 \\
-1&0&0&0&0 \\
0&0&0&0&0 \\
0&0&0&0&-1 \\
0&0&0&1&0
\end{pmatrix}
, \nonumber\\ 
T_6 =\ & \frac{1}{\sqrt{2}}( b^\dagger_\uparrow + b_\uparrow) = \frac{1}{\sqrt{2}}
\begin{pmatrix}
0&0&0&1&0 \\
0&0&0&0&-1 \\
0&0&0&0&0 \\
1&0&0&0&0 \\
0&-1&0&0&0
\end{pmatrix}
, \ \ T_7 =  \frac{1}{i\sqrt{2}}( b^\dagger_\uparrow - b_\uparrow) =  \frac{i}{\sqrt{2}}
\begin{pmatrix}
0&0&0&1&0 \\
0&0&0&0&-1\\
0&0&0&0&0 \\
-1&0&0&0&0 \\
0&1&0&0&0
\end{pmatrix}
, \nonumber\\ 
T_8 =\ & \frac{1}{\sqrt{2}}( b^\dagger_z + b_z ) = \frac{1}{\sqrt{2}}
\begin{pmatrix}
0&0&1&0&0 \\
0&0&0&0&0 \\
1&0&0&0&1 \\
0&0&0&0&0 \\
0&0&1&0&0
\end{pmatrix}
, \ \ T_9 =  \frac{1}{i\sqrt{2}}( b^\dagger_z - b_z ) =\frac{i}{\sqrt{2}}
\begin{pmatrix}
0&0&1&0&0 \\
0&0&0&0&0 \\
-1&0&0&0&1 \\
0&0&0&0&0 \\
0&0&-1&0&0
\end{pmatrix}
, \nonumber\\ 
T_{10} =\ & \frac{1}{2}\left[N - \Omega_5 \right] = 
\begin{pmatrix}
-1&0&0&0&0 \\
0&0&0&0&0 \\
0&0&0&0&0 \\
0&0&0&0&0 \\
0&0&0&0&1
\end{pmatrix} ,
\label{eq:SO5SF}
\end{align}
with their commutators given in Table~\ref{tab:commut2}.
\begin{table}[!ht]
\centering
\begin{tabular}{c|c|c|c|c|c|c|c|c|c|c} 
\hline
\hline
         & $T_1$   & $T_2$   & $T_3$     & $T_4$               & $T_5$                     &  $T_6$             &  $T_7$                   &  $T_8$                &  $T_9$                  &  $T_{10}$    \\
\hline       
$T_1$   & 0        & $i T_3$ & $-i T_2$  & $i T_9 / \sqrt{2}$ & $-i T_8 / \sqrt{2}$        & $i T_9 / \sqrt{2}$ & $-i T_8 / \sqrt{2}$      & $i(T_7+T_5)/\sqrt{2}$ & $-i(T_6+T_4)/\sqrt{2}$ &     0        \\
$T_2$   &          &  0      & $i T_1$   & $-i T_8 / \sqrt{2}$ & $-i T_9 / \sqrt{2}$        & $i T_8 / \sqrt{2}$ & $i T_9 / \sqrt{2}$      &   $i(T_4-T_6)/\sqrt{2}$  & $i(T_5-T_7)/\sqrt{2}$ & 0            \\        
$T_3$   &          &        & 0         & $-i T_5$           & $i T_4$                    & $i T_7$            & $-i T_6$                 & 0                     & 0                      & 0            \\
$T_4$   &          &        &          & 0                  & $i\widetilde{N}_\downarrow$& 0                  & 0                        & $-i T_2 / \sqrt{2}$    & $i T_1 / \sqrt{2}$     & $-i T_5$     \\
$T_5$   &          &        &          &                   & 0                          & 0                  & 0                        & $-i T_1 / \sqrt{2}$   &  $-i T_2 / \sqrt{2}$   & $i T_4$      \\
$T_6$   &          &        &          &                   &                           & 0                  & $i\widetilde{N}_\uparrow$ & $i T_2 / \sqrt{2}$    & $i T_1 / \sqrt{2}$     & $-i T_7$      \\
$T_7$   &          &        &          &                   &                           &                   & 0                        & $-i T_1 / \sqrt{2}$   & $i T_2 / \sqrt{2}$     & $i T_6$       \\
$T_8$   &          &        &          &                   &                           &                   &                         & 0                     & $i T_{10}$             & $-i T_9$      \\  
$T_9$   &          &        &          &                   &                           &                   &                         &                      & 0                      & $i T_8$       \\ 
$T_{10}$  &          &        &          &                   &                           &                   &                        &                      &                       &  0            \\ 
\hline
\hline
\end{tabular}
\caption{
The commutation relations  between the ten SO(5) generators with matrix representations given in Eq.~\eqref{eq:SO5SF}.
The commutator at the intersection of row(a) and column(b) is ordered as [row(a),column(b)].
}
\label{tab:commut2}
\end{table}

Further, a set of $L_{ij}$s can be defined as linear combinations of the $T_i$s, 
\begin{align}
    T_1 & \ =\ L_{12} \ ,\ 
    T_2 \ =\  L_{23} \ ,\ 
    T_3 \ =\  L_{31} \ ,\ 
    T_4  \ =\  - \frac{1}{\sqrt{2}} \left(L_{15} + L_{34}\right) \ , \ 
    T_5  \ =\   \frac{1}{\sqrt{2}} \left(L_{14} - L_{35}\right) \ ,
    \nonumber\\ 
    T_6  & \ =\   \frac{1}{\sqrt{2}} \left(L_{34} - L_{15}\right)\ ,\ 
    T_7  \ =\   \frac{1}{\sqrt{2}} \left(L_{14} + L_{35}\right)\ ,\ 
    T_8 \ =\  L_{24} \ ,\ 
    T_9 \ =\  L_{25} \ ,\ 
    T_{10} \ =\  L_{54} \ ,
\end{align}
such that
(with $L_{ij} = - L_{ji} $),
\begin{equation}
    \left[ L_{ij}, L_{kl} \right] \ =\ i \left( \delta_{jk} L_{il} + \delta_{il} L_{jk} - \delta_{jl} L_{ik} - \delta_{ik} L_{jl} \right) 
\; .
\end{equation}
%

\section{Spectrum of the Agassi Model} 
\label{app:spectrum}
\noindent
The energies of the lowest-lying states in the Agassi model for the sets of couplings in Eq.~(\ref{eq:setdefs})
and for a selection of mode-pair number and occupancy are given in Table~\ref{tab:EvalsEVGsets}.
\begin{table}[!ht]
\centering
\begin{tabularx}{0.95\columnwidth}{cccYYYYY} 
\hline\hline
$\Omega$ & $N$ & Quantity & set-0 & set-1 & set-2 & set-3 & set-4 \\
\hline
2 & 0 & ${\cal E}_0$ & 0.000 & 0.000 & 0.000 & 0.000 & 0.000 \\
\hline
2 & 2 & ${\cal E}_0$ & -0.500 & -0.957 & -1.368 & -1.868 & -2.331 \\
2 & 2 & ${\cal E}_1$ & 0.000 & 0.000 & 0.000 & 0.000 & 0.000 \\
2 & 2 & ${\cal E}_2$ & 0.500 & 0.457 & 0.868 & 0.368 & 0.831 \\
\hline
2 & 4 & ${\cal E}_0$ & 0.000 & -0.500 & -0.500 & -1.500 & -1.500 \\
\hline
\hline
4 & 0 & ${\cal E}_0$ & 0.000 & 0.000 & 0.000 & 0.000 & 0.000 \\
\hline
4 & 2 & ${\cal E}_0$ & -0.250 & -0.701 & -0.923 & -1.660 & -1.902 \\
4 & 2 & ${\cal E}_1$ & -0.250 & -0.280 & -0.451 & -0.280 & -0.451 \\
4 & 2 & ${\cal E}_2$ & 0.000 & 0.000 & 0.000 & 0.000 & 0.000 \\
\hline
4 & 4 & ${\cal E}_0$ & -0.500 & -1.125 & -1.797 & -2.495 & -3.007 \\
4 & 4 & ${\cal E}_1$ & -0.250 & -0.684 & -1.400 & -1.166 & -1.896 \\
4 & 4 & ${\cal E}_2$ & -0.250 & -0.376 & -0.480 & -0.858 & -0.855 \\
\hline
4 & 6 & ${\cal E}_0$ & -0.250 & -0.951 & -1.173 & -2.410 & -2.652 \\
4 & 6 & ${\cal E}_1$ & -0.250 & -0.530 & -0.701 & -1.030 & -1.201 \\
4 & 6 & ${\cal E}_2$ & 0.000 & -0.250 & -0.250 & -0.750 & -0.750 \\
\hline
4 & 8 & ${\cal E}_0$ & 0.000 & -0.500 & -0.500 & -1.500 & -1.500 \\
\hline
\hline
6 & 0 & ${\cal E}_0$ & 0.000 & 0.000 & 0.000 & 0.000 & 0.000 \\
\hline
6 & 2 & ${\cal E}_0$ & -0.167 & -0.623 & -0.777 & -1.600 & -1.764 \\
6 & 2 & ${\cal E}_1$ & -0.167 & -0.186 & -0.300 & -0.186 & -0.300 \\
6 & 2 & ${\cal E}_2$ & -0.167 & -0.186 & -0.300 & -0.186 & -0.300 \\
\hline
6 & 4 & ${\cal E}_0$ & -0.333 & -1.068 & -1.492 & -2.672 & -3.012 \\
6 & 4 & ${\cal E}_1$ & -0.333 & -0.615 & -1.098 & -1.268 & -1.761 \\
6 & 4 & ${\cal E}_2$ & -0.333 & -0.607 & -1.060 & -1.183 & -1.500 \\
\hline
6 & 6 & ${\cal E}_0$ & -0.500 & -1.325 & -2.363 & -3.201 & -3.723 \\
6 & 6 & ${\cal E}_1$ & -0.333 & -1.052 & -2.257 & -1.934 & -2.991 \\
6 & 6 & ${\cal E}_2$ & -0.333 & -0.766 & -1.100 & -1.607 & -2.049 \\
\hline
6 & 8 & ${\cal E}_0$ & -0.333 & -1.234 & -1.659 & -3.171 & -3.512 \\
6 & 8 & ${\cal E}_1$ & -0.333 & -0.782 & -1.265 & -1.768 & -2.261 \\
6 & 8 & ${\cal E}_2$ & -0.333 & -0.773 & -1.227 & -1.683 & -2.000 \\
\hline
6 & 10 & ${\cal E}_0$ & -0.167 & -0.956 & -1.110 & -2.600 & -2.764 \\
6 & 10 & ${\cal E}_1$ & -0.167 & -0.520 & -0.634 & -1.186 & -1.300 \\
6 & 10 & ${\cal E}_2$ & -0.167 & -0.520 & -0.634 & -1.186 & -1.300 \\
\hline
6 & 12 & ${\cal E}_0$ & 0.000 & -0.500 & -0.500 & -1.500 & -1.500 \\
\hline
\hline
8 & 0 & ${\cal E}_0$ & 0.000 & 0.000 & 0.000 & 0.000 & 0.000 \\
\hline
8 & 2 & ${\cal E}_0$ & -0.125 & -0.587 & -0.705 & -1.572 & -1.696 \\
8 & 2 & ${\cal E}_1$ & -0.125 & -0.140 & -0.225 & -0.140 & -0.225 \\
8 & 2 & ${\cal E}_2$ & -0.125 & -0.140 & -0.225 & -0.140 & -0.225 \\
\hline
8 & 4 & ${\cal E}_0$ & -0.250 & -1.042 & -1.350 & -2.755 & -3.010 \\
8 & 4 & ${\cal E}_1$ & -0.250 & -0.583 & -0.948 & -1.323 & -1.694 \\
8 & 4 & ${\cal E}_2$ & -0.250 & -0.580 & -0.901 & -1.285 & -1.515 \\
\hline
8 & 6 & ${\cal E}_0$ & -0.375 & -1.365 & -2.057 & -3.543 & -3.936 \\
8 & 6 & ${\cal E}_1$ & -0.375 & -1.034 & -1.919 & -2.222 & -3.000 \\
8 & 6 & ${\cal E}_2$ & -0.375 & -0.878 & -1.688 & -2.028 & -2.400 \\
\hline
8 & 8 & ${\cal E}_0$ & -0.500 & -1.547 & -3.041 & -3.933 & -4.456 \\
8 & 8 & ${\cal E}_1$ & -0.375 & -1.349 & -3.017 & -2.669 & -3.900 \\
8 & 8 & ${\cal E}_2$ & -0.375 & -1.104 & -1.735 & -2.379 & -3.168 \\
\hline
8 & 10 & ${\cal E}_0$ & -0.375 & -1.490 & -2.182 & -3.918 & -4.311 \\
8 & 10 & ${\cal E}_1$ & -0.375 & -1.159 & -2.044 & -2.597 & -3.375 \\
8 & 10 & ${\cal E}_2$ & -0.375 & -1.003 & -1.813 & -2.403 & -2.775 \\
\hline
8 & 12 & ${\cal E}_0$ & -0.250 & -1.292 & -1.600 & -3.505 & -3.760 \\
8 & 12 & ${\cal E}_1$ & -0.250 & -0.833 & -1.198 & -2.073 & -2.444 \\
8 & 12 & ${\cal E}_2$ & -0.250 & -0.830 & -1.151 & -2.035 & -2.265 \\
\hline
8 & 14 & ${\cal E}_0$ & -0.125 & -0.962 & -1.080 & -2.697 & -2.821 \\
8 & 14 & ${\cal E}_1$ &  -0.125 & -0.515 & -0.600 & -1.265 & -1.350 \\
8 & 14 & ${\cal E}_2$ &  -0.125 & -0.515 & -0.600 & -1.265 & -1.350 \\
\hline
8 & 16 & ${\cal E}_0$ & 0.000 & -0.500 & -0.500 & -1.500 & -1.500 \\
\hline\hline
\end{tabularx}
\caption{
The energy density, ${\cal E}_i = E_i/\Omega$
(average energy per mode),
of the ground state, first- and second-excited states of the Agassi model for the sets of parameters defined in the text, $(\varepsilon , V , g)$:
set-0 $ = (1.0,0.0,0.0)$, 
set-1 $ = (1.0,0.5,0.5)$, 
set-2 $ = (1.0,1.5,0.5)$, 
set-3 $ = (1.0,0.5,1.5)$, 
set-4 $ = (1.0,1.5,1.5)$. 
}
\label{tab:EvalsEVGsets}
\end{table}
%

\section{Matrix Representation of Single-Qu5it Givens Operators} 
\label{app:repGIVENS}
\noindent
The matrix representations of selected ${\cal X}_{ij}$ Givens operators for a single qu5it are
\begin{align}
{\cal X}_{01} = & 
\begin{pmatrix}
0&1&0&0&0 \\
1&0&0&0&0 \\
0&0&0&0&0 \\
0&0&0&0&0 \\
0&0&0&0&0
\end{pmatrix}
, \ 
{\cal X}_{03} =  
\begin{pmatrix}
0&0&0&1&0 \\
0&0&0&0&0 \\
0&0&0&0&0 \\
1&0&0&0&0 \\
0&0&0&0&0
\end{pmatrix}
, \
{\cal X}_{12} = 
\begin{pmatrix}
0&0&0&0&0 \\
0&0&1&0&0 \\
0&1&0&0&0 \\
0&0&0&0&0 \\
0&0&0&0&0
\end{pmatrix}, 
\nonumber\\
{\cal X}_{14} =  &
\begin{pmatrix}
0&0&0&0&0 \\
0&0&0&0&1 \\
0&0&0&0&0 \\
0&0&0&0&0 \\
0&1&0&0&0
\end{pmatrix}, \
{\cal X}_{23} =
\begin{pmatrix}
0&0&0&0&0 \\
0&0&0&0&0 \\
0&0&0&1&0 \\
0&0&1&0&0 \\
0&0&0&0&0
\end{pmatrix}, \ 
{\cal X}_{34} =  
\begin{pmatrix}
0&0&0&0&0 \\
0&0&0&0&0 \\
0&0&0&0&0 \\
0&0&0&0&1 \\
0&0&0&1&0
\end{pmatrix}
 ,
\label{eq:Xij}
\end{align}
and selected 
${\cal Y}_{ij}$ Givens operators are
\begin{align}
{\cal Y}_{01} =\; & 
i \begin{pmatrix}
0&-1&0&0&0 \\
1&0&0&0&0 \\
0&0&0&0&0 \\
0&0&0&0&0 \\
0&0&0&0&0
\end{pmatrix}
, \ 
{\cal Y}_{03} =  
i\begin{pmatrix}
0&0&0&-1&0 \\
0&0&0&0&0 \\
0&0&0&0&0 \\
1&0&0&0&0 \\
0&0&0&0&0
\end{pmatrix}, \
{\cal Y}_{12} =
i\begin{pmatrix}
0&0&0&0&0 \\
0&0&-1&0&0 \\
0&1&0&0&0 \\
0&0&0&0&0 \\
0&0&0&0&0
\end{pmatrix}, \nonumber \\ 
{\cal Y}_{14} = \; & 
i\begin{pmatrix}
0&0&0&0&0 \\
0&0&0&0&-1 \\
0&0&0&0&0 \\
0&0&0&0&0 \\
0&1&0&0&0
\end{pmatrix},  \
{\cal Y}_{23} =
i\begin{pmatrix}
0&0&0&0&0 \\
0&0&0&0&0 \\
0&0&0&-1&0 \\
0&0&1&0&0 \\
0&0&0&0&0
\end{pmatrix}, \ 
{\cal Y}_{34} =  
i\begin{pmatrix}
0&0&0&0&0 \\
0&0&0&0&0 \\
0&0&0&0&0 \\
0&0&0&0&-1 \\
0&0&0&1&0
\end{pmatrix}
.
\label{eq:Yij}
\end{align}

\section{Long-Time Evolution} 
\label{app:longtime}
\noindent
In this section, time evolution of modest-size systems are performed over longer time intervals.
This is to look for behaviors and periodicities that are not made apparent over shorter time intervals.
\begin{figure}[!tb]
\centering
\includegraphics[width=\columnwidth]{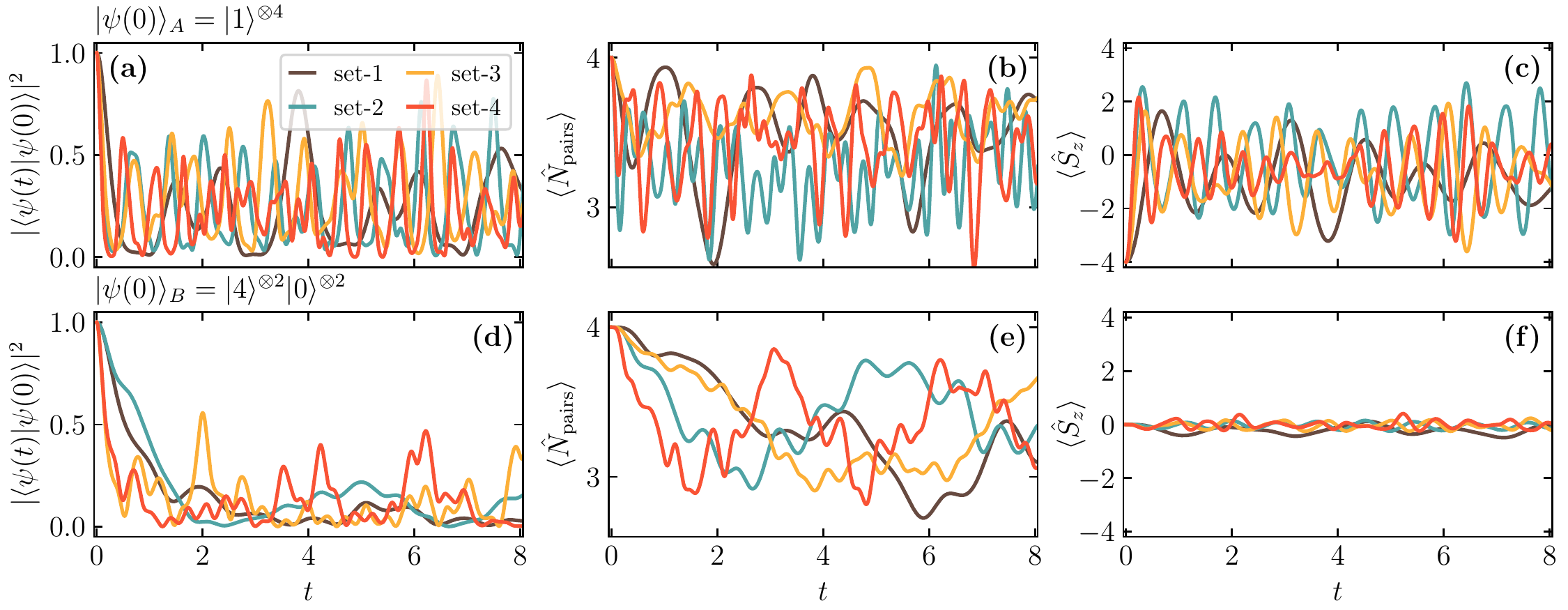} 
\includegraphics[width=\columnwidth]{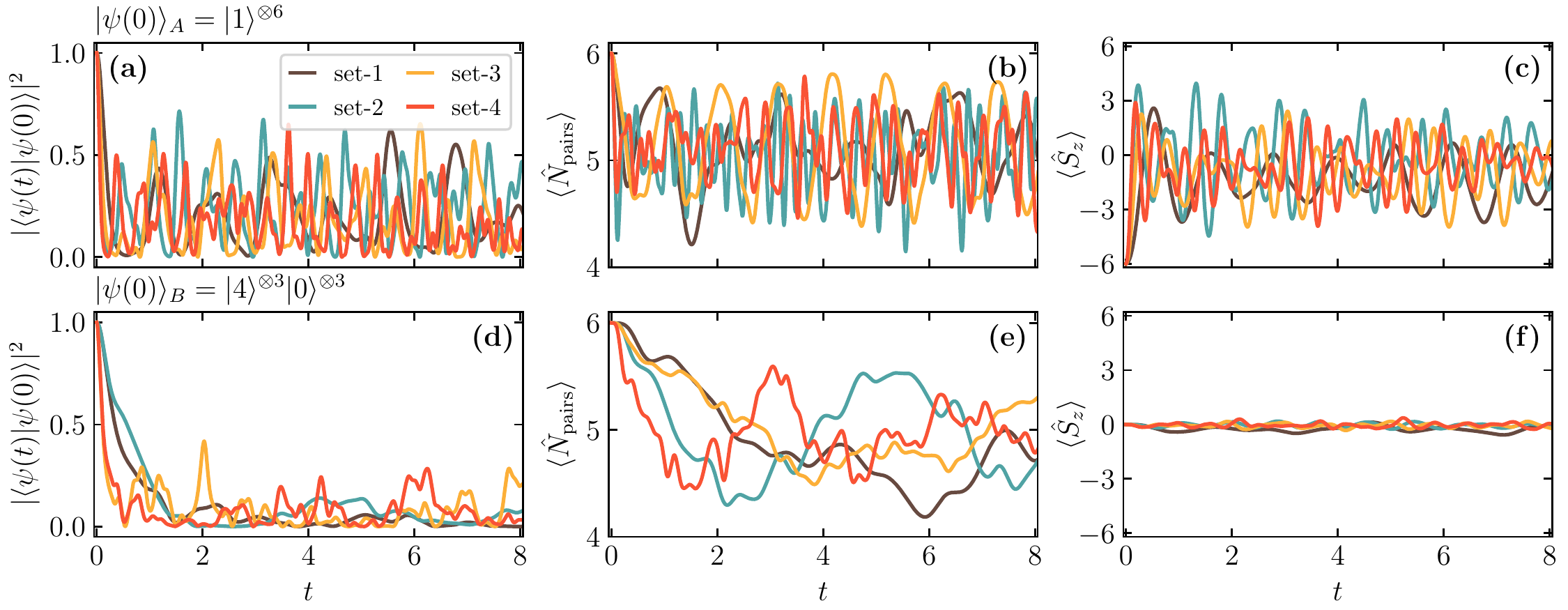} 
\includegraphics[width=\columnwidth]{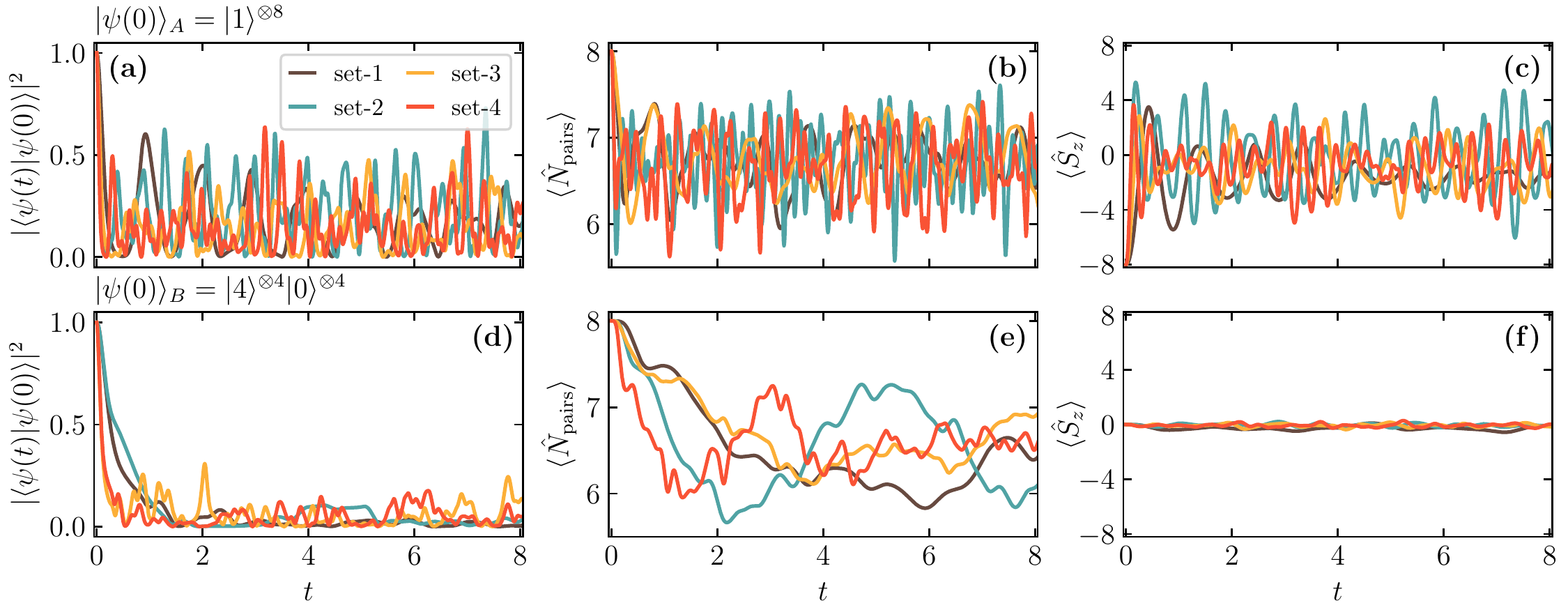} 
\caption{The exact time dependence of \textbf{(a,d)} $|\langle \psi (t)|\psi (0)\rangle|^2$, \textbf{(b,e)} number of pairs, 
and \textbf{(c,f)} 
$\langle \hat S_z \rangle$
for a system with 
$N=\Omega=8$ particles (rows 1-2),
$N=\Omega=12$ particles (rows 3-4),
and $N=\Omega=16$ particles (rows 5-6),
for four sets of couplings, set-1 to set-4, given in Eq.~(\ref{eq:setdefs}).
For the different panels, \textbf{(a-c)} are associated with the initial state 
$|\psi(0)\rangle_A$,
while \textbf{(d-f)} with 
$|\psi(0)\rangle_B$, as given in Eq.~(\ref{eq:psi0s}).
}
\label{fig:N8alllongT}
\end{figure}
Starting from $|\psi(0)\rangle_A$, the observables continue to oscillate at long times, even as the interaction terms increase in strength.
In contrast, starting from $|\psi(0)\rangle_B$, the observables rapidly converge to their average late time values,
with fluctuations that decrease with increasing strength of interactions.

\section{Circuits for the Qubit-Based Mappings} 
\label{app:qubitcircuits}
\noindent

\subsection{A Physics-Aware Jordan-Wigner Mapping}
\noindent
In this Appendix, we present a physics-aware JW mapping of the Agassi model to qubits,
which requires four qubits per mode-pair.
The $s_z=\pm \frac{1}{2}$ ($\sigma =\; \uparrow, \downarrow$) states of a given mode 
are arranged to be adjacent to minimize the length of JW $Z$-strings. 
Specifically, for one mode, we set
\begin{equation}
    \hat{c}_0=\hat{c}_{1,\downarrow}\ , \quad 
    \hat{c}_1=\hat{c}_{1,\uparrow}\ , \quad 
    \hat{c}_2=\hat{c}_{-1,\downarrow}\ , \quad 
    \hat{c}_3=\hat{c}_{-1,\uparrow}
    \ ,
\end{equation}
and the operators defined in Eq.~\eqref{eq:SO5gensops} can be written in terms of spin operators 
via the JW transformation,
\begin{equation}
    \hat{c}_n=\bigotimes_{l<n}(-\hat{Z}_l) \hat{\sigma}_n^- \ , \quad 
    \hat{c}^\dag_n=\bigotimes_{l<n}(-\hat{Z}_l)\hat{\sigma}_n^+ \ ,
\end{equation}
with $\hat{\sigma}^{\pm}=(\hat{X}\pm i\hat{Y})/2$. Therefore,
\begin{align}
    \hat{T}_{1,+} \ & = \ \hat{\sigma}_0^-\hat{\sigma}_1^+ + \hat{\sigma}_2^-\hat{\sigma}_3^+ \ , \quad
    \hat{T}_{1,-} \ = \ \hat{\sigma}_0^+\hat{\sigma}_1^- + \hat{\sigma}_2^+\hat{\sigma}_3^- \ , \quad
    \hat{T}_{1,z} \ = \ \frac{1}{2}(\hat{\Lambda}^{(0)}_1+\hat{\Lambda}^{(0)}_3-\hat{\Lambda}^{(0)}_0-\hat{\Lambda}^{(0)}_2) \ , \nonumber \\
    \hat{b}_{1,\uparrow} \ & = \ -\hat{\sigma}_1^- \hat{Z}_2 \hat{\sigma}_3^- \ , \quad 
    \hat{b}_{1,\downarrow} \ = \ -\hat{\sigma}_0^- \hat{Z}_1 \hat{\sigma}_2^- \ , \quad 
    \hat{b}_{1,z} \ = \ \frac{1}{\sqrt{2}}(\hat{\sigma}_0^- \hat{Z}_1\hat{Z}_2 \hat{\sigma}_3^- + \hat{\sigma}_1^-\hat{\sigma}_2^-) \ ,
\end{align}
with $\hat{\Lambda}^{(0)/(1)}_n=(1\pm\hat{Z}_n)/2$. 
Throughout this appendix, subscripts are used to denote the target qubit, for example,
$\hat{\sigma}_0^-\hat{\sigma}_1^+$  means
$\hat{\sigma}^-\otimes \hat{\sigma}^+ \otimes \hat{I} \otimes \hat{I}$.

\subsubsection{One Mode-Pair}
\label{app:JW1mp}
\begin{figure}[!ht]
\centering
\includegraphics[width=\columnwidth]{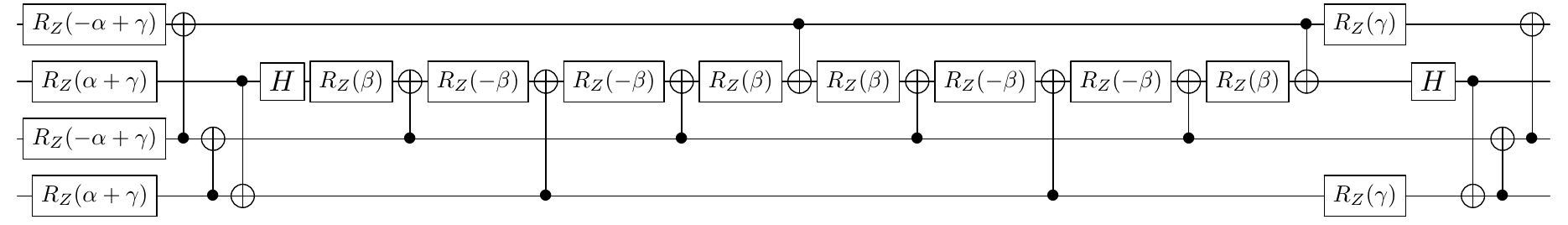}
\caption{A quantum circuit required for one LO Trotter step of time evolution for a single mode-pair 
using the paJW mapping to four qubits, from the Hamiltonian given in Eq.~\eqref{eq:JWH2}, with $\{\alpha=t\epsilon/2, \beta=-t(V+g)/4, \gamma=-tg/2\}$.
}
\label{fig:JW1pair}
\end{figure}

\noindent
The Hamiltonian for a single mode pair is given by
\begin{equation}
    \hat{H}_1 \ = \ \frac{\epsilon}{2} (\hat{\Lambda}^{(0)}_1+\hat{\Lambda}^{(0)}_3-\hat{\Lambda}^{(0)}_0-\hat{\Lambda}^{(0)}_2) \ - \ (V+g) (\hat{\sigma}_0^-\hat{\sigma}_1^+\hat{\sigma}_2^-\hat{\sigma}_3^+ + \text{H.c.}) \ - \ g(\hat{\Lambda}^{(0)}_0\hat{\Lambda}^{(0)}_2 + \hat{\Lambda}^{(0)}_1\hat{\Lambda}^{(0)}_3)
\ .
\label{eq:JWH2}
\end{equation}
In order to implement the (Trotterized) time evolution operator for this Hamiltonian, 
 the strategies from Refs.~\cite{Stetina:2020abi,Farrell:2022wyt} will be followed, which introduce the $G$ operator that diagonalizes the $(\hat{\sigma}_0^-\hat{\sigma}_1^+\hat{\sigma}_2^-\hat{\sigma}_3^+ + \rm{H.c.})$ operators. 
For our present purposes,
\begin{equation}
    G \ = \begin{gathered}
    \includegraphics{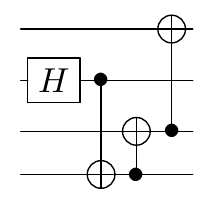}
  \end{gathered} ,
  \label{eq:Gop}
\end{equation}
which leads to 
\begin{align}
  G^\dag &(\hat{\sigma}_0^-\hat{\sigma}_1^+\hat{\sigma}_2^-\hat{\sigma}_3^+ + \text{H.c.}) G \nonumber \\
  &= \frac{1}{8} G^\dag (\hat{X}_0\hat{X}_1\hat{X}_2\hat{X}_3
   +\hat{X}_0\hat{X}_1\hat{Y}_2\hat{Y}_3
   -\hat{X}_0\hat{Y}_1\hat{X}_2\hat{Y}_3
   +\hat{X}_0\hat{Y}_1\hat{Y}_2\hat{X}_3
   +\hat{Y}_0\hat{X}_1\hat{X}_2\hat{Y}_3
   -\hat{Y}_0\hat{X}_1\hat{Y}_2\hat{X}_3
   +\hat{Y}_0\hat{Y}_1\hat{X}_2\hat{X}_3
   +\hat{Y}_0\hat{Y}_1\hat{Y}_2\hat{Y}_3)G  \nonumber\\
   &=\frac{1}{8}(\hat{I}_0\hat{Z}_1\hat{I}_2\hat{I}_3
   -\hat{I}_0\hat{Z}_1\hat{Z}_2\hat{I}_3
   +\hat{I}_0\hat{Z}_1\hat{I}_2\hat{Z}_3
   -\hat{I}_0\hat{Z}_1\hat{Z}_2\hat{Z}_3
   -\hat{Z}_0\hat{Z}_1\hat{Z}_2\hat{I}_3
   +\hat{Z}_0\hat{Z}_1\hat{I}_2\hat{I}_3
   -\hat{Z}_0\hat{Z}_1\hat{Z}_2\hat{Z}_3
   +\hat{Z}_0\hat{Z}_1\hat{I}_2\hat{Z}_3) \ .
\end{align}
Also, the $\hat{Z}\hat{Z}$ terms in $(\hat{\Lambda}^{(0)}_0\hat{\Lambda}^{(0)}_2 + \hat{\Lambda}^{(0)}_1\hat{\Lambda}^{(0)}_3)$ become a single $\hat{Z}$ operator,
\begin{equation}
    G^\dag (\hat{Z}_0\hat{I}_1\hat{Z}_2\hat{I}_3
    +\hat{I}_0\hat{Z}_1\hat{I}_2\hat{Z}_3)G \ = \ \hat{Z}_0\hat{I}_1\hat{I}_2\hat{I}_3 + \hat{I}_0\hat{I}_1\hat{I}_2\hat{Z}_3 \ .
\end{equation}
The quantum circuit required for one LO Trotter step is shown in Fig.~\ref{fig:JW1pair}.
This circuit reproduces the desired evolution up to an overall phase, $e^{-i g t/2}$.

\subsubsection{Two Mode-Pairs}
\label{app:JW2mp}
\begin{figure}[!ht]
\centering
\includegraphics{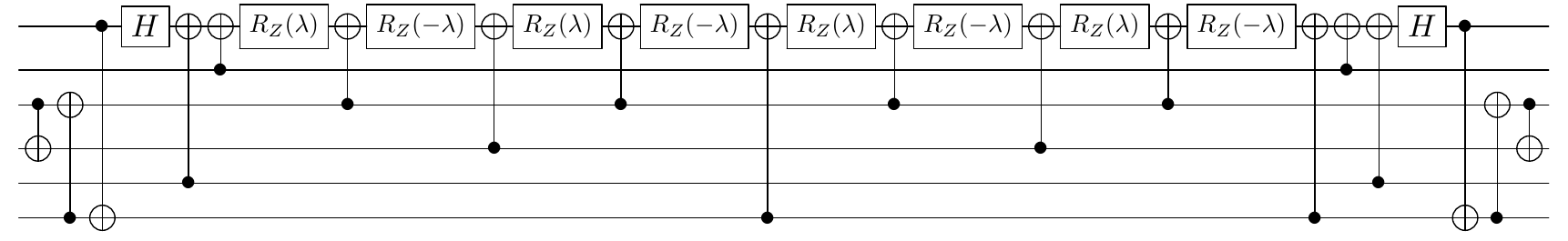}
\caption{A quantum circuit required for one Trotter step of time evolution for two  mode-pairs 
using the paJW mapping to four qubits per mode, from the term $\exp{[itg(\hat{\sigma}_0^+ \hat Z_1 \hat{\sigma}_2^+\hat{\sigma}_4^- \hat Z_5 \hat{\sigma}_6^- + \text{H.c.})]}$ in the Hamiltonian given in Eq.~\eqref{eq:JWH4}, with $\lambda=-tg/4$ (the lines representing qubits 3 and 7 are omitted for clarity).
}
\label{fig:JW2pair}
\end{figure}

\noindent
The paJW mapping that we have chosen, and the nature of the system, 
means that the required operator structure of the Hamiltonian does 
not have any ``dangling'' Pauli strings between and across sites. For example, the operator $c_4$, in terms of spin operators, is $\hat{\sigma}^-_4$ and not $\hat{Z}_0\hat{Z}_1\hat{Z}_2\hat{Z}_3\hat{\sigma}^-_4$ (the spins between positions 0 to 3 will always be up or down in even numbers).
This greatly simplifies the form of the Hamiltonian.
The terms of the Hamiltonian acting within one mode pair are simply tensor-product operators, while the interactions acting across mode pairs  can be compactly written, leading to 
\begin{align}
    \hat{H}_2 \ = \ 
    \hat{H}_1\otimes \hat I^{\otimes 4} + \hat I^{\otimes 4}\otimes  \hat{H}_1 
    \ & - \ 2V  \left[\left(\hat{\sigma}_0^-\hat{\sigma}_1^+ + 
    \hat{\sigma}_2^-\hat{\sigma}_3^+ \right)\left(\hat{\sigma}_4^-\hat{\sigma}_5^+ + \hat{\sigma}_6^-\hat{\sigma}_7^+ \right) \ + \ \rm{H.c.}\right] \nonumber \\
    &-\ g \left[ \left(
    \hat{\sigma}_0^+ \hat Z_1 \hat{\sigma}_2^+ + 
    \hat{\sigma}_1^+ \hat Z_2 \hat{\sigma}_3^+ \right)
    \left(\hat{\sigma}_4^- \hat Z_5 \hat{\sigma}_6^- + 
    \hat{\sigma}_5^- \hat Z_6 \hat{\sigma}_7^- \right)
    \ +\ {\rm H.c.}\right]
\ .
\label{eq:JWH4}
\end{align}
There are eight multi-qubit operators (and hermitian conjugates) required to time evolve the wavefunction beyond the circuits required to evolve one mode-pair. The terms with coefficient $2V$ (four-qubit terms) use the same circuit as the one shown in Fig.~\ref{fig:JW1pair} with $\{\alpha=0,\beta=-tV/2,\gamma=0\}$, and for the ones with coefficient $g$ (six-qubit terms), an example is shown in Fig.~\ref{fig:JW2pair}, where a similar operator as the one in Eq.~\eqref{eq:Gop} is introduced to diagonalize the $(\hat{\sigma}_0^+ \hat Z_1 \hat{\sigma}_2^+\hat{\sigma}_4^- \hat Z_5 \hat{\sigma}_6^- + \text{H.c.})$-type operators. In this case,
\begin{equation}
    \tilde{G} \ = \begin{gathered}
    \includegraphics{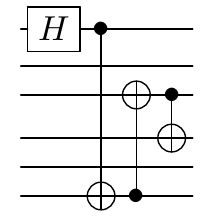}
  \end{gathered},
  \label{eq:tGop}
\end{equation}
where the ``idling qubits'' (lines with no gates applied) have to match with the index $n$ of the operator $\hat{Z}_n$. This leads to
\begin{align}
   \tilde{G}^\dag &(\hat{\sigma}_0^+ \hat Z_1 \hat{\sigma}_2^+\hat{\sigma}_4^- \hat Z_5 \hat{\sigma}_6^- + \text{H.c.}) \tilde{G} \nonumber \\
   &= \frac{1}{8} \tilde{G}^\dag (\hat{X}_0\hat{Z}_1\hat{X}_2\hat{X}_3\hat{Z}_4\hat{X}_5
    -\hat{X}_0\hat{Z}_1\hat{X}_2\hat{Y}_3\hat{Z}_4\hat{Y}_5
    +\hat{X}_0\hat{Z}_1\hat{Y}_2\hat{X}_3\hat{Z}_4\hat{Y}_5
    +\hat{X}_0\hat{Z}_1\hat{Y}_2\hat{Y}_3\hat{Z}_4\hat{X}_5
    +\hat{Y}_0\hat{Z}_1\hat{X}_2\hat{X}_3\hat{Z}_4\hat{Y}_5 \nonumber \\
    & \quad \quad \quad \quad +\hat{Y}_0\hat{Z}_1\hat{X}_2\hat{Y}_3\hat{Z}_4\hat{X}_5
    -\hat{Y}_0\hat{Z}_1\hat{Y}_2\hat{X}_3\hat{Z}_4\hat{X}_5
    +\hat{Y}_0\hat{Z}_1\hat{Y}_2\hat{Y}_3\hat{Z}_4\hat{Y}_5) \tilde{G}  \nonumber\\
    &= \frac{1}{8} (\hat{Z}_0\hat{Z}_1\hat{I}_2\hat{I}_3\hat{Z}_4\hat{I}_5
    +\hat{Z}_0\hat{Z}_1\hat{Z}_2\hat{Z}_3\hat{Z}_4\hat{I}_5
    -\hat{Z}_0\hat{Z}_1\hat{Z}_2\hat{I}_3\hat{Z}_4\hat{I}_5
    -\hat{Z}_0\hat{Z}_1\hat{I}_2\hat{Z}_3\hat{Z}_4\hat{I}_5
    -\hat{Z}_0\hat{Z}_1\hat{I}_2\hat{I}_3\hat{Z}_4\hat{Z}_5 \nonumber \\
    & \quad \quad \quad \quad -\hat{Z}_0\hat{Z}_1\hat{Z}_2\hat{Z}_3\hat{Z}_4\hat{Z}_5
    +\hat{Z}_0\hat{Z}_1\hat{Z}_2\hat{I}_3\hat{Z}_4\hat{Z}_5
    +\hat{Z}_0\hat{Z}_1\hat{I}_2\hat{Z}_3\hat{Z}_4\hat{Z}_5) \ .
\end{align}
%

\subsection{State-To-State Qubit-Qu5it Mapping}
\label{app:Q2Q5}
\noindent
In this  Appendix, we present the mapping of the Agassi model to qubits, 
with three qubits per mode-pair containing 
the five dimensional Hilbert space.
Using the basis defined in the main text in Eq.~\eqref{eq:3qmap1},
\begin{equation}
|0\rangle = |000\rangle \ , \quad |1\rangle = |001\rangle \ , \quad |2\rangle = |010\rangle \ , \quad |3\rangle = |011\rangle \ , \quad |4\rangle = |100\rangle \ ,
\label{eq:1mpPHYS}
\end{equation}
the Hamiltonian and corresponding circuits to perform time evolution can be constructed.
While other mappings have also been pursued, 
the number of entangling gates is found to be minimized with this choice. 
For example, if instead of its binary representation, identifying
$|4\rangle = |101\rangle$
leads to a reduced number (by 2) of CNOT gates for the single mode-pair terms, but an increase for the two mode-pair terms (by 32). 
Another explored mapping is the Gray encoding, which leads an increase in CNOT gates for both the single and two mode-pair terms.

The states in Eq.~(\ref{eq:1mpPHYS}) constitute the physical space, mapping to the five states of a mode-pair, 
while states $|101\rangle , |110\rangle , |111\rangle$ 
are unphysical and only populated through errors in simulation.
In this basis, the operators in Eq.~\eqref{eq:SO5gensops} take the following form,
\begin{align}
    \hat{T}_{1,+} \ &= \ \hat{\Lambda}^{(0)}_0(\hat{\sigma}_1^-\hat{\sigma}_2^+ + \hat{\Lambda}^{(1)}_1\hat{\sigma}_2^-) \ , \quad 
    \hat{T}_{1,-} \ = \ \hat{\Lambda}^{(0)}_0(\hat{\sigma}_1^+\hat{\sigma}_2^- + \hat{\Lambda}^{(1)}_1\hat{\sigma}_2^+) \ , \quad 
    \hat{T}_{z} \ = \ \hat{\Lambda}^{(0)}_0\hat{Z}_1\hat{\Lambda}^{(1)}_2 \ , \nonumber \\
    \hat{b}_{1,\uparrow} \ & = \ \hat{\Lambda}^{(0)}_0\hat{\sigma}_1^+\hat{\sigma}_2^+ - \hat{\sigma}_0^+\hat{\Lambda}^{(0)}_1\hat{\sigma}_2^- \ , \quad
    \hat{b}_{1,\downarrow} \ = \ \hat{\Lambda}^{(0)}_0\hat{\Lambda}^{(0)}_1\hat{\sigma}_2^+ - \hat{\sigma}_0^+\hat{\sigma}_1^-\hat{\sigma}_2^- \ .
\end{align}
%

\subsubsection{One Mode-Pair}
\label{app:Q2Q51mp}
\noindent
The single mode-pair Hamiltonian in Eq.~\eqref{eq:onemodepairH} is reproduced by,
\begin{align}
    \hat{H}_1\ &= \ \varepsilon\; \hat{\Lambda}^{(0)}_0 \hat{Z}_1 \hat{\Lambda}^{(1)}_2 
    \  - \ (V+g)\; \hat{\Lambda}^{(0)}_0\hat{X}_1 \hat{\Lambda}^{(1)}_2
    \ - \ g\; (\hat{\Lambda}^{(0)}_0\hat{I}_1\hat{\Lambda}^{(1)}_2 + 2 \hat{\Lambda}^{(1)}_0 \hat{\Lambda}^{(0)}_1 \hat{\Lambda}^{(0)}_2)
    \nonumber \\
    &= \ \varepsilon\; \hat{\Lambda}^{(0)}_0 \hat{Z}_1 \hat{\Lambda}^{(1)}_2 
    \  - \ (V+g)\; \hat{\Lambda}^{(0)}_0\hat{X}_1 \hat{\Lambda}^{(1)}_2
    \ - \ \frac{g}{2}\; (\hat{I}_0\hat{I}_1\hat{I}_2 - \hat{Z}_0\hat{I}_1\hat{Z}_2 + 2\hat{\Lambda}^{(1)}_0 \hat{Z}_1 \hat{\Lambda}^{(0)}_2) \ ,
    \label{eq:Q2Q5H2}
\end{align}
and the (LO Trotterized) time evolution can be accomplished by,
\begin{equation}
    \hat{U}_2(t) \ \approx \ e^{-it(\varepsilon \hat{\Lambda}^{(0)}_0 \hat{Z}_1 \hat{\Lambda}^{(1)}_2 -g \hat{\Lambda}^{(1)}_0 \hat{Z}_1 \hat{\Lambda}^{(0)}_2)}\;
    e^{it(V+g)\hat{\Lambda}^{(0)}_0 \hat{X}_1 \hat{\Lambda}^{(1)}_2}\;
    e^{it\frac{g}{2}\hat{Z}_0\hat{I}_1\hat{Z}_2} 
    \ .
\end{equation}
These terms can be  implemented with the quantum circuits:
\begin{subequations}
\begin{align} 
   e^{-it(\varepsilon \hat{\Lambda}^{(0)}_0 \hat{Z}_1 \hat{\Lambda}^{(1)}_2 -g \hat{\Lambda}^{(1)}_0 \hat{Z}_1 \hat{\Lambda}^{(0)}_2)}\ & : \begin{gathered}
    \includegraphics{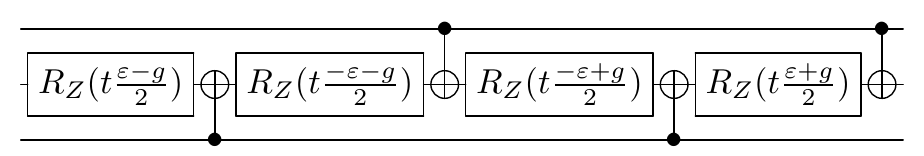}
  \end{gathered}, \\
  e^{it(V+g)\hat{\Lambda}^{(0)}_0 \hat{X}_1 \hat{\Lambda}^{(1)}_2}\ & : \begin{gathered}
    \includegraphics{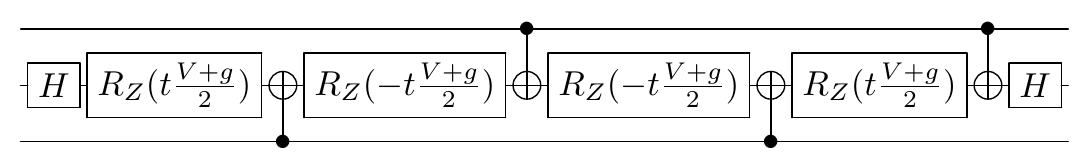}
  \end{gathered} \label{eq:Vgcircuit}, \\
  e^{it\frac{g}{2}\hat{Z}_0\hat{I}_1\hat{Z}_2}\ & : \begin{gathered}
    \includegraphics{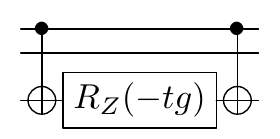}
  \end{gathered},
\end{align}
\end{subequations}
where the global phase from the last term in Eq.~\eqref{eq:Q2Q5H2}, $\hat{I}_0\hat{I}_1\hat{I}_2$, is not included.

\subsubsection{Two Mode-Pairs}
\label{app:Q2Q52mp}
\noindent

For two mode-pairs, the Hamiltonian becomes,
\begin{align}
    \hat{H}_2 \ = & \ 
    \hat{H}_1\otimes \hat I^{\otimes 4} + \hat I^{\otimes 4}\otimes  \hat{H}_1 
    \  - \ 2V  \left[\hat{\Lambda}^{(0)}_0(\hat{\sigma}_1^+\hat{\sigma}_2^-+\hat{\Lambda}^{(1)}_1\hat{\sigma}_2^+)\hat{\Lambda}^{(0)}_3(\hat{\sigma}_4^+\hat{\sigma}_5^- + \hat{\Lambda}^{(1)}_4\hat{\sigma}_5^+) \ + \ \rm{H.c.}\right] \nonumber \\
    &- \ g \left\lbrace \left[ \hat{\Lambda}^{(0)}_0(\hat{\sigma}_1^- + \hat{\Lambda}^{(0)}_1)\hat{\sigma}_2^- - \hat{\sigma}_0^-(\hat{\Lambda}^{(0)}_1+\hat{\sigma}_1^+)\hat{\sigma}_2^+\right]
    \left[ 
    \hat{\Lambda}^{(0)}_3(\hat{\sigma}_4^+ + \hat{\Lambda}^{(0)}_4)\hat{\sigma}_5^+ - \hat{\sigma}_3^+(\hat{\Lambda}^{(0)}_4+\hat{\sigma}_4^-)\hat{\sigma}_5^-\right]
    \ +\ {\rm H.c.}\right\rbrace
\ .
\label{eq:Q2Q5H4}
\end{align}
Upon inspection,
there are found to be four contributions from the $2V$ terms,
\begin{align}
    &(a)\quad \hat{\Lambda}^{(0)}_0 \hat{\sigma}_1^+ \hat{\sigma}_2^-\hat{\Lambda}^{(0)}_3 \hat{\sigma}_4^+ \hat{\sigma}_5^- + \text{H.c.} \ , \quad
    (b)\quad \hat{\Lambda}^{(0)}_0 \hat{\sigma}_1^+ \hat{\sigma}_2^-\hat{\Lambda}^{(0)}_3 \hat{\Lambda}^{(1)}_4 \hat{\sigma}_5^+ + \text{H.c.}\ , \nonumber \\
    &(c)\quad \hat{\Lambda}^{(0)}_0 \hat{\Lambda}^{(1)}_1 \hat{\sigma}_2^+ \hat{\Lambda}^{(0)}_3 \hat{\sigma}_4^+ \hat{\sigma}_5^- + \text{H.c.}\ , \quad
    (d)\quad \hat{\Lambda}^{(0)}_0 \hat{\Lambda}^{(1)}_1 \hat{\sigma}_2^+ \hat{\Lambda}^{(0)}_3 \hat{\Lambda}^{(1)}_4 \hat{\sigma}_5^+ + \text{H.c.}\ ,
\end{align}
with each term diagonalized by a different $G_i$ operator,
\begin{equation}
    G_{(a)} \ = \ \begin{gathered}
    \includegraphics{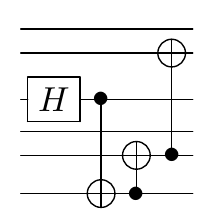}
  \end{gathered} \ , \quad G_{(b)} \ = \ \begin{gathered}
    \includegraphics{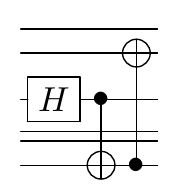}
  \end{gathered} \ ,\quad G_{(c)} \ = \ \begin{gathered}
    \includegraphics{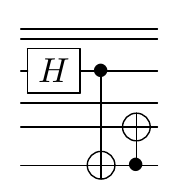}
  \end{gathered} \ , \quad G_{(d)} \ = \ \begin{gathered}
    \includegraphics{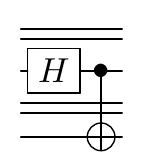}
  \end{gathered} \ .
  \label{eq:Gabcdop}
\end{equation}
The circuits that implement the time evolution induced by each term are the following,
{\allowdisplaybreaks \begin{subequations}
\begin{align}
    (a) \ : &\begin{gathered}
    \includegraphics{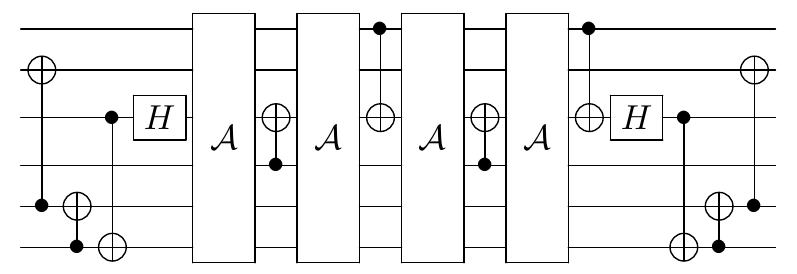}
  \end{gathered}, \nonumber  \\
  &\begin{gathered}
    \includegraphics{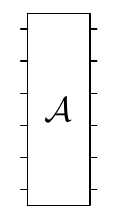}
  \end{gathered} = \begin{gathered}
    \includegraphics{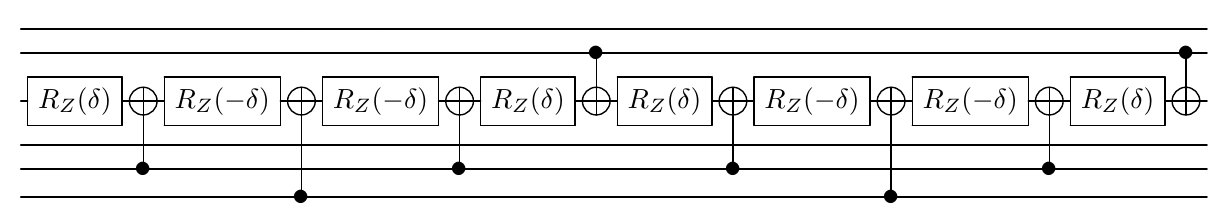}
  \end{gathered}, \\[0.5em]
    (b) \ : & \begin{gathered}
    \includegraphics{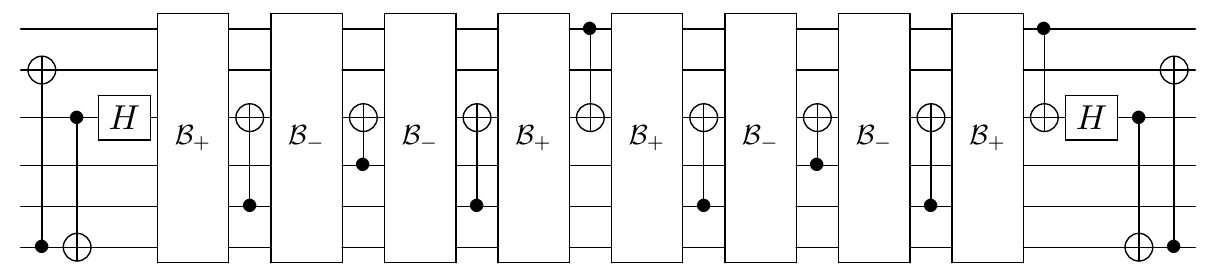}
  \end{gathered}, \nonumber  \\
  & \begin{gathered}
    \includegraphics{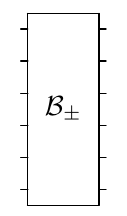}
  \end{gathered} = \begin{gathered}
    \includegraphics{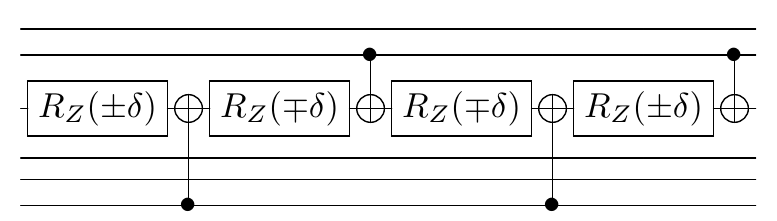}
  \end{gathered}, \\[0.5em]
    (c) \ : &\begin{gathered}
    \includegraphics{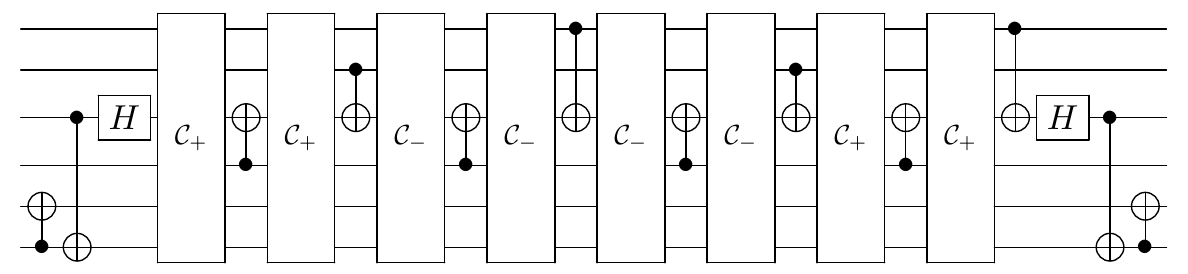}
  \end{gathered}, \nonumber  \\
  & \begin{gathered}
    \includegraphics{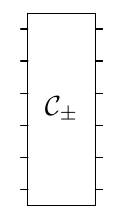}
  \end{gathered} = \begin{gathered}
    \includegraphics{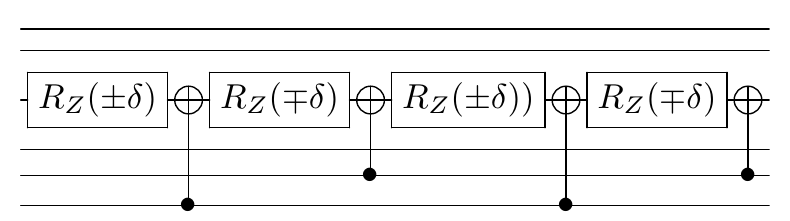}
  \end{gathered}, \\[0.5em]
    (d) \ : & \begin{gathered}
     \includegraphics{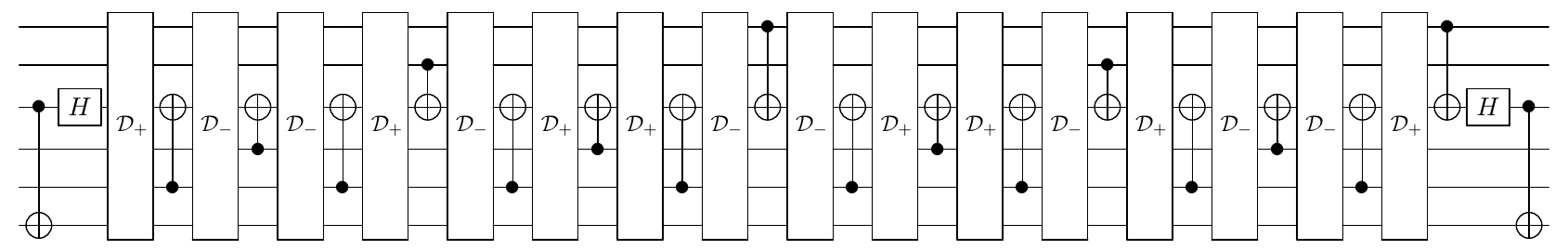}
  \end{gathered}, \nonumber \\
  & \begin{gathered}
     \includegraphics{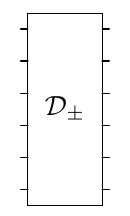}
  \end{gathered} = \begin{gathered}
     \includegraphics{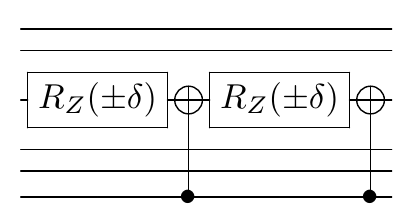}
  \end{gathered},
\end{align}
\end{subequations}}
with $\delta=-tV/8$.
The resources required to execute these circuits are shown in Table~\ref{tab:costsQ2Q5}, 
where the $g$ terms are also included, as they have a similar structure. There are two exceptions, since there are terms with five and six $\hat{\sigma}^{\pm}$. For example,
\begin{subequations}
\begin{align}
    & e^{itg(\hat{\Lambda}^{(0)}_0\hat{\sigma}_1^-\hat{\sigma}_2^-\hat{\sigma}_3^+\hat{\sigma}_4^-\hat{\sigma}_5^- + \text{H.c.})} = \begin{gathered}
    \includegraphics{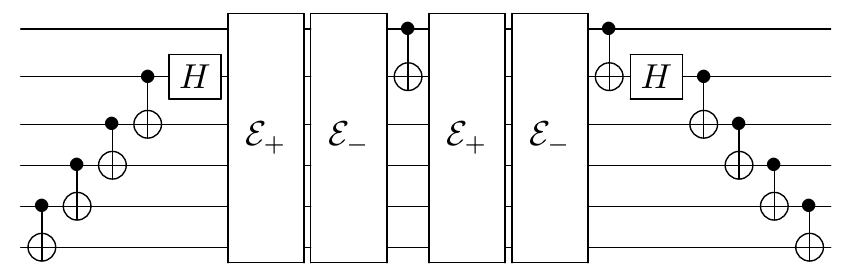}
  \end{gathered}, \nonumber  \\
  &\begin{gathered}
    \includegraphics{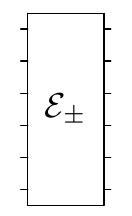}
  \end{gathered} = \begin{gathered}
  \includegraphics{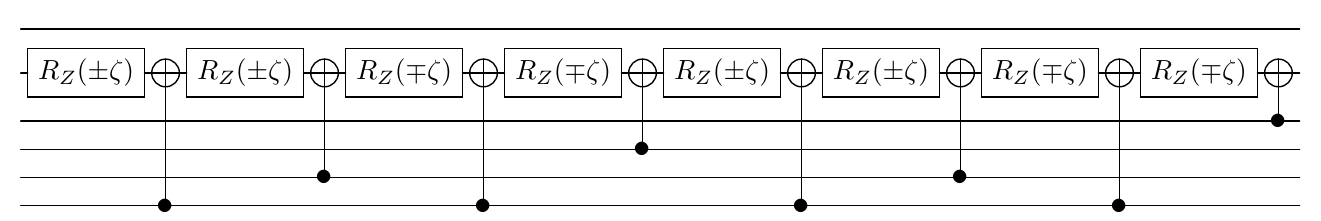}
    \end{gathered}, \\[0.5em]
    &e^{itg(\hat{\sigma}_0^-\hat{\sigma}_1^+\hat{\sigma}_2^+\hat{\sigma}_3^+\hat{\sigma}_4^-\hat{\sigma}_5^- + \text{H.c.})} = \begin{gathered}
    \includegraphics{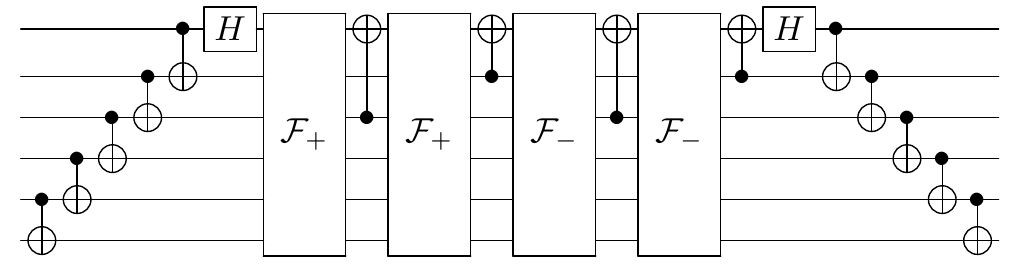}
  \end{gathered}, \nonumber \\
  &\begin{gathered}
    \includegraphics{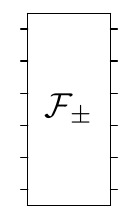}
  \end{gathered} = \begin{gathered}
  \includegraphics{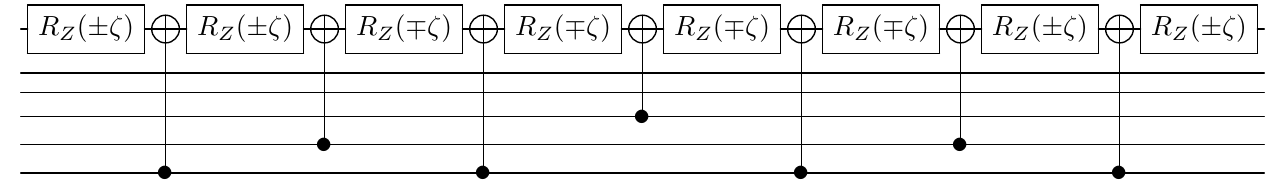}
    \end{gathered},
\end{align}
\end{subequations}
with $\zeta=-tg/16$. 
Note that these decompositions are not unique, as different definitions of the $G$ operators are possible, and other forms may be more efficient
depending on the connectivity of the quantum computer.
\begin{table}[!ht]
\centering
\begin{tabularx}{0.5\columnwidth}{rrYYY} 
\hline\hline
 & & $H$ & $R_Z$ & CNOT \\
\hline
& $\hat{\Lambda}^{(0)}_0 \hat{\sigma}_1^+ \hat{\sigma}_2^-\hat{\Lambda}^{(0)}_3 \hat{\sigma}_4^+ \hat{\sigma}_5^- + \text{H.c.}$ & 2 & 32 & 42 \\
& $\hat{\Lambda}^{(0)}_0 \hat{\sigma}_1^+ \hat{\sigma}_2^-\hat{\Lambda}^{(0)}_3 \hat{\Lambda}^{(1)}_4 \hat{\sigma}_5^+ + \text{H.c.}$ & 2 & 32 & 44 \\
& $\hat{\Lambda}^{(0)}_0 \hat{\Lambda}^{(1)}_1 \hat{\sigma}_2^+ \hat{\Lambda}^{(0)}_3 \hat{\sigma}_4^+ \hat{\sigma}_5^- + \text{H.c.}$ & 2 & 32 & 44 \\
& $\hat{\Lambda}^{(0)}_0 \hat{\Lambda}^{(1)}_1 \hat{\sigma}_2^+ \hat{\Lambda}^{(0)}_3 \hat{\Lambda}^{(1)}_4 \hat{\sigma}_5^+ + \text{H.c.}$ & 2 & 32 & 50 \\ \hline
(*) & $\hat{\Lambda}^{(0)}_0\hat{\sigma}_1^-\hat{\sigma}_2^-\hat{\Lambda}^{(0)}_3\hat{\sigma}_4^+\hat{\sigma}_5^+ + \text{H.c.}$ & 2 & 32 & 42 \\
(*) & $\hat{\Lambda}^{(0)}_0\hat{\sigma}_1^-\hat{\sigma}_2^-\hat{\Lambda}^{(0)}_3\hat{\Lambda}^{(0)}_4\hat{\sigma}_5^+ + \text{H.c.}$ & 2 & 32 & 44 \\
 & $\hat{\Lambda}^{(0)}_0\hat{\sigma}_1^-\hat{\sigma}_2^-\hat{\sigma}_3^+\hat{\Lambda}^{(0)}_4\hat{\sigma}_5^- + \text{H.c.}$ & 2 & 32 & 42 \\
 & $\hat{\Lambda}^{(0)}_0\hat{\sigma}_1^-\hat{\sigma}_2^-\hat{\sigma}_3^+\hat{\sigma}_4^-\hat{\sigma}_5^- + \text{H.c.}$ & 2 & 32 & 42 \\
(*) & $\hat{\Lambda}^{(0)}_0\hat{\Lambda}^{(0)}_1\hat{\sigma}_2^-\hat{\Lambda}^{(0)}_3\hat{\sigma}_4^+\hat{\sigma}_5^+ + \text{H.c.}$ & 2 & 32 & 44 \\
(*) & $\hat{\Lambda}^{(0)}_0\hat{\Lambda}^{(0)}_1\hat{\sigma}_2^-\hat{\Lambda}^{(0)}_3\hat{\Lambda}^{(0)}_4\hat{\sigma}_5^+ + \text{H.c.}$ & 2 & 32 & 50 \\
& $\hat{\Lambda}^{(0)}_0\hat{\Lambda}^{(0)}_1\hat{\sigma}_2^-\hat{\sigma}_3^+\hat{\Lambda}^{(0)}_4\hat{\sigma}_5^- + \text{H.c.}$ & 2 & 32 & 44 \\
& $\hat{\Lambda}^{(0)}_0\hat{\Lambda}^{(0)}_1\hat{\sigma}_2^-\hat{\sigma}_3^+\hat{\sigma}_4^-\hat{\sigma}_5^- + \text{H.c.}$ & 2 & 32 & 42 \\
& $\hat{\sigma}_0^-\hat{\Lambda}^{(0)}_1\hat{\sigma}_2^+\hat{\Lambda}^{(0)}_3\hat{\sigma}_4^+\hat{\sigma}_5^+ + \text{H.c.}$ & 2 & 32 & 42 \\
& $\hat{\sigma}_0^-\hat{\Lambda}^{(0)}_1\hat{\sigma}_2^+\hat{\Lambda}^{(0)}_3\hat{\Lambda}^{(0)}_4\hat{\sigma}_5^+ + \text{H.c.}$ & 2 & 32 & 44 \\
& $\hat{\sigma}_0^-\hat{\Lambda}^{(0)}_1\hat{\sigma}_2^+\hat{\sigma}_3^+\hat{\Lambda}^{(0)}_4\hat{\sigma}_5^- + \text{H.c.}$ & 2 & 32 & 42 \\
& $\hat{\sigma}_0^-\hat{\Lambda}^{(0)}_1\hat{\sigma}_2^+\hat{\sigma}_3^+\hat{\sigma}_4^-\hat{\sigma}_5^- + \text{H.c.}$ & 2 & 32 & 42 \\
& $\hat{\sigma}_0^-\hat{\sigma}_1^+\hat{\sigma}_2^+\hat{\Lambda}^{(0)}_3\hat{\sigma}_4^+\hat{\sigma}_5^+ + \text{H.c.}$ & 2 & 32 & 42 \\
& $\hat{\sigma}_0^-\hat{\sigma}_1^+\hat{\sigma}_2^+\hat{\Lambda}^{(0)}_3\hat{\Lambda}^{(0)}_4\hat{\sigma}_5^+ + \text{H.c.}$ & 2 & 32 & 42 \\
& $\hat{\sigma}_0^-\hat{\sigma}_1^+\hat{\sigma}_2^+\hat{\sigma}_3^+\hat{\Lambda}^{(0)}_4\hat{\sigma}_5^- + \text{H.c.}$ & 2 & 32 & 42 \\
& $\hat{\sigma}_0^-\hat{\sigma}_1^+\hat{\sigma}_2^+\hat{\sigma}_3^+\hat{\sigma}_4^-\hat{\sigma}_5^- + \text{H.c.}$ & 2 & 32 & 42 \\ \hline
& Total & 32 & 512 & 688 \\ \hline\hline
\end{tabularx}
\caption{Resource requirements to perform one Trotter step of time evolution for each two-mode-pair term in the Hamiltonian given in Eq.~\eqref{eq:Q2Q5H4}. 
The ``(*)''  items from the $g$ terms have the same structure as those from the $2V$ terms,
and therefore can be grouped together.
}
\label{tab:costsQ2Q5}
\end{table}

\section{A Sign Problem in Quantum Simulations: Trotterized Time Evolution and Device Noise} 
\label{app:stnprob}
\noindent
By considering Trotterized time evolution, 
we have identified a potential sign problem in quantum simulations 
of observables determined from many contributing states.
Disturbances to cancellations between contributing amplitudes 
results from a combination of device errors, 
which can be Pauli-twirled towards incoherence (contributing statistical noise)~\cite{Wallman:2016}, 
and systematic algorithmic errors, such as from Trotterization.
In the case of statistical noise, this sign problem will manifest as a signal-to-noise (StN) problem.
Sign and StN problems plague classical lattice gauge theory calculations of systems at 
finite density or with finite baryon number~\cite{Parisi:1983ae,Lepage:1989hd,Beane:2009kya,Beane:2009gs,Beane:2009py,Endres:2011jm,NPLQCD:2012mex,Grabowska:2012ik,Beane:2014oea,Wagman:2016bam,Wagman:2017xfh}. 
The hope continues to be that some or all of these types of problems will be absent or mitigated in quantum simulations that will allow for progress to be made that is not possible for classical lattice QCD, 
including ground- and excited-state preparation, as well as time-evolution, of nuclei 
(even modestly sized nuclei).
A similar hope is present for simulation of high-energy fragmentation of hadrons and nuclei,
as created experimentally in high-energy colliders,
where a large number of interfering coupled channels limits robust classical simulation capabilities.

In this appendix, the sign problem 
we have identified in Trotter evolution, 
which can be seen, for example, in Fig.~\ref{fig:SimulationN10wvf1}c,
is dissected further in the system with $\Omega=N=8$ modes, 
and for the set-4 couplings.
\begin{figure}[!thb]
\centering
\includegraphics[width=\columnwidth]{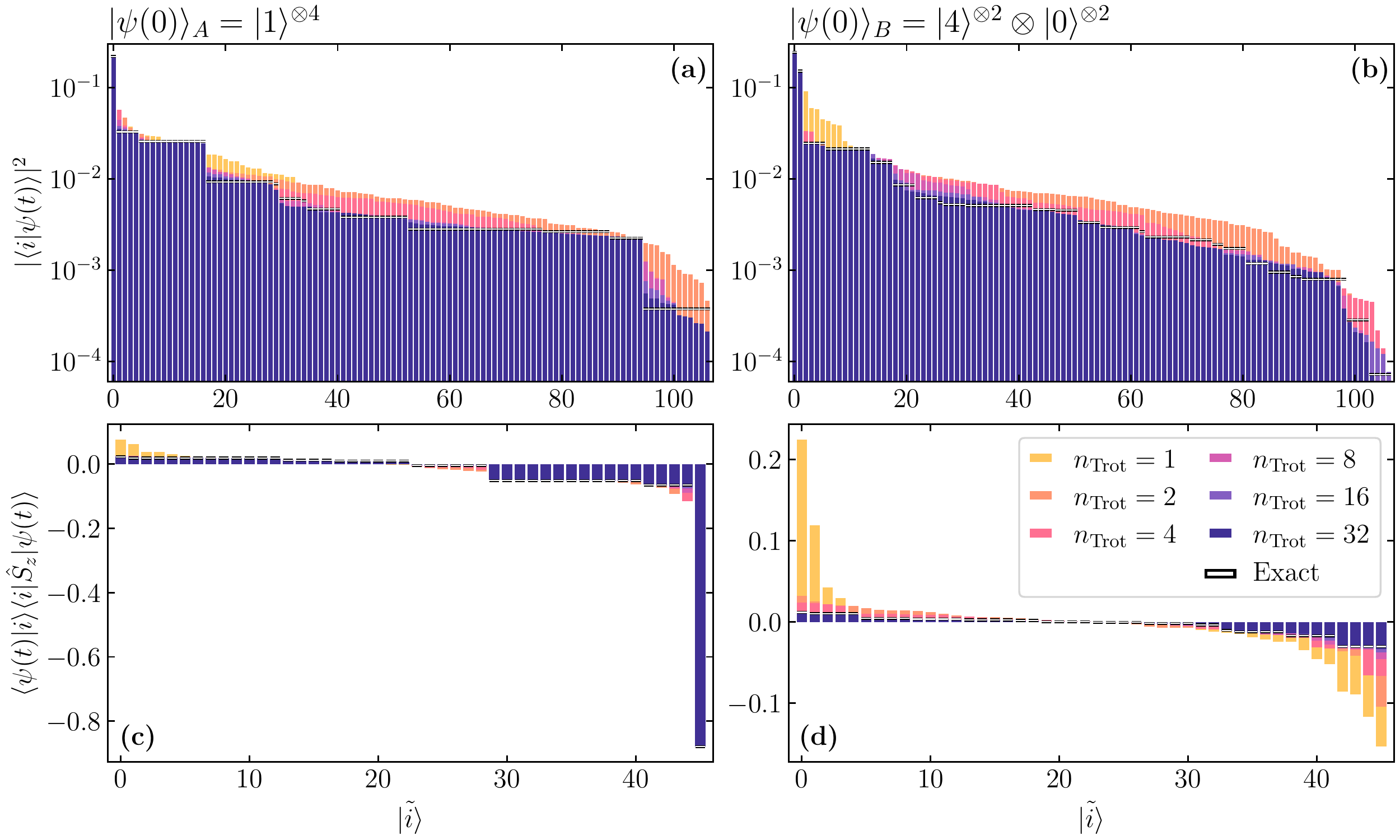} 
\caption{{\bf(a,b)} Sorted non-zero values of probability density, $|\langle i | \psi\rangle|^2$, for each basis state in the wavefunction evolved to $t=0.4$ with set-4 couplings (not all states $| i \rangle$ are shown, only those with non-zero contribution, labeled as $\tilde{| i \rangle}$).
{\bf(c,d)} Sorted non-zero values of $z$-component of spin density, $\langle \psi | i \rangle\langle i |\hat{S}_z| \psi\rangle$.
The left panels {\bf(a,c)} correspond to the $|\psi(0)\rangle_A$ initial state, while the right panels {\bf(b,d)} correspond to the $|\psi(0)\rangle_B$ initial state.
}
\label{fig:N8StN}
\end{figure}
The panels (a,b) of Fig.~\ref{fig:N8StN} show the sorted non-zero probability densities, $|\langle i | \psi\rangle|^2$ (with $|i\rangle$ being all the states in the computational basis, which, in base-10, are 
$\{0,\ldots,5^{\Omega/2}-1\}$), associated with time-evolved states to $t=0.4$, starting from $|\psi(0)\rangle_A$ and $|\psi(0)\rangle_B$.
While somewhat different, as expected, the range of values are similar.
The panels (c,d) show the sorted non-zero $z$-component spin densities, $\langle \psi | i \rangle\langle i |\hat{S}_z| \psi\rangle$, which are markedly different for the two initial states.  
There is a single large contribution when starting from $|\psi(0)\rangle_A$ that is not present when starting from $|\psi(0)\rangle_B$.
The mean value and standard deviations of each element of $\langle \psi | i \rangle\langle i |\hat{S}_z| \psi\rangle$ are $\bar x_A= -0.03$ and $\sigma_A=0.13$, providing an expectation value of $\langle \hat S_z \rangle_A = -1.4$, and 
$\bar x_B= -0.004$ and $\sigma_B=0.011$, giving $\langle \hat S_z \rangle_B = -0.17$.

Given that the range and magnitude of the probability densities in the wavefunction are comparable between the two initial condition, systematic errors introduced by Trotter evolution, or whatever algorithm is used for time evolution, and by the fidelity of the quantum device, 
in the probability distributions will be comparable.
The same is true for the $z$-component spin density.  
However, the order of magnitude difference between the two spin densities,
and particularly cancellations among contributions,
means that spin densities from $|\psi(0)\rangle_B$ 
will be significantly more impacted than those from $|\psi(0)\rangle_A$ by time-evolution inaccuracies.
The differing relative impacts of Trotter errors 
on the spin densities can be seen clearly in the lower panels in Fig.~\ref{fig:N8StN}, in particular, in the contrast between the lower-left (c) and lower-right (d) panels.
This effect is responsible for the differing convergence 
of the spin density (and other observables) between 
$|\psi(0)\rangle_B$ and $|\psi(0)\rangle_A$
with decreasing Trotter step size that is seen in Fig.~\ref{fig:SimulationN10wvf1} and other such figures.

The behavior of Trotter errors in simulations starting from different initial states have been previously considered in the context of complexity bounds~\cite{Burak:2021npj,Su:2020gzf,Dong:2021Quant,Changhao:2022npj}.
It was found that the general error bounds can be improved for time evolution from states in the low-energy sector.

The identified sign problem is different in its nature to the StN problem 
in classical lattice QCD calculations at finite baryon number.
There, the spectrum resulting from spontaneous breaking of the global chiral symmetries 
furnishes light pseudo-Goldstone bosons that contribute low-lying states to variance correlation functions 
associated with (multi-)baryon correlation functions. 
In the present case, 
the cancellations between (small) amplitudes that produce small expectation values are disturbed by simulation errors, either systematic algorithm errors or device errors.
Improving the StN problem  at late times in baryon correlation functions determined with lattice QCD calculations requires exponentially large classical resources with increasing time.
In contrast, the sign problem we have identified here originating from Trotterized time evolution can be systematically reduced by using smaller Trotter time intervals, which requires resources that scale with the number of Trotter steps.
In the context of quantum simulations, 
this sign problem requires more extensive study to estimate its impact on observables of interest, including the impact of transitions to chaotic evolution~\cite{2019npjQI...5...78S}.

\clearpage
\bibliography{bib}

\end{document}